\documentclass[11pt]{article}

\usepackage[margin=1in]{geometry}
\usepackage{setspace}
\usepackage{amsmath, amssymb, amsthm}
\usepackage{mathtools}
\usepackage{bm}
\usepackage{authblk}
\numberwithin{equation}{section}

\usepackage{graphicx}
\usepackage{float}
\usepackage{tikz}
\usetikzlibrary{calc, arrows.meta, positioning}
\usepackage{booktabs,multirow,threeparttable}

\usepackage[authoryear]{natbib}
\usepackage[colorlinks=true, linkcolor=blue, citecolor=blue, urlcolor=blue]{hyperref}

\newtheorem{assumption}{Assumption}[section]
\newtheorem{proposition}{Proposition}[section]
\newtheorem{lemma}{Lemma}[section]
\newtheorem{theorem}{Theorem}[section]
\newtheorem{remark}{Remark}[section]

\title{Cause-specific structural nested cumulative survival time models for competing risks, with application to the Antibody Mediated Prevention trials}
\author[1]{Yuxiao Li\thanks{The first two authors contributed equally.}}
\author[2]{Yuhao Deng$^\ast$}
\author[2]{Bo Zhang\thanks{Assistant Professor of Biostatistics, Vaccine and Infectious Disease Division, Fred Hutchinson Cancer Center. Email: {\tt bzhang3@fredhutch.org}. }}

\affil[1]{Department of Statistics, University of Washington}
\affil[2]{Vaccine and Infectious Disease Division, Fred Hutchinson Cancer Center}
\date{}

\begin{document}

\maketitle    

\begin{abstract}
The Antibody Mediated Prevention (AMP) trials randomized participants to VRC01, a broadly neutralizing antibody, or placebo; the primary analysis estimated prevention efficacy against overall HIV-1 acquisition and against acquisition of VRC01-sensitive strains. All participants were offered daily oral pre-exposure prophylaxis (PrEP), and PrEP uptake in a trial may differ substantially from uptake in other settings. It is therefore of interest to estimate the effect of oral PrEP on acquisition of VRC01-sensitive and VRC01-insensitive strains, as well as the counterfactual cumulative incidence of such acquisition under no or limited PrEP uptake. To address these questions, we extend the structural nested cumulative survival time model (SNCSTM), a causal survival model for time-varying treatments, to a cause-specific version. We show that the cause-specific SNCSTM is closely related to the Fine--Gray subdistribution hazard model, and we study identification under two frameworks. The first, based on sequential ignorability, applies when the time-varying treatment is distinct from the randomized assignment, as with oral PrEP use in AMP. The second, based on an instrumental variable, applies when the time-varying treatment represents compliance with the randomized assignment, and it accommodates unmeasured confounding between the time-varying treatment and event time. Under both frameworks, we propose consistent and asymptotically normal estimators of the structural parameters. We apply the methods to HVTN 704/HPTN 085, one of the two AMP trials, to estimate the effects of oral PrEP use and of VRC01 infusions administered every 8 weeks on acquisition of VRC01-sensitive and VRC01-insensitive HIV-1 strains. We also estimate the counterfactual strain-specific cumulative incidence curves under no oral PrEP use and under full adherence to the scheduled VRC01 infusions.
\end{abstract}

\noindent\textbf{Keywords:} Fine--Gray competing risk model; G-estimation; Subdistribution hazard; Sequential ignorability; Instrumental variable.

\section{Introduction}
\label{sec: intro}

\subsection{The Antibody Mediated Prevention trials}
\label{subsec: intro AMP}
The Antibody Mediated Prevention (AMP) trials (HVTN 703/HPTN 081 and HVTN 704/HPTN 085) were a pair of landmark Phase 2b, randomized, placebo-controlled HIV prevention studies designed to test whether VRC01, a CD4-binding site (CD4bs)-targeting broadly neutralizing monoclonal antibody (bnAb), could prevent HIV-1 acquisition. VRC01 was administered as an intravenous infusion every eight weeks to HIV-uninfected individuals at elevated risk of infection \citep{corey2021two}. HVTN 704/HPTN 085 enrolled cisgender men and transgender women who have sex with men in the Americas and Europe, whereas HVTN 703/HPTN 081 enrolled cisgender women in sub-Saharan Africa. Participants were randomized to one of three arms: placebo, VRC01 at 10 mg per kilogram (low-dose arm), or VRC01 at 30 mg per kilogram (high-dose arm). The study found VRC01 did not significantly reduce overall HIV-1 acquisition compared with placebo (prevention efficacy [PE]: 26.6\%; 95\% confidence interval [CI]: -11.7\% to 51.8\%; $P=0.15$ in HVTN 704/HPTN 085, and 8.8\%; 95\% CI: -45.1\% to 42.6\%; $P=0.70$ in HVTN 703/HPTN 081). 

%sieve analyses provided proof-of-concept that bnAbs can be effective against viral strains that are sensitive to them (estimated pooled PE = 75.4\%; 95\% CI, 45.5\% to 88.9\%).

\subsection{Layered design; cause-specific risk and prevention efficacy}
\label{subsec: intro problem}
The AMP trials have two distinct features. First, infecting HIV-1 strains were isolated and tested for sensitivity to the study product, VRC01. HIV-1 strains are highly heterogeneous and exhibit a wide range of sensitivity to VRC01, so in addition to overall prevention efficacy, \citet[Figure~2]{corey2021two} reported the prevention efficacy of the pooled VRC01 groups versus placebo against sensitive strains (estimated pooled PE = 75.4\%; 95\% CI: 45.5\% to 88.9\%), defined by an in vitro 80\% inhibitory concentration $\mathrm{IC}_{80} < 1\,\mu$g/mL. Because a participant can acquire HIV-1 only once during follow-up, acquisition of a VRC01-sensitive strain and acquisition of an insensitive strain are mutually exclusive, so each is a competing event for the other.

Second, in AMP and many other efficacy trials of HIV-1 vaccines and bnAbs, all participants receive standard-of-care HIV prevention regardless of randomization arm, including facilitated access to pre-exposure prophylaxis (PrEP), in accordance with ethics committee and institutional review board recommendations and national PrEP guidelines \citep{bailey2013social,gilbert2017basis,gilbert2019ongoing,janes2019taking}. Daily oral PrEP effectively prevents HIV-1 acquisition, though high efficacy requires strict adherence \citep{baeten2012antiretroviral,fonner2016effectiveness,mccormack2016pre}. We refer to such a design, in which participants in all arms receive a background intervention and an investigational intervention is added on top of this common layer, as a \emph{layered design}. Primary analysis of the AMP trials compared the bnAb arm and the placebo arm as randomized; this intention-to-treat (ITT) analysis reflects the efficacy of bnAbs plus PrEP versus PrEP alone in a trial environment, where PrEP use may not reflect typical use outside the trial.

\subsection{A cause-specific structural nested cumulative survival time model}
\label{subsec: intro our contribution}
Our goal is twofold. First, we estimate the causal effect of oral PrEP uptake on HIV-1 acquisition with a VRC01-sensitive strain in the AMP trials, where acquisition with a VRC01-insensitive strain is a competing event. Second, we estimate the counterfactual cumulative incidence of acquiring a VRC01-sensitive virus under a hypothetical scenario in which daily oral PrEP is withheld, or its uptake is substantially lower than observed in the AMP trial. Contrasting these counterfactual cause-specific cumulative incidence curves in the low- and high-dose arms with that in the placebo arm yields the prevention efficacy of VRC01 when daily oral PrEP is withheld, or its uptake is low. This counterfactual PE complements the original intention-to-treat analysis and helps assess the potential utility of VRC01 where PrEP uptake is low.

Structural nested cumulative survival time models (SNCSTMs) provide a principled approach to estimating the effect of a time-varying treatment process---here, oral PrEP uptake---on a time-to-event endpoint. Given an estimate of this effect, one can ``blip down'' the treatment process to recover the counterfactual cumulative incidence curve under a hypothetical scenario in which treatment is withheld \citep{robins2005structural,seaman2020adjusting,shi2021instrumental,ying2023structural,michiels2025adjusting}. Inference for these models has been developed under both sequential ignorability and instrumental variable assumptions \citep{seaman2020adjusting,ying2023structural,michiels2025adjusting}. To our knowledge, however, no variant of the SNCSTM accommodates competing events.

In this paper, we propose a cause-specific structural nested cumulative survival time model (cs-SNCSTM) and show that it relates to the Fine--Gray subdistribution hazards model in much the same way that the SNCSTM relates to Aalen's additive hazards model. We study identification and estimation under two complementary frameworks: one based on sequential ignorability (SI) and the other based on instrumental variables (IVs). For each framework, we establish the theoretical properties of the proposed estimators of the structural parameters and evaluate their finite-sample performance through extensive simulation studies.

The SI framework suits our analysis of oral PrEP. Because participants were blinded to VRC01 assignment, PrEP uptake decisions were unlikely to depend on the randomization arm, so VRC01 assignment cannot serve as an instrument for PrEP uptake; we instead adjust for measured covariates that may affect both PrEP uptake and HIV-1 acquisition. Applying the cs-SNCSTM to HVTN 704/HPTN 085, we use SI-based methods to estimate the effect of oral PrEP on the acquisition of VRC01-sensitive and VRC01-insensitive HIV-1 strains, together with the counterfactual cause-specific cumulative incidence functions under no PrEP uptake.

The IV framework suits settings in which randomized assignment drives subsequent treatment. In the AMP trials, VRC01 infusions were administered every 8 weeks and are themselves a time-varying treatment strongly associated with the initial randomized assignment, as in the treatment-switching and time-varying compliance settings of \citet{ying2023structural} and \citet{michiels2025adjusting}. Unlike the SI framework, the IV framework accommodates unmeasured confounding between treatment uptake and the outcome. Returning to HVTN 704/HPTN 085, we use IV-based methods to estimate the effect of VRC01 infusions on the acquisition of VRC01-sensitive and VRC01-insensitive strains, together with the cause-specific cumulative incidence functions under full adherence to the infusion schedule.

Code accompanying the paper is available via GitHub:
\url{https://github.com/Jacobs-Lee/cause-specific-SNCSTMs-simulation}.
\section{Models}
\label{sec: models}

\subsection{Structural nested cumulative survival time model}
\label{subsec: review of SNCSTM}
We begin by reviewing the SNCSTM in the absence of competing events \citep{seaman2020adjusting, shi2021instrumental, ying2023structural,michiels2025adjusting}. We treat the data for subjects $i = 1,\ldots,n$ as independent and identically distributed. For each subject, let $\tilde T_i$ denote the time-to-event endpoint of interest (e.g., time to HIV-1 acquisition) and $C_i$ the censoring time, and define the observed event time as $T_i = \min\{\tilde T_i, C_i\}$. The event indicator is $\Delta_i = \mathbf{1}\{\tilde T_i \leq C_i\}$, so that $\Delta_i = 1$ indicates the event was observed and $\Delta_i = 0$ indicates censoring. We define the counting process for the underlying event time as $\tilde N_i(t) = \mathbf{1}\{\tilde T_i \leq t\}$ and the observed counting process as $N_i(t) = \mathbf{1}\{T_i \leq t, \Delta_i = 1\}$. The associated at-risk processes are $\tilde Y_i(t) = \mathbf{1}\{\tilde T_i \geq t\}$ and $Y_i(t) = \mathbf{1}\{T_i \geq t\}$.

In addition, we assume the time-varying treatment process is measured at discrete, scheduled follow-up times $0 = t_1 < \cdots < t_{K+1} = \tau$, where $\tau$ denotes the end of follow-up. Let $D_{i}(t) \in \mathbb{R}$ denote the time-varying treatment uptake, and $D_{i,k}=D_{i}(t_k)$ denote the treatment received at time $t_k$. By convention, we define $D_{i,k} = 0$ if $t_k > \tilde T_i$. We assume $D_i(t) = D_{i,k}$ for $t \in [t_k, t_{k+1})$. The treatment uptake history up to time $t_k$ is denoted $\bar D_{i,k} = \{D_{i,l} : 1 \leq l \leq k\}$, and the full treatment uptake history is denoted $\bar D_i = \{D_{i,l} : 1 \leq l \leq K\}$.  Finally, we let $Z_i$ denote randomized treatment assignment at baseline and $L_i$ denote baseline covariates, such as prognostic factors. In the treatment-switching example of \citet{ying2023structural} and \citet{michiels2025adjusting}, $Z_i$ denotes the initial randomized treatment assignment and $D_{i,k}$ the treatment actually received at time $t_k$. In our oral PrEP analysis of the AMP trials, $Z_i$ denotes the randomized bnAb assignment and $D_{i,k}$ indicates oral PrEP uptake at time $t_k$. Although both examples fit within the SNCSTM framework, the relationship between $Z_i$ and $D_{i,k}$ differs: in the treatment switching example, $Z_i$ strongly influences $D_{i,k}$, whereas in the oral PrEP example, $Z_i$ is unlikely to be associated with $D_{i,k}$. As discussed in Sections~\ref{sec: identificaito and estimation SI} and \ref{sec: inference under IV}, these settings therefore call for different identification strategies and inferential methods. Throughout the remainder of the article, we drop the subscript $i$ when no confusion arises.

We next introduce the potential outcomes (PO) framework underpinning the
SNCSTM. Let $\tilde T(z, \bar d_k, \bar 0)$ denote the potential time to
the outcome had the person been assigned $Z = z$, with treatment uptake
set to $\bar d_k$ up to time $t_k$ and to $0$ thereafter. By the
consistency assumption, $\tilde T = \tilde T(z, \bar d_k, \bar 0)$ with
probability one for individuals assigned $Z = z$ whose observed treatment
uptake history satisfies $\bar D_k = \bar d_k$ up to time $t_k$ and
equals $0$ thereafter. We similarly write $\tilde N(t; z, \bar d_k, \bar
0)$ and $\tilde Y(t; z, \bar d_k, \bar 0)$ for the associated potential
counting process and potential at-risk process. Under the SNCSTM
framework, intervening on treatment uptake affects the time-to-event
outcome only after the intervention time. Consequently, the events $\{\tilde{T} \geq t_k\}$ and $\{\tilde{T}(Z, \bar{D}_{k-1}, \bar{0}) \geq t_k\}$ coincide, and for all $t \in [t_k, t_{k+1})$, the events $\{\tilde{T} \geq t\}$ and $\{\tilde{T}(Z, \bar{D}_k, \bar{0}) \geq t\}$ coincide.

For notational convenience, we define $\bar D_0=\emptyset$. A structural nested cumulative survival time model, or SNCSTM, contrasts the counterfactual survival probability at time $t$ under the observed treatment uptake history up to time $t_k$ followed by zero treatment uptake thereafter, with that under the observed treatment uptake history only up to time $t_{k-1}$ followed by zero treatment uptake thereafter, among participants who have not yet experienced the event at time $t_k$, with treatment uptake history $\bar D_k$, assigned treatment $Z$, and with baseline covariates $L$. Specifically, for $t\geq t_k$, model $\mathcal M_k$ is
\begin{equation}
\frac{
\mathbb{P}\left[
\tilde T(Z, \bar D_k, \bar 0)\geq t
\mid
\bar D_k, Z, L,\tilde T\geq t_k
\right]
}{
\mathbb{P}\left[
\tilde T(Z, \bar D_{k-1},\bar 0)\geq t
\mid
\bar D_k, Z, L, \tilde T\geq t_k
\right]
}
=
\exp\left\{
-\int_{t_k}^{t \wedge t_{k+1}} D_k dB_D(s)
\right\},
\label{eq:SNCSTM}
\end{equation}
where $dB_D(s)$ represents the instantaneous causal effect of treatment uptake, with positive values indicating increased risk. A simple model could take $B_D(t) = \beta t$, and Equation \eqref{eq:SNCSTM} reduces to
\begin{equation}
\frac{
\mathbb{P}\left[
\tilde T(Z, \bar D_k, \bar 0)\geq t
\mid
\bar D_k, Z, L,\tilde T\geq t_k
\right]
}{
\mathbb{P}\left[
\tilde T(Z, \bar D_{k-1},\bar 0)\geq t
\mid
\bar D_k, Z, L, \tilde T\geq t_k
\right]
}
=
\exp\left\{
-\beta D_k (t \wedge t_{k+1} - t_k)
\right\}.
\end{equation}

Taking logarithms on both sides of \eqref{eq:SNCSTM} and differentiating with respect to $t$, we can rewrite the model $\mathcal{M}_k$ equivalently as the following additive hazard difference model:
\begin{equation}
\begin{aligned}
&\mathbb{E}\left[
d\tilde N(t;Z, \bar D_k, \bar 0)
\mid
\bar D_k,Z,L,\tilde T\geq t_k,
\tilde T(Z, \bar D_k, \bar 0)\geq t
\right]- D_k dB_D(t) \\
=
&\mathbb{E}\left[
d\tilde N(t;Z, \bar D_{k-1}, \bar 0)
\mid
\bar D_{k},Z,L,\tilde T\geq t_k,
\tilde T(Z, \bar D_{k-1}, \bar 0)\geq t
\right]
,
\qquad t\in[t_k,t_{k+1}),
\end{aligned}
\label{eq:additive-hazard-difference}
\end{equation}
where $d\tilde N(t;Z, \bar D_k, \bar 0)$ and $d\tilde N(t;Z, \bar D_{k-1}, \bar 0)$ denote the jump of the potential counting process, under $(Z, \bar D_k, \bar 0)$ and $(Z, \bar D_{k-1}, \bar 0)$, respectively. Equation \eqref{eq:additive-hazard-difference} states that the current treatment uptake $D_k$ has an additive effect on the hazard over the interval $[t_k,t_{k+1})$ conditional on $\bar D_k$, $Z$, $L$,
$\tilde T\geq t_k$, and the respective counterfactual survival to $t$.

\subsection{A cause-specific generalization of the SNCSTM and its connection to Fine--Gray subdistribution hazards model}
\label{subsec: introduce SNCS^2TM}
We generalize the SNCSTM to a cause-specific version that targets a cause-specific time-to-event endpoint in the presence of competing events. Suppose there are $J$ types of events, and let $V \in \{1, \ldots, J\}$ denote the type of the event that occurs. Without loss of generality, we focus on the cumulative incidence of the type-$1$ event. In the AMP trial, for instance, there are two types of HIV-1 acquisition ($J = 2$): $V = 1$ denotes acquisition of a VRC01-sensitive strain, and $V = 2$ denotes acquisition of a VRC01-insensitive strain.

For $j = 1, \ldots, J$, let $d\tilde{N}^j(t; z, \bar{d})$ denote the increment over $[t, t + dt)$ of the potential counting process that jumps at the time of the first event if that event is of type $j$, when $Z$ is set to $z$ and the treatment uptake process $\bar{D}$ is set to $\bar{d}$. The
corresponding potential counting process is
$\tilde N^j(t;z,\bar d)
=
\int_0^t d\tilde N^j(s;z,\bar d)$,
and the potential subdistribution risk indicator for type-$j$ event is
$\tilde Y^j(t;z,\bar d)=1 - \tilde N^j(t^-;z,\bar d)$,
where $t^-$ denotes the instant immediately preceding time $t$. Note that $\tilde Y^j(t;z,\bar d)$ retains subjects who have already experienced a competing event ($V \neq j$) in the risk set, consistent with the Fine--Gray subdistribution hazards framework. Following this notation, the potential event time and the potential event type are
$$
\begin{aligned}
%\tilde N(t;z,\bar d)=\sum_{j=1}^J\tilde N^j(t;z,\bar d),\quad
\tilde T(z,\bar d)=\inf \left\{t:\sum_{j=1}^J \tilde N^j(t;z,\bar d)=1\right\},\qquad
V(z,\bar d)=j ~\text{ if }~ d \tilde N^j(\tilde T(z,\bar d);z,\bar d) =1,
\end{aligned}
$$
respectively.
Under the observed treatment assignment \(Z\) and treatment uptake history \(\bar D\), we write the corresponding realized quantities as
\(\tilde N^j(t), \tilde Y^j(t), \tilde N(t), \tilde T\) and \(V\). Naturally, treatment received after an event cannot affect its occurrence or type.

\begin{assumption}[Consistency]
\label{assumption:consistency}
%For any $z$ and treatment history $\bar d$, if $Z=z$ and the observed treatment history satisfies $\bar D=\bar d$, then $d\tilde N^j(t)=d\tilde N^j(t;z,\bar d)$ for every cause $j$ and all $t\in[0,\tau]$, with probability one.
For every $t \in [t_k,t_{k+1})$ and every cause $j\in\{1,\ldots,J\}$, $d\tilde N^j(t)=d\tilde N^j(t;Z,\bar D_k,\underline{d}_k)$ with probability one for every future treatment trajectory, $\underline{d}_k$, after $t_k$.
\end{assumption}

For each $k$ and $t\geq t_k$, we propose the following cause-specific SNCSTM for the type-1 event under this potential counting process notation:
\begin{equation}
\frac{P(\tilde Y^1(t;Z,\bar D_k, \bar 0)=1\mid \bar D_k, Z, L,\tilde T\geq t_k )}{P(\tilde Y^1 (t;Z,\bar D_{k-1}, \bar 0)=1\mid \bar D_k, Z, L,\tilde T\geq t_k)}
=\exp\left\{-\int_{t_k}^{t\wedge t_{k+1}} D_kdB_D(s) \right\}.
\label{eq:cause_specific_SNCSTM}
\end{equation}
Model \eqref{eq:cause_specific_SNCSTM} assumes that switching treatment uptake from $0$ to $D_k$ changes the potential subdistribution survival function of the type-1 event, among participants who have not yet experienced any event by $t_k$, given treatment uptake history $\bar D_k$, assigned treatment $Z$, and with baseline covariates $L$. Conditioning on $\{\tilde T\geq t_k\}$ at $t_k$ is essential here, because $D_k$ is defined only when $\tilde{T} \geq t_k$, and this conditioning ensures the ratio of potential subdistribution survival functions remains well defined \citep{austin2020review}.

Just as the original SNCSTM is closely connected to the additive hazards model, the cause-specific SNCSTM is closely related to the Fine--Gray subdistribution hazards model. To see this, note that 
\[
\begin{aligned}
&-d\log \Bigl(
P\bigl(
\tilde Y^1(t;Z,\bar D_k,\bar 0)=1
\mid
\bar D_k,Z,L,\tilde T\geq t_k
\bigr)
\Bigr)
\\
%={}&
%\frac{
%P\bigl(
%d\tilde N^1(t;Z,\bar D_k,\bar 0)=1
%\mid
%\bar D_k,Z,L,\tilde T\geq t_k
%\bigr)
%}{
%P\bigl(
%\tilde Y^1(t;Z,\bar D_k,\bar 0)=1
%\mid
%\bar D_k,Z,L,\tilde T\geq t_k
%\bigr)
%}
%\\
%={}&
%P\left(
%d\tilde N^1(t;Z,\bar D_k,\bar 0)=1
%\;\middle|\;
%\begin{array}{l}
%\tilde Y^1(t;Z,\bar D_k,\bar 0)=1,\\
%\bar D_k,Z,L,\tilde T\geq t_k
%\end{array}
%\right)
%\\
={}&
\mathbb E\left[
d\tilde N^1(t;Z,\bar D_k,\bar 0)
\;\middle|\;
\bar D_k,Z,L,\tilde T\geq t_k,
\tilde Y^1(t;Z,\bar D_k,\bar 0)=1
\right].
\end{aligned}
\]

Taking logarithms of each side of equation \eqref{eq:cause_specific_SNCSTM} and differentiating with respect to $t$ yields
\begin{equation}
\begin{aligned}
&\mathbb{E}[d\tilde N^1(t;Z, \bar D_k, \bar 0)\mid \bar D_k, Z, L,\tilde T\geq t_k,\tilde Y^1(t;Z, \bar D_k, \bar 0)=1] -D_kdB_D(t) \\
=&\mathbb{E}[d\tilde N^1(t;Z, \bar D_{k-1}, \bar 0)\mid \bar D_k, Z, L,\tilde T\geq t_k,\tilde Y^1(t;Z, \bar D_{k-1}, \bar 0)=1 ],
\qquad t\in[t_k,t_{k+1}),
\end{aligned} \label{eq:core_1}
\end{equation}
which states that the current treatment uptake $D_k$ has an additive effect on the potential subdistribution hazard over the interval $[t_k, t_{k+1})$.

\section{Sequential ignorability-based methods}
\label{sec: identificaito and estimation SI}

\subsection{Identification assumptions and moment equations}
\label{subsec: SI identify beta}
The sequential ignorability-based approach treats $Z$ and $L$ jointly as baseline covariates, denoted $L^* = (Z, L)$. We nevertheless retain $Z$ in the potential outcome notation. As discussed in Section~\ref{subsec: intro our contribution}, VRC01 assignment ($Z$) in the AMP trial is not expected to be associated with subsequent oral PrEP uptake ($D_k$) and therefore cannot serve as an instrument for it.

To fix ideas, we first study identification in a complete-data setting with no censoring. 

\iffalse
\begin{assumption}[Consistency]
\label{assumption:consistency}
For any $z$ and treatment history $\bar d$, if $Z=z$ and the observed treatment history satisfies $\bar D=\bar d$, then
\[
\tilde T=\tilde T(z,\bar d)
\]
with probability one.
\end{assumption}
\fi

\begin{assumption}[Sequential Ignorability]
\label{assumption:SI}
Among individuals who remain event-free at time $t_k$, and conditional on the treatment uptake history $\bar D_{k-1}$ and baseline covariates $L^*=(Z,L)$, the exposure $D_k$ at time $t_k$ is independent of the jump of the potential counting process for the type-1 event that would arise if all future exposures were set to zero. That is,
\begin{equation*}
D_k\perp
\left\{
d\tilde N^1(t;Z,\bar D_{k-1},\bar 0):t\geq t_k
\right\}
\mid
\bar D_{k-1},L^*,\tilde T\geq t_k.
\end{equation*}
\end{assumption}

%Since $\left(
%d\tilde N^1(t;Z,\bar D_{k-1},\bar 0),
%\tilde Y^1(t;Z,\bar D_{k-1},\bar 0)
%\right)$
%are derived from $\left\{
%d\tilde N^1(t;Z,\bar D_{k-1},\bar 0):t\geq t_k
%\right\}$, by Assumption~\ref{assumption:SI}, we obtain that for $t\geq t_k$,
%\[
%D_k \perp
%\left(
%d\tilde N^1(t;Z,\bar D_{k-1},\bar 0),
%\tilde Y^1(t;Z,\bar D_{k-1},\bar 0)
%\right)
%\mid
%\bar D_{k-1},L^*,\tilde T\geq t_k.
%\]
Assumption~\ref{assumption:SI} enables us to drop $D_k$ from the conditioning set,
\begin{equation}
    \begin{aligned}
\label{eq: eliminate Dk in conditioning}
&\mathbb{E}\left[
d\tilde N^1(t;Z,\bar D_{k-1},\bar 0)
\;\middle|\;
\bar D_k,L^*,\tilde T\geq t_k,
\tilde Y^1(t;Z,\bar D_{k-1},\bar 0)=1
\right]
\\
={}&
\mathbb{E}\left[
d\tilde N^1(t;Z,\bar D_{k-1},\bar 0)
\;\middle|\;
\bar D_{k-1},L^*,\tilde T\geq t_k,
\tilde Y^1(t;Z,\bar D_{k-1},\bar 0)=1
\right].
\end{aligned}
\end{equation}
Combining Equation \eqref{eq: eliminate Dk in conditioning} with Equation \eqref{eq:core_1}, we obtain the following core equation
\begin{equation}
\begin{aligned}
\label{eq: SI core equation}
&\mathbb{E}\left[
d\tilde N^1(t;Z,\bar D_k,\bar 0)
\mid
\bar D_k,L^*,\tilde T\geq t_k,
\tilde Y^1(t;Z,\bar D_k,\bar 0)=1
\right]
- D_kdB_D(t)
\\
={}&
\mathbb{E}\left[
d\tilde N^1(t;Z,\bar D_{k-1},\bar 0)
\mid
\bar D_{k-1},L^*,\tilde T\geq t_k,
\tilde Y^1(t;Z,\bar D_{k-1},\bar 0)=1
\right].
\end{aligned}
\end{equation}
Equation \eqref{eq: SI core equation} then motivates the following moment equation for $t\in[t_k,t_{k+1})$:
\begin{equation}
\begin{aligned}
\mathbb E\Bigl[
&\mathbf 1\{\tilde T\geq t_k\}
\left(
D_k-
\mathbb E[
D_k
\mid
\bar D_{k-1},L^*,\tilde T\geq t_k,
\tilde Y^1(t;Z,\bar D_k,\bar 0)=1
]
\right)
\\
&\qquad\qquad\times
\tilde Y^1(t)
\left(
d\tilde N^1(t)-D_kdB_D(t)
\right)
\Bigr]
=0,
\end{aligned}
\label{eq:SI_moment1}
\end{equation}
which we prove formally in Supplementary Material~\ref{app:proof_SI_moment_validity}. 
%The parameter $\beta$ can be estimated after modeling $\mathbb E\left[D_k\mid \bar D_{k-1},L^*,\tilde T\geq t_k,
%\tilde Y^1(t;Z,\bar D_k,\bar 0)=1
%\right].
%$
However, estimating $B_D(t)$ from the moment equation \eqref{eq:SI_moment1} faces two impediments. First, the conditioning event in the nuisance function $\mathbb E\left[D_k\mid \bar D_{k-1},L^*,\tilde T\geq t_k, \tilde Y^1(t;Z,\bar D_k,\bar 0)=1 \right]$ involves the potential outcome $\tilde Y^1(t;Z,\bar D_k,\bar 0)$. Second, quantities such as $\tilde Y^1(t)$ and $\tilde N^1(t)$ are not always observable in the presence of censoring. We address these two impediments in turn.

First, for $t \in [t_k, t_{k+1})$, we define the interval-wise cumulative effect $A_k(t) = \int_{t_k}^{t} dB_D(s)$ and the following treatment residual:
\begin{equation*}
e_k(t \mid \bar D_{k-1},L^*, T\geq t_k)  = D_k-
\frac{
\mathbb E\left[
D_k
\exp\left\{
-D_kA_k(t^-)
\right\}
\Bigm|
\bar D_{k-1},L^*, T\geq t_k
\right]
}{
\mathbb E\left[
\exp\left\{
-D_k A_k(t^-)
\right\}
\Bigm|
\bar D_{k-1},L^*, T\geq t_k
\right]
},
\end{equation*}
which depends only on the observed-data distribution. Proposition~\ref{prop:conditional_treatment_mean} shows that the residual $D_k - \mathbb{E}\left[ D_k \mid \bar D_{k-1}, L^*, \widetilde T \geq t_k, \widetilde Y^1(t; \bar D_k, \bar 0) = 1 \right]$ appearing in the moment equation~\eqref{eq:SI_moment1} equals $e_k(t \mid \bar D_{k-1}, L^*, T \geq t_k)$, and is therefore identified from the observed data.

\begin{proposition}
\label{prop:conditional_treatment_mean}
Suppose that the cause-specific SNCSTM
\eqref{eq:cause_specific_SNCSTM}, and
Assumptions~\ref{assumption:consistency} and \ref{assumption:SI} hold. For
$k=1,\ldots,K$ and $t\in[t_k,t_{k+1})$, we have
{\small
\begin{equation}
\label{eq:SI nuisance expression}
D_k - \mathbb E\left[
D_k
\mid
\bar D_{k-1},L^*,\widetilde T\geq t_k,
\widetilde Y^1(t;\bar D_k,\bar 0)=1
\right]
=
e_k(t \mid \bar D_{k-1},L^*, T\geq t_k).
\end{equation}
}
\end{proposition}

Second, let $C$ be the censoring time and assume conditional independent censoring, as formalized in Assumption \ref{assumpion:conditional_independent_censoring_si}.

\begin{assumption}[Conditional independent censoring]
\begin{equation*}
    C\perp (\{d\tilde N^j(t):t\in [0,\tau],j=1,\cdots,J \},\bar D_K) \mid L^*
\end{equation*}
\label{assumpion:conditional_independent_censoring_si}
\end{assumption}

Assumption \ref{assumpion:conditional_independent_censoring_si} requires the censoring time variable $C$ to be independent of the event processes and the full treatment history, conditional on the baseline covariates \(L^*\). For $t\in[0,\tau]$, define
\[
r(t)=\mathbf 1\{C\geq \tilde{T}\wedge t\},
\qquad
G(t\mid L^*)=P(C\geq t\mid L^*),
\]
\[
w(t)=\frac{r(t)G(t\mid L^*)}
{G(T\wedge t\mid L^*)},
\qquad
Y^{1,w}(t)=w(t)Y^1(t),
\]
so that $Y^{1,w}(t)$ is always observable, and 
$$
\check Y^1(t)=\mathbf{1}\{T\geq t_k \}Y^{1,w}(t),\quad t\in [t_k,t_{k+1}).
$$ Theorem \ref{thm:SI_moment_validity} adapts inverse probability of censoring weighting (IPCW) to construct a valid moment equation under covariate-dependent censoring.

%the observed
%counting-process increment satisfies
%\[
%dN^1(t)
%=
%\mathbf 1\{C\geq t\}\,d\tilde N^1(t).
%\]
%However,  we generally have
%\[
%Y^1(t)
%\neq
%\mathbf 1\{C\geq t\}\tilde Y^1(t).
%\]
%In particular, after censoring and in the absence of an observed event,
%\(N^1(t)\) remains zero and hence \(Y^1(t)\) remains one, whereas
%\(\mathbf 1\{C\geq t\}\tilde Y^1(t)=0\).

\begin{theorem}[Moment equation in the presence of censoring]
\label{thm:SI_moment_validity}

Suppose that the cause-specific SNCSTM \eqref{eq:cause_specific_SNCSTM}, and
Assumptions~\ref{assumption:consistency}--\ref{assumpion:conditional_independent_censoring_si}
hold. 
Assume further that $G(t\mid L^*)>0$ for all $t\in[0,\tau]$ and $L$. Then, for each
$k=1,\ldots,K$ and all $t\in[t_k,t_{k+1})$,
\begin{equation}
\mathbb E \left[
{e_k(t \mid \bar D_{k-1},L^*, T\geq t_k)}
\check Y^{1}(t)
\left(
dN^1(t)- D_k\,dB_D(t)
\right)
\right]
=0.
\label{eq:SI_moment}
\end{equation}
\end{theorem}

We will refer to \eqref{eq:SI_moment} as the SI moment equation. Equation \eqref{eq:SI_moment} involves only observed quantities. Constructing an estimating equation from it requires estimating two nuisance functions: the conditional survival function $G(t \mid L^\ast)$ of the censoring time, and the conditional distribution of $D_k$ given the observed history among those with $T \geq t_k$.

\subsection{Estimation and inference for parameters in cause-specific SNCSTMs}
\label{subsec: SI est and inference}

%Because the SI moment condition \eqref{eq:SI_moment} depends on $B_D(t)$ only through its increments $dB_D(t)$, on each interval $[t_k, t_{k+1})$ the parameter of interest can equivalently be taken to be $A_k(t) = \int_{t_k}^{t} dB_D(s),t \in [t_k, t_{k+1}).$ Proposition~\ref{prop: closed_form_Ak} gives a closed-form expression for $A_k(t)$ that solves \eqref{eq:SI_moment}. 

\begin{proposition}
\label{prop: closed_form_Ak}
    For $t\in [t_k,t_{k+1})$, assume that $B_D(t)$ is continuous. Under Assumptions~\ref{assumption:consistency}--\ref{assumpion:conditional_independent_censoring_si}, and the cause-specific SNCSTM
\eqref{eq:cause_specific_SNCSTM}, the parameter of interest $A_k(t)$ that solves the SI moment equation \eqref{eq:SI_moment} admits the following closed-form solution
\begin{equation*}
    A_k(t) = \int_{t_k}^t\frac{\mathbb{E}[ e_k(s \mid \bar D_{k-1},L^*, T\geq t_k)dN^1(s) ] }{\mathbb{E}[e_k(s \mid \bar D_{k-1},L^*, T\geq t_k)\check Y^1(s)D_k]},
\end{equation*}
provided that $\mathbb{E}[e_k(t \mid \bar D_{k-1},L^*, T\geq t_k)\check Y^1(t)D_k] \neq 0$ for all $t\in [t_k,t_{k+1})$.
\end{proposition}

Let $\theta$ denote the parameters indexing the treatment assignment model $f(D_k \mid \bar D_{k-1}, L^*, T \geq t_k; \theta)$ and the censoring model $G(t \mid L; \theta)$, and let $\hat\theta$ be a consistent and asymptotically linear estimator of $\theta$. We order the cause-1 event times $s_1<\cdots< s_M$, where $\{s_1,\cdots,s_M \}=\{T_i:\Delta_i=1, V_i=1 \}$. Also define $\Delta N^1_i(s_m)= \mathbf{1}\{T_i=s_m,\Delta_i=1,V_i=1 \}$. An estimator $\hat A_k(t)$ for $A_k(t)$ can be obtained recursively as follows:
\begin{equation*}
   \hat A_k(t)= \sum_{t_k\leq s_m\leq t}d \hat A_k(s_m)= \sum_{t_k\leq s_m\leq t}\frac{\sum_{i=1}^n e_{i,k}(s_m;\hat \theta)\Delta N_i^1(s_m) }{\sum_{i=1}^n e_{i,k}(s_m;\hat \theta)\check Y^1_i(s_m;\hat \theta)D_{i,k}},
\end{equation*}
with initial value $\hat A_k(t_k^-)=0$. An estimator of $B_D(t)$ for $t\in[t_k,t_{k+1})$, $k=1,\cdots,K$, is then \[
\hat B_D(t) = \sum_{j=1}^k \hat A_j(t\wedge t_{j+1}^-).
\]
%Under the regularity conditions in Supplemental Materials \ref{app:SI_uniformly_consistent} and \ref{app:SI_weakly_convergence}, the estimator $\hat A_k(t)$ is uniformly consistent for $A_k(t)$, and $\{\sqrt n (\hat A_k(t)-A_k(t)):t\in[t_k,t_{k+1}) \}$ converges weakly to a mean-zero Gaussian process. 
Propositions \ref{prop:SI_uniformly_consistent} and \ref{prop:SI_weakly_convergence} establish the theoretical properties of  $\hat B_D(t)$.

\begin{proposition}
\label{prop:SI_uniformly_consistent}
Under the regularity conditions in Supplemental Material \ref{app:SI_uniformly_consistent}, the estimator $\hat B_D(t)$ is uniformly consistent for $B_D(t)$; that is,
\begin{equation*}
    \sup_{t\in [0,\tau]}|\hat B_D(t)-B_D(t)| \xrightarrow{P} 0
    \quad \text{as } n\to\infty .
\end{equation*}
\end{proposition}

\begin{proposition}
\label{prop:SI_weakly_convergence}
Under the regularity conditions in Supplemental Material \ref{app:SI_weakly_convergence}, the normalized process $\sqrt{n}\{\hat B_D(t)-B_D(t)\}$ converges weakly to a zero-mean Gaussian process with variance function that can be consistently estimated by 
\begin{equation*}
    \frac 1n \sum_{i=1}^n \hat \psi_i^2(t;\hat \theta),
\end{equation*}
where $\hat \psi_i(t;\hat \theta)$ is defined in Supplemental Material \ref{app:SI_weakly_convergence}.
\end{proposition}

%For implementation, we construct $\hat \phi_i(t;\hat\theta)$ recursively and further derived $\hat \psi_i(t;\hat \theta)$ for computational convenience.

\begin{remark}\label{remark: estimate beta}
If one further assumes a constant-effect model, that is, $B_D(t) = \beta t$, then $\beta$ can be estimated by (see, e.g., \citealp{ying2023structural})
\[
\hat{\beta} = \int_0^{\tau} W(t)\, d\hat{B}_D(t),
\]
where $W(t) = \tilde{W}(t) \big/ \int_0^{\tau} \tilde{W}(s)\, ds$ and $\tilde{W}(t) = \sum_{i=1}^n \mathbf{1}\{T_i \geq t_k\}\, Y_i^1(t)$ for $t \in [t_k, t_{k+1})$. If $t_{k+1}-t_k$ is constant for every $k=1,\ldots,K$, then the periodic effect $A_k(t) \equiv A(t)$ can be estimated by 
\[
\hat{A}(t) = \int_0^t \frac{\sum_{k=1}^{K} \tilde{W}(t_k+s)d\hat{B}_D(t_k+s)}{\sum_{k=1}^{K}\tilde{W}(t_k+s)}.
\]
\end{remark}

\section{Instrumental variable-based methods}
\label{sec: inference under IV}
\subsection{Identification assumptions and moment equations}
\label{subsec: IV estimate beta}
In applications where the time-varying treatment process $\bar D$ denotes time-varying adherence to the randomized treatment assignment $Z$, one may instead leverage the fact that $Z$ is a valid instrumental variable to identify the structural parameters in Equation \eqref{eq:cause_specific_SNCSTM}, analogous to the identification of structural parameters in SNCSTMs in the absence of competing events \citep{ying2023structural,michiels2025adjusting}. 

We assume that $Z$ satisfies the standard instrumental variable assumptions formalized in Assumptions~\ref{assumption:IV relevance}--\ref{assumption:IV independence}.

\begin{assumption}[IV relevance]
\begin{equation*}
    Z\not\perp D_k\mid \bar D_{k-1},L,\tilde T\geq t_k
\end{equation*}
\label{assumption:IV relevance}
\end{assumption}
Assumption~\ref{assumption:IV relevance} requires that the instrument $Z$ be associated with the treatment received at time $t_k$, conditional on the observed treatment history, baseline covariates, and survival to $t_k$. This assumption may fail when $D_k$ represents a subsequent treatment decision that is not directly influenced by $Z$. For instance, in the AMP trial, assignment to VRC01 versus placebo is unlikely to be associated with a participant's oral PrEP uptake, since the trial is blinded. By contrast, the {\color{blue}IV relevance} assumption typically holds when $D_k$ represents adherence to the treatment assigned by $Z$. In the AMP trial, VRC01 infusions were administered every $8$ weeks, so if $D_k$ denotes the $k$-th infusion, the initial randomization $Z$ serves as an instrument for $D_k$. 

\begin{assumption}[Exclusion restriction]
\label{assumption:Exclusion restriction}
For any fixed treatment history $\bar d$, the instrumental variable $Z$ has no direct effect on the potential counting process of any event type. That is, for any $z,z'$ and $j\in\{1,\ldots,J\}$,
\begin{equation*}
    d\tilde N^j(t;z,\bar d)=
    d\tilde N^j(t;z',\bar d),\quad \forall t\in[0,\tau].
\end{equation*}
\end{assumption}
Assumption~\ref{assumption:Exclusion restriction} requires that $Z$ affect the potential counting processes only through the treatment uptake process, with no direct effect once the treatment history is fixed. We can therefore write $d\tilde N^j(t;z,\bar d)$ as $d\tilde N^j(t;\bar d)$. Assumption~\ref{assumption:Exclusion restriction} is most plausible in a blinded trial, where participants are unaware of their treatment assignment.

\begin{assumption}[IV independence]
\begin{equation*}
    Z\perp \{d\tilde N^j(t;\bar 0):t\in [0,\tau],j=1,\cdots,J \}\mid L
\end{equation*}
\label{assumption:IV independence}
\end{assumption}
Assumption~\ref{assumption:IV independence} states that, conditional on $L$, $Z$ is independent of the potential counting process for any event type under no treatment uptake. This assumption holds by design in randomized trials. 
%By Assumption~\ref{assumption:Exclusion restriction}, the potential counting process under a fixed treatment history does not depend on the value of $z$. Therefore, this condition may equivalently be stated using $d\tilde N^j(t;z,\bar 0)$ for any fixed $z$.

\begin{remark}[Role of unmeasured confounders]
    Unmeasured confounding of the relationship between treatment received and the outcome violates sequential ignorability but is permitted under the IV framework, provided the instrumental variable assumptions hold. The IV approach is therefore a useful alternative when sequential ignorability is implausible.
\end{remark}

We next impose the conditional independent censoring assumption.

\begin{assumption}[Conditional independent censoring]
\begin{equation*}
    C\perp (\{d\tilde N^j(t):t\in [0,\tau],j=1,\cdots,J \},\bar D_K,Z) \mid L
\end{equation*}
\label{assumpion:conditional_independent_censoring_iv}
\end{assumption}

Theorem \ref{thm:IV_moment_validity} establishes an IV moment equation under Assumptions \ref{assumption:IV relevance}--\ref{assumpion:conditional_independent_censoring_iv}.

\begin{theorem}[IV moment equation]
\label{thm:IV_moment_validity}

Suppose that the cause-specific SNCSTM
\eqref{eq:cause_specific_SNCSTM}, Assumptions~\ref{assumption:consistency} and ~\ref{assumption:IV relevance}--\ref{assumpion:conditional_independent_censoring_iv}
hold. %For $t\in[0,\tau]$, define
%\[
%r(t)=\mathbf 1\{C\geq T\wedge t\},
%\qquad
%G(t\mid L)=P(C\geq t\mid L),
%\]
%\[
%w(t)=\frac{r(t)G(t\mid L)}
%{G(T\wedge t\mid L)},
%\qquad
%Y^{1,w}(t)=w(t)Y^1(t),
%\]
%and 
%$$
%\check Y^1(t)=\mathbf{1}\{T\geq t_k %\}Y^{1,w}(t),\quad t\in [t_k,t_{k+1}).
%$$
Assume further that $G(t\mid L)>0$ for all $t\in (0,\tau]$ and $L$. Then for $t\in [0,\tau]$, the following moment condition holds:
\begin{equation}
\mathbb E\left[
\left(
Z-\mathbb E[Z\mid L]
\right)
\check Y^{1}(t)
\exp\left\{\int_{0}^{t^-}
 D(s)\,dB_D(s)\right\}
\left(
dN^1(t)-D(t)\,dB_D(t)
\right)
\right]
=0.
\label{eq:IV_moment}
\end{equation}
\label{thm:IV_moment}
\end{theorem}
We will refer to \eqref{eq:IV_moment} as the IV moment equation.

%The proof is provided in Appendix~\ref{app:proof_IV_moment_validity}. 
%Write $m(L)=\mathbb E[Z\mid L]$ and let $\widehat m(L)$ denote its estimator. Moreover, let $\widehat G(t\mid L)$ and $\widehat Y_i^{1,w}(t)$ denote estimators of $G(t\mid L)$ and $Y_i^{1,w}(t)$, respectively. Under the model that $B_D(t)=\beta t$ , we can integrate the moment condition over the treatment intervals
%$[t_k,t_{k+1})$, sum over $k=1,\ldots,K$, and obtain a root-based estimator of $\beta$ by solving the estimating equation:
%\begin{equation}
%\sum_{i=1}^{n}\sum_{k=1}^K
%\int_{t_k}^{t_{k+1}}
%\left\{
%Z_i-\widehat m(L_i)
%\right\}
%\mathbf 1\{T_i\geq t_k\}
%\widehat Y_i^{1,w}(t)
%\exp\left\{B_k(t)\right\}
%\left\{
%dN_i^1(t)-D_{i,k}\beta\,dt
%\right\}
%=0.
%\label{eq:IV_estimation}
%\end{equation}

\subsection{Estimation and inference for parameters in cause-specific SNCSTMs}
\label{subsec: IV estimation and inference}

Proposition~\ref{prop: closed_form_B_D} gives a closed-form expression for $B_D(t)$ that solves the IV moment equation.

\begin{proposition}
\label{prop: closed_form_B_D}
    Write $Z^c = Z - \mathbb E[Z \mid L]$, and assume that $B_D(t)$ is continuous in $t$. Under the cause-specific SNCSTM \eqref{eq:cause_specific_SNCSTM} and Assumptions~\ref{assumption:consistency} and \ref{assumption:IV relevance}--\ref{assumpion:conditional_independent_censoring_iv}, the parameter of interest $B_D(t)$ that solves the IV moment equation \eqref{eq:IV_moment} admits the following closed-form solution
\begin{equation}
    B_D(t)=\int_0^t \frac{\mathbb E\left[
Z^c
\exp\{\int_0^{s^-} D(u)dB_D(u) \}
dN^1(s)
\right] }{\mathbb E\left[
Z^c
\check Y^1(s)
\exp\{\int_0^{s^-} D(u)dB_D(u) \}
D(s)
\right] },
\end{equation}
provided that $\mathbb E\left[
Z^c
\check Y^1(t)
\exp\{\int_0^{t^-} D(s)dB_D(s) \}
D(t)
\right]\neq 0$ for all $t\in (0,\tau]$.
\end{proposition}

%It is worth mentioning that IV relevance is merely a necessary but not sufficient condition for $\mathbb E\Biggl[
%Z^c
%\check Y^1(t)
%\exp\{\int_0^{t^-} D(s)dB_D(s) \}
%D(t)
%\Biggr]\neq 0$. However, in trials where $Z$ is treatment randomization and has strong positive associations with treatment $D(t)$, it is reasonable to assume this non-zero expectation. 

Let $\theta$ denote parameters indexing the treatment assignment model $\mathbb E[Z\mid L;\theta]$ and the censoring model $G(t\mid L;\theta)$, and let $\hat{\theta}$ denote a consistent and asymptotically linear estimator of $\theta$. Let $s_1<\cdots< s_M$ and $\Delta N^1_i(s_m)$ be defined the same as in Section \ref{subsec: SI est and inference}. An estimator $\hat B_D(t)$ for $B_D(t)$ can be obtained using the following recursive form:
\begin{equation}
   \hat B_D(t)= \sum_{s_m\leq t}d \hat B_D(s_m)= \sum_{s_m\leq t}\frac{\sum_i
Z_i^c(\hat \theta)
\exp\left\{\sum_{1\leq l<m} D_i(s_l)d\hat B_D(s_l) \right\}
\Delta N_i^1(s_m) }{\sum_i
Z_i^c(\hat \theta)
\check Y^1_i(s_m;\hat \theta)
\exp\left\{\sum_{1\leq l<m} D_i(s_l)d\hat B_D(s_l) \right\}
D_i(s_m)},
\end{equation}
with initial value $\hat B_D(0)=0$.

Finally, Propositions \ref{prop:IV_uniformly_consistent} and \ref{prop:IV_weakly_convergence} establish the properties of $\hat B_D(t)$.

\begin{proposition}
\label{prop:IV_uniformly_consistent}
Under the regularity conditions in Supplemental Material \ref{app:IV_uniformly_consistent}, the estimator $\hat B_D(t)$ is uniformly consistent for $B_D(t)$; that is,
\begin{equation}
    \sup_{t\in[0,\tau]}|\hat B_D(t)-B_D(t)| \xrightarrow{P} 0
    \quad \text{as } n\to\infty .
\end{equation}
\end{proposition}

\begin{proposition}
\label{prop:IV_weakly_convergence}
Under conditions in Supplemental Material \ref{app:IV_weakly_convergence}, the normalized process $\sqrt{n}\{\hat B_D(t)-B_D(t)\}$ converges weakly to a mean-zero Gaussian process with variance function that can be consistently estimated by 
\begin{equation}
    \frac 1n \sum_{i=1}^n \hat \epsilon_i^2(t;\hat \theta),
\end{equation}
where $\hat \epsilon_i^2(t;\hat \theta)$ is defined in Supplemental Material \ref{app:IV_weakly_convergence}.
\end{proposition}

\section{Counterfactual cause-specific cumulative incidence function}
\label{sec: SI counterfactual CIF}
With the estimated treatment effect, $\hat{B}_D(t)$, we can blip down the observed treatment uptake $\bar D$ to $\bar 0$, yielding the counterfactual cause-specific cumulative incidence that would have been observed in the absence of any treatment uptake. %{\color{blue}Define the cumulative effect due to treatment
%\[
%\Gamma_j(t)=\sum_{k=1}^j \int_{t_k}^{t\wedge t_{k+1}}D_kdB_D(s)=\int_0^{t\wedge t_{j+1}}D(s)dB_D(s), \quad t\ge t_j.
%\]}
Theorem \ref{thm:observed_blip_down} derives the expected value of the subdistribution risk indicator for the type-$1$ event under treatment assignment $Z = z$ and no treatment uptake.

\begin{theorem}
\label{thm:observed_blip_down}
Suppose that the cause-specific SNCSTM
\eqref{eq:cause_specific_SNCSTM},
and the assumptions underpinning Theorem \ref{thm:SI_moment_validity} or \ref{thm:IV_moment} hold (where we replace $L$ with $L^*$ in the following under sequential ignorability). Then, for each $k=1,\ldots,K$ and all
$t\in[t_k,t_{k+1})$,
\begin{equation}
%\begin{aligned}
%\mathbb E\Bigl[
%\tilde Y^1(t;Z,\bar 0)
%\Bigr]
%=&
%\begin{cases}
%\displaystyle
%\mathbb E\Biggl[
%\frac{
%Y^{1,w}(t)\exp\{\int_0^{t\wedge t_{2}}D(s)dB_D(s)\}
%}{
%G(t\mid L^*)
%}
%\Biggr],
%& k=1,
%\\[1.2em]
%\displaystyle
%\mathbb E\Biggl[
%\sum_{j=1}^{k-1}
%\mathbf 1\{t_j\leq T<t_{j+1} ,\,\Delta=1 ,\,V\neq 1 \}
%\frac{\exp\{\int_0^{t\wedge t_{j+1}}D(s)dB_D(s)\}}{G(T\mid L^*)}
%\\
%\displaystyle\qquad\qquad
%+
%\mathbf 1\{T\geq t_k\}
%\frac{
%Y^{1,w}(t)\exp\{\int_0^{t\wedge t_{k+1}}D(s)dB_D(s)\}
%}{
%G(t\mid L^*)
%}
%\Biggr],
%& k\geq 2.
%\end{cases}
%\end{aligned}
\begin{aligned}
\mathbb E\Bigl[\tilde Y^1(t;Z,\bar 0)\Bigr]
&= \mathbb E\Biggl[
\sum_{j=1}^{k-1}
\mathbf 1\{t_j\leq T<t_{j+1} ,\,\Delta=1 ,\,V\neq 1 \}
\frac{\exp\{\int_0^{t\wedge t_{j+1}}D(s)dB_D(s)\}}{G(T\mid L)}
\\
&\qquad\qquad
+
\mathbf 1\{T\geq t_k\}
\frac{
Y^{1,w}(t)\exp\{\int_0^{t\wedge t_{k+1}}D(s)dB_D(s)\}
}{
G(t\mid L)
}
\Biggr].
\end{aligned}
\end{equation}
\label{thm:observed_blip_down2}
\end{theorem}

In particular, only the second term in Equation \eqref{thm:observed_blip_down2} remains if $t\in[0,t_2)$. The counterfactual cause-1-specific cumulative incidence under the treatment assignment $Z$ in the absence of time-varying treatment uptake is
\begin{equation}
F^1(t) := P\{\tilde T(Z,\bar 0)\leq t, V(Z,\bar 0) = 1\} = 1 -\mathbb{E}[\tilde Y^1(t;Z,\bar 0)],
\end{equation}
which can be estimated by plugging in fitted nuisance models and estimated $B_D(\cdot)$. Following Propositions \ref{prop:SI_weakly_convergence} and \ref{prop:IV_weakly_convergence}, if the estimated parameter $\hat\theta$ in nuisance functions is regular and asymptotically linear, then the plug-in estimator of $F^1(t)$ is also regular and asymptotically linear. 
%and, for $t\in[t_k,t_{k+1})$, the corresponding plug-in estimator is
%\begin{equation*}
%\widehat P(\tilde T(Z,\bar 0)\leq t,V=1)
%=
%\begin{cases}
%\displaystyle
%1-
%\frac{1}{n}
%\sum_{i=1}^n
%\frac{
%\widehat Y_i^{1,w}(t)\exp\{\widehat \Gamma_{i,1}(t)\}
%}{
%\widehat G(t\mid L_i^*)
%},
%& k=1,
%\\[1.2em]
%\displaystyle
%1-
%\frac{1}{n}
%\sum_{i=1}^n
%\Biggl[
%\sum_{j=1}^{k-1}
%\mathbf 1\{t_j\leq T_i<t_{j+1},\,\Delta_i=1 ,\,V_i\neq 1\}
%\frac{
%\exp\{\widehat \Gamma_{i,j}(t)\}
%}{
%\widehat G(T_i\mid L_i^*)
%}
%\\
%\displaystyle\qquad\qquad\qquad\qquad
%+
%\mathbf 1\{T_i\geq t_k\}
%\frac{
%\widehat Y_i^{1,w}(t)\exp\{\widehat \Gamma_{i,k}(t)\}
%}{
%\widehat G(t\mid L_i^*)
%}
%\Biggr],
%& k\geq 2,
%\end{cases}
%\end{equation*}
%where $\widehat Y_i^{1,w}(t)$ is obtained by replacing $G(t\mid L_i^*)$ in
%$Y_i^{1,w}(t)$ with $\widehat G(t\mid L_i^*)$, and
%$\widehat \Gamma_{i,k}(t)$ is obtained by replacing $B_D(t)$ in $\Gamma_{i,k}(t)$ with $\widehat B_D(t)$. 
Nonparametric bootstrap can then be used to obtain confidence intervals for each fixed $t$.
\section{Simulation}
\label{sec: simulation studies}
Our goal is to evaluate the finite-sample performance of the proposed estimators for $B_D(t)$ in cause-specific SNCSTMs under both sequential ignorability-based and instrumental variable-based frameworks. In all settings, we consider three treatment uptake intervals,
$[t_1,t_2)$, $[t_2,t_3)$, and $[t_3,t_4)$, with
$(t_1,t_2,t_3,t_4)=(0,0.5,1,1.5)$ and administrative
follow-up ending at $\tau=t_4$. We consider three values of sample size \(n\): \(1000\), \(5000\), and \(10000\).

We generate baseline covariates differently in the SI and IV settings:
\begin{description}
    \item[SI settings] We independently generate two observed baseline
    covariates: $L_1\sim N(0,1)$ truncated to $[-3,3]$ and
    $L_2\sim \operatorname{Bernoulli}(0.5)$. Using the notation in Section \ref{sec: identificaito and estimation SI}, $L^\ast = (L_1, L_2)$.
    \item[IV settings] In addition to $L_1$ and $L_2$ generated as in the SI
    settings, we generate an unmeasured baseline covariate $U$ with
    $P(U=-1)=P(U=1)=1/2$, and a randomized three-level instrument
    $Z\in\{0,1,2\}$ with $P(Z=z)=1/3$ for $z=0,1,2$. Both $U$ and $Z$ are
    independent of each other and of $(L_1,L_2)$.
\end{description}

Next, we generate the treatment uptake $(D_1,D_2,D_3)$ in the SI and IV
settings as follows:
\begin{description}
    \item[SI settings] We let $\operatorname{logit}P(D_1=1\mid L)
    = -1+0.3L_1+0.3(L_2-0.5),$
    and, for $k=2,3$,
    {\small\[
    \operatorname{logit}P(D_k=1\mid \bar D_{k-1},L)
    = -1.7+2D_{k-1}+0.6\,\widetilde D_{k-1}
      +0.3L_1+0.3(L_2-0.5),
    \]}
    where $\widetilde D_{k-1}=(k-1)^{-1}\sum_{s=1}^{k-1}D_s$ is the average of past treatment uptake. This data generating process satisfies sequential ignorability.

    \item[IV settings] We let $\operatorname{logit}P(D_1=1\mid L,U,Z)
    = -1+0.3L_1+0.3(L_2-0.5)+0.8Z+0.3U,$
    and, for $k=2,3$,
    {\small\[
    \begin{aligned}
    \operatorname{logit}P(D_k=1\mid \bar D_{k-1},L,U,Z)
    ={}& -1.7+2D_{k-1}+0.6\,\widetilde D_{k-1}\\
       &+0.3L_1+0.3(L_2-0.5)+0.8Z+0.3U.
    \end{aligned}
    \]}
    %Because $U$ affects $D_k$ and cause-specific cumulative incidence functions, sequential ignorability fails here.
\end{description}

For notational convenience, let $X$ denote the full set of baseline variables in the potential-outcome model: $X = (L_1, L_2)$ in the SI setting and $X = (L_1, L_2, U)$ in the IV setting. By the exclusion restriction, $Z$ has no direct effect on the outcome, so $X$ does not include $Z$.  We specify the following cause-specific cumulative incidence functions when there is no treatment uptake (i.e., $\bar D = \bar 0$) as
{\small\[
F_1(t;\bar 0\mid X)=0.6m(X)H_{0.025}(t),
\qquad
F_2(t;\bar 0\mid X)=0.4m(X)H_{0.035}(t),
\]
}
where, for $0\le t\le\tau$, $H_a(t)=\frac{1-e^{-at}}{1-e^{-a\tau}}$, $m(X)=\operatorname{expit}\{b+\eta(X)\}$, and
{\small\[
\eta(X)=
\begin{cases}
0.05L_1+0.05(L_2-0.5), & \text{SI setting},\\
0.05L_1+0.05(L_2-0.5)+0.15U, & \text{IV setting}.
\end{cases}
\]}

We consider four overall event probabilities by the end of follow-up under
$\bar D=\bar 0$ and in the absence of censoring,
$r\in\{7\%,15\%,30\%,60\%\}$, and calibrate the parameter $b$ separately
in the SI and IV settings to achieve each target.

Write $F_j(t;\bar d_k,\bar 0\mid X)
=
P\left\{
\tilde T(\bar d_k,\bar 0)\le t,\,
V(\bar d_k,\bar 0)=j
\mid X
\right\},~k=0,1,2,3.$
We consider the following cause-$j$-specific SNCSTM:
\begin{equation*}
\frac{P(\tilde Y^j(t;Z,\bar D_k, \bar 0)=1\mid \bar D_k, Z, L,\tilde T\geq t_k )}{P(\tilde Y^j (t;Z,\bar D_{k-1}, \bar 0)=1\mid \bar D_k, Z, L,\tilde T\geq t_k)}
=\exp\left\{
-\beta^{(k)}_jD_k\bigl(\min(t,t_{k+1})-t_k\bigr)
\right\}
\end{equation*}
and a time-homogeneous setting in which $\beta_1^{(1)} = \beta_1^{(2)} = \beta_1^{(3)} = \beta_1 = - 0.25r$ and $\beta_2^{(1)} = \beta_2^{(2)} = \beta_2^{(3)} = \beta_2 = - 0.125r$. As shown in the Supplemental Material~\ref{app:Recursive rules for CIFs},  $F_j(t;\bar d_k,\bar 0\mid X)$ satisfies the following recursive relationship:
{\small\[
\begin{aligned}
F_j(t;\bar d_k,\bar 0\mid X)
&={}F_j(t;\bar d_{k-1},\bar 0\mid X) + \\
&\left\{
1-F_{3-j}(t_k;\bar d_{k-1},\bar 0\mid X)
-F_j(t;\bar d_{k-1},\bar 0\mid X)
\right\}\times
\left[
1-\exp\left\{
-\beta_jd_k\bigl(\min(t,t_{k+1})-t_k\bigr)
\right\}
\right],
\end{aligned}
\]}
for $t_k\le t\le\tau$, while
$F_j(t;\bar d_k,\bar 0\mid X)=F_j(t;\bar d_{k-1},\bar 0\mid X)$ for $t<t_k$. Hence, starting from $F_j(t;\bar 0\mid X)$, we apply this recursive relationship sequentially to obtain the cause-specific cumulative incidence under each treatment uptake trajectory $\bar d_k$.

%following cause-$j$-specific SNCSTM:
%{\small\[
%\frac{
%1-
%P\left\{
%\tilde T(\bar d_k,\bar 0)\le t,\,
%V(\bar d_k,\bar 0)=j
%\mid
%\tilde T(\bar d_k,\bar 0)\ge t_k,X
%\right\}
%}{
%1-
%P\left\{
%\tilde T(\bar d_{k-1},\bar 0)\le t,\,
%V(\bar d_{k-1},\bar 0)=j
%\mid
%\tilde T(\bar d_{k-1},\bar 0)\ge t_k,X
%\right\}
%}
%=
%\exp\left\{
%-\beta^{(k)}_jd_k\bigl(\min(t,t_{k+1})-t_k\bigr)
%\right\},
%\]}

Finally, latent censoring time $C^*$ is generated independently
of the event process and treatment assignments conditional
on $L$, using a Cox proportional hazards model with hazard
\[
h_C(t\mid L_1, L_2)
=
2\lambda_{C0}t
\exp\left\{
0.3L_1+0.3(L_2-0.5)
\right\}.
\]
We set $\lambda_{C0}=0.045$
so that $P(C^*<\tau)\approx0.10$ and let $C=\min(C^*,\tau)$.

In both the SI and IV settings, for each data-generating process (specified by the sample size $n$ and event rate $r$) and each cause $j$, we computed (i) the proposed nonparametric estimator $\hat{B}_D(t)$ of $B_D(t)$, evaluated at $t \in \{t_2, t_3, t_4\}$, and (ii) the averaged estimator $\hat{\beta}$ of $\beta$. For each estimator, we report the absolute relative bias, Monte Carlo standard deviation (MCSD), mean estimated standard error (SE), and coverage percentage (CP) of the 95\% confidence interval. Results are based on 10,000 simulation replications per setting.

Table~\ref{tab:sim_30} summarizes the simulation results for an event rate of 30\%. Under both the SI and IV settings, the relative bias was small across all sample sizes, remaining below 1\% in most cases and decreasing as the sample size increased. The mean estimated standard errors closely matched the Monte Carlo standard deviations, and the empirical coverage of the 95\% confidence intervals was close to the nominal level. Results for other event rates, reported in Section~\ref{app:More_simulation_results} of the Supplementary Material, are qualitatively similar to those in Table~\ref{tab:sim_30}.

\begin{table}[htbp]
\centering
\small
\setlength{\tabcolsep}{4pt}
\caption{Simulation results under the SI and IV settings when the event rate $r = 30\%$. Results are based on 10,000 Monte Carlo replications. \textbf{Perc. Bias}: absolute relative bias. \textbf{MCSD}: Monte Carlo standard deviation of point estimates. \textbf{Ave. SE}: mean estimated standard error. \textbf{95\% CP}: empirical coverage of pointwise 95\% confidence intervals.}
\label{tab:sim_30}
\begin{tabular}{@{}lll cccc @{\hspace{1.5em}} cccc@{}}
\toprule
 & & & \multicolumn{4}{c}{SI setting} & \multicolumn{4}{c}{IV setting} \\
\cmidrule(lr){4-7} \cmidrule(l){8-11}
$n$ & Cause & Metric
 & $\hat B_D(0.5)$ & $\hat B_D(1)$ & $\hat B_D(1.5)$ & $\hat\beta$
 & $\hat B_D(0.5)$ & $\hat B_D(1)$ & $\hat B_D(1.5)$ & $\hat\beta$ \\
\midrule
\multirow{8}{*}{$1{,}000$} & \multirow{4}{*}{1}
   & Perc. Bias & 0.42\% & 0.26\% & 0.30\% & 0.31\% & 0.40\% & 0.39\% & 0.20\% & 0.27\% \\
 & & MCSD       & 0.016 & 0.026 & 0.036 & 0.023 & 0.055 & 0.075 & 0.093 & 0.063 \\
 & & Ave. SE    & 0.015 & 0.026 & 0.036 & 0.023 & 0.054 & 0.074 & 0.093 & 0.062 \\
 & & 95\% CP    & 94.9\% & 95.0\% & 94.7\% & 94.8\% & 95.6\% & 95.3\% & 95.3\% & 95.2\% \\
\addlinespace[2pt]
 & \multirow{4}{*}{2}
   & Perc. Bias & 0.28\% & 0.99\% & 0.45\% & 0.41\% & 0.40\% & 0.63\% & 1.10\% & 1.09\% \\
 & & MCSD       & 0.013 & 0.022 & 0.030 & 0.020 & 0.046 & 0.062 & 0.077 & 0.052 \\
 & & Ave. SE    & 0.013 & 0.022 & 0.030 & 0.019 & 0.045 & 0.062 & 0.076 & 0.051 \\
 & & 95\% CP    & 94.9\% & 95.3\% & 94.9\% & 94.8\% & 95.2\% & 95.3\% & 95.4\% & 95.3\% \\
\midrule
\multirow{8}{*}{$5{,}000$} & \multirow{4}{*}{1}
   & Perc. Bias & 0.11\% & 0.03\% & 0.06\% & 0.05\% & 1.22\% & 0.98\% & 0.68\% & 0.73\% \\
 & & MCSD       & 0.007 & 0.012 & 0.016 & 0.010 & 0.024 & 0.033 & 0.041 & 0.028 \\
 & & Ave. SE    & 0.007 & 0.012 & 0.016 & 0.010 & 0.024 & 0.033 & 0.041 & 0.027 \\
 & & 95\% CP    & 95.0\% & 94.6\% & 94.8\% & 94.5\% & 95.5\% & 95.0\% & 94.9\% & 95.0\% \\
\addlinespace[2pt]
 & \multirow{4}{*}{2}
   & Perc. Bias & 0.73\% & 0.21\% & 0.41\% & 0.40\% & 0.12\% & 0.18\% & 0.20\% & 0.18\% \\
 & & MCSD       & 0.006 & 0.010 & 0.014 & 0.009 & 0.020 & 0.027 & 0.034 & 0.023 \\
 & & Ave. SE    & 0.006 & 0.010 & 0.014 & 0.009 & 0.020 & 0.027 & 0.033 & 0.023 \\
 & & 95\% CP    & 95.2\% & 95.0\% & 94.7\% & 94.8\% & 95.0\% & 94.9\% & 94.8\% & 94.9\% \\
\midrule
\multirow{8}{*}{$10{,}000$} & \multirow{4}{*}{1}
   & Perc. Bias & 0.04\% & 0.01\% & 0.02\% & 0.02\% & 0.49\% & 0.32\% & 0.30\% & 0.30\% \\
 & & MCSD       & 0.005 & 0.008 & 0.011 & 0.007 & 0.017 & 0.023 & 0.029 & 0.019 \\
 & & Ave. SE    & 0.005 & 0.008 & 0.011 & 0.007 & 0.017 & 0.023 & 0.029 & 0.019 \\
 & & 95\% CP    & 95.0\% & 95.1\% & 95.2\% & 95.2\% & 95.0\% & 94.9\% & 95.1\% & 95.2\% \\
\addlinespace[2pt]
 & \multirow{4}{*}{2}
   & Perc. Bias & 0.07\% & 0.16\% & 0.03\% & 0.02\% & 0.32\% & 0.73\% & 0.50\% & 0.48\% \\
 & & MCSD       & 0.004 & 0.007 & 0.010 & 0.006 & 0.014 & 0.019 & 0.024 & 0.016 \\
 & & Ave. SE    & 0.004 & 0.007 & 0.010 & 0.006 & 0.014 & 0.019 & 0.024 & 0.016 \\
 & & 95\% CP    & 95.0\% & 95.1\% & 94.9\% & 95.0\% & 94.7\% & 95.1\% & 94.9\% & 94.9\% \\
\bottomrule
\end{tabular}
\end{table}

\section{HVTN 704/HPTN 085}
\label{sec: case study}

\subsection{Effect of oral PrEP and ``oral PrEP-free" cause-specific CIFs}
\label{sec: case study sequential ignorability}
We first study the causal effect of oral PrEP on HIV-1 acquisition with a VRC01-sensitive strain and with a VRC01-insensitive strain, as well as the counterfactual cumulative incidence of each in the absence of oral PrEP. 

In the AMP trials, oral PrEP use was quantified using two biomarkers measured in dried blood spots (DBS): emtricitabine triphosphate (FTC-TP) and tenofovir diphosphate (TFV-DP), the intracellular active metabolites of emtricitabine and tenofovir, respectively, the two components of oral PrEP. Both were measured in all cases and in a subset of non-cases selected by sampling stratified by VRC01 arm and self-reported oral PrEP uptake. The resulting longitudinal DBS cohort comprised 174 participants (54 cases and 120 non-cases), whose biomarkers were measured at study visits approximately every 8 weeks. FTC-TP has a half-life of approximately 35 hours in DBS, whereas TFV-DP has a half-life of approximately 17 days \citep{castillo2013tenofovir}. Because TFV-DP may accumulate over weeks, TFV-DP detected during a period $[t_k, t_{k+1})$ may partly reflect oral PrEP use in earlier periods. FTC-TP, by contrast, mainly reflects recent use \citep{castillo2013tenofovir, anderson2018intracellular}, so we used FTC-TP in our analysis.

We consider $K = 10$ time intervals and administrative censoring at $\tau = 560$ days (corresponding to $8$ weeks post the last scheduled infusion), with $[t_k, t_{k+1}) = [56(k-1), 56k)$ for $k = 1, \ldots, 9$ and $[t_{10}, \tau] = [504, 560]$. Within each interval $[t_k, t_{k+1})$, we let $D_k = 1$ if the first FTC-TP measurement between $[t_k,t_{k+1})$ was at or above the lower limit of quantification (LLOQ), and $D_k = 0$ otherwise; a detectable FTC-TP level indicates recent use of daily oral PrEP. The primary outcome of interest is HIV-1 acquisition with a VRC01-sensitive strain, defined as a strain with $\mathrm{IC}_{80} \leq 3\,\mu\mathrm{g/mL}$, and the competing event is HIV-1 acquisition with a VRC01-insensitive strain, defined as a strain with $\mathrm{IC}_{80} > 3\,\mu\mathrm{g/mL}$.

We estimated $B_D(t)$ in the cause-specific SNCSTM under the sequential ignorability assumption. When constructing the estimator, we fitted the conditional censoring distribution using Cox regression adjusting for age, and the propensity score $g_k = f(D_k \mid \bar{D}_{k-1}, L^*, \tilde{T} \geq t_k)$ using logistic regression adjusting for age and average oral PrEP uptake up to time $t_k$.

\begin{figure}[h]
    \centering
\includegraphics[width=0.95\linewidth]{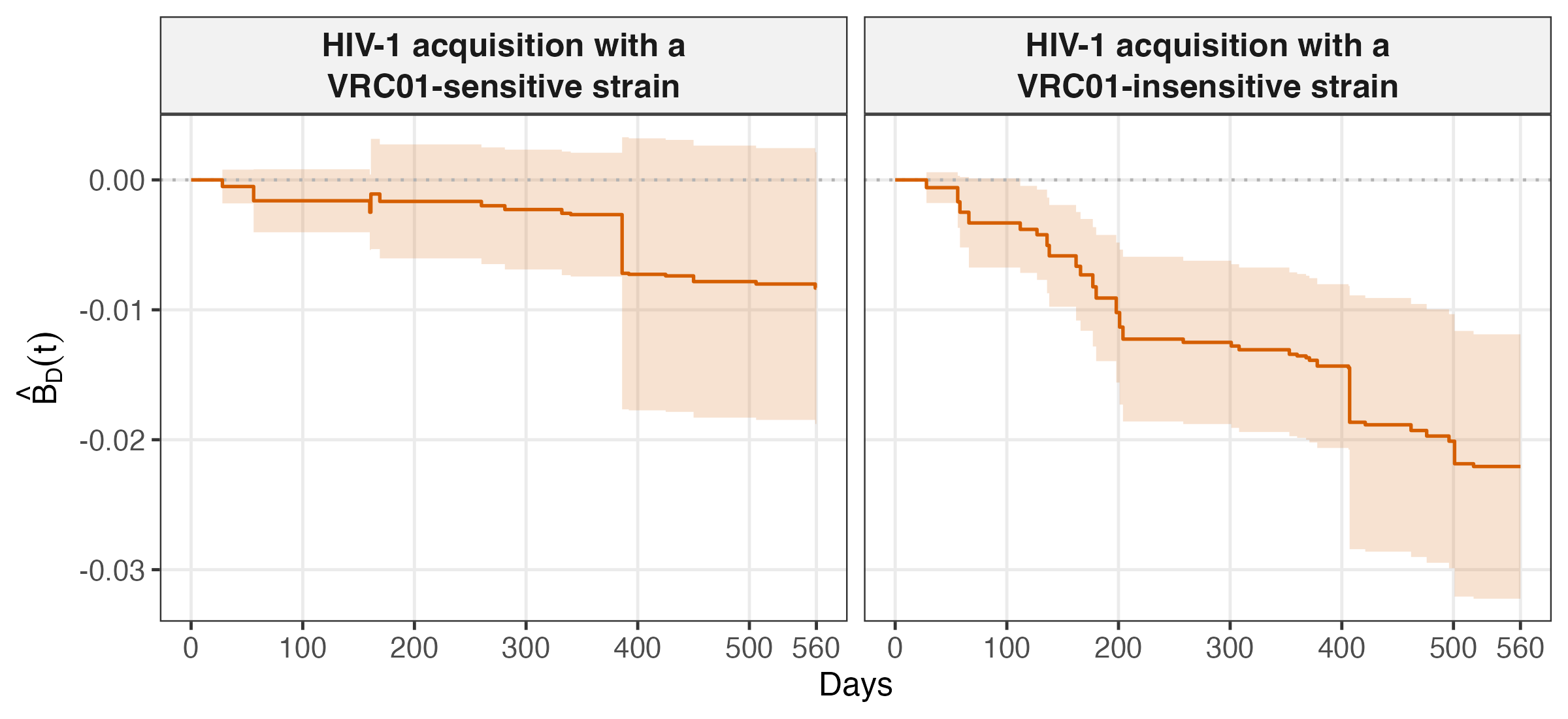}
    \caption{Estimated cumulative effects of oral PrEP on HIV-1 acquisition with a VRC01-sensitive strain (left panel) and VRC01-insensitive strain (right panel). Shaded region indicates pointwise 95\% confidence intervals. A negative value of $\hat{B}_D(t)$ indicates that oral PrEP reduces the hazard of acquiring HIV-1 of the given strain type.}
    \label{fig:B_D_si}
\end{figure}

Figure~\ref{fig:B_D_si} shows $\hat{B}_D(t)$, the estimated cumulative effect of oral PrEP on acquisition of VRC01-sensitive (left panel) and VRC01-insensitive (right panel) HIV-1 strains, with shaded bands indicating 95\% pointwise confidence intervals. Here, $\hat{B}_D(t)$ is the cumulative effect through time $t$ on the subdistribution hazard scale. For example, $\hat{B}_D(t) = -0.01$ corresponds roughly to one fewer HIV-1 acquisition of the given strain type per 100 participants who would remain free of that strain type by time $t$ under continuous oral PrEP use through $t$, compared with no use. For both strain types, the estimates become progressively more negative over follow-up, suggesting a protective effect of oral PrEP \citep[see, e.g.,][]{grant2010preexposure}.

If we further assume a constant additive effect on the subdistribution hazard, that is, $B_D(t) = \beta t$, the estimated effect of oral PrEP is $-1.49 \times 10^{-5}$ per day (95\% CI: $-3.38 \times 10^{-5}$ to $0.41 \times 10^{-5}$) on the acquisition of VRC01-sensitive strains and $-4.00 \times 10^{-5}$ per day (95\% CI: $-5.84 \times 10^{-5}$ to $-2.16 \times 10^{-5}$) on the acquisition of VRC01-insensitive strains. Both estimates are negative, but only the 95\% confidence interval for VRC01-insensitive strains excludes zero.  %Because oral PrEP acts by inhibiting viral reverse transcription, %whereas VRC01 targets the CD4-binding site of the HIV-1 envelope \citep{wu2010rational},
%its protection should not depend on a strain's sensitivity to VRC01 \citep{anderson2012emtricitabine}.

\begin{figure}[h]
    \centering
\includegraphics[width=0.95\linewidth]{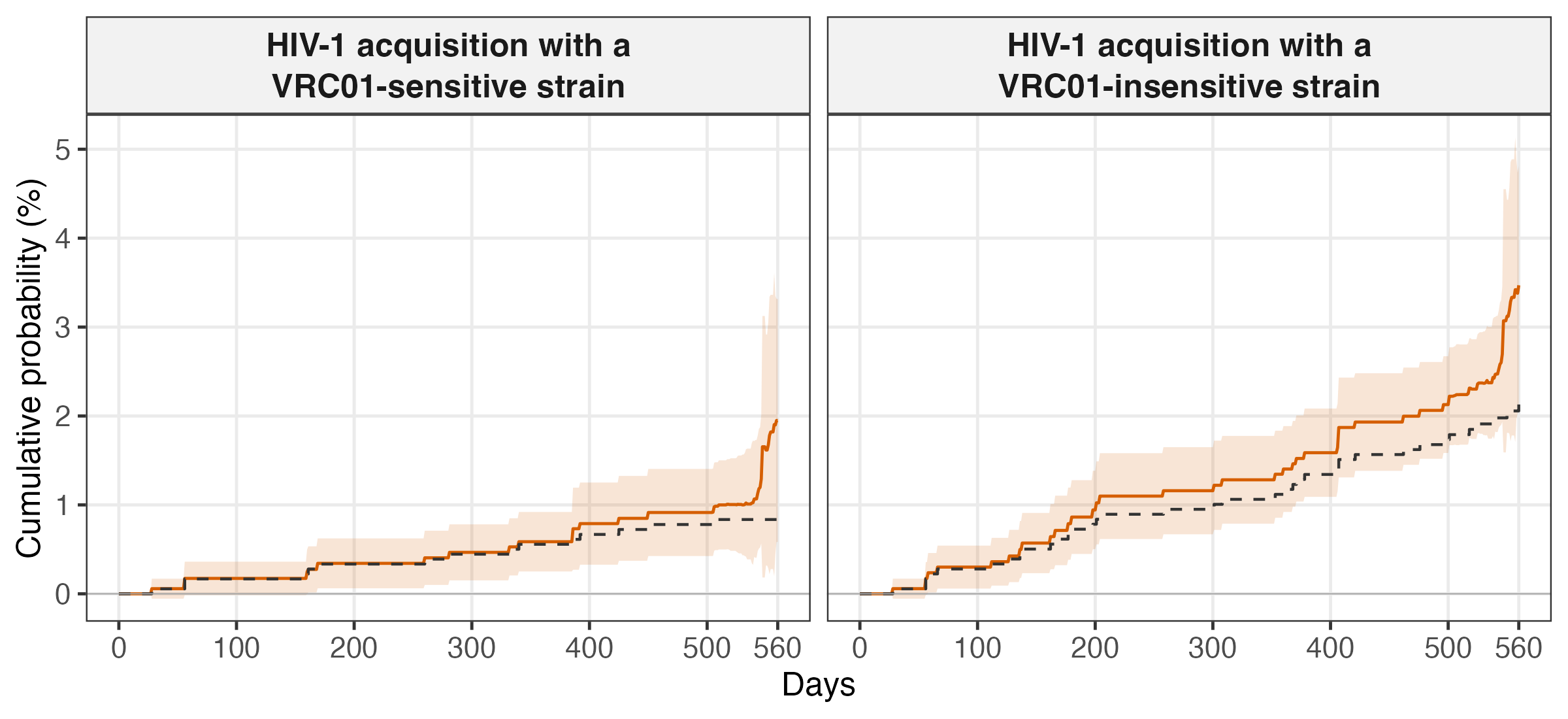}
    \caption{Orange solid lines: estimated counterfactual cause-specific cumulative incidence functions under no oral PrEP uptake. Shaded region indicates pointwise 95\% confidence intervals. Black dashed lines: observed cause-specific cumulative incidence functions estimated using the Fine--Gray model without removing the oral PrEP uptake.}
    \label{fig:cumulative_si}
\end{figure}

Finally, Figure~\ref{fig:cumulative_si} compares two sets of strain-specific CIFs, both pooled across the VRC01 arms: the estimated counterfactual CIFs under no oral PrEP uptake (orange solid lines), and the intention-to-treat CIFs estimated from a Fine--Gray model, which reflect oral PrEP uptake as it actually occurred (black dashed lines). For both causes, the ``oral PrEP-free'' cumulative incidences are higher than the observed ones, again consistent with a protective effect of oral PrEP.

To assess the robustness of our findings, we conducted a sensitivity analysis where we defined $D_1$ using the FTC-TP measurement on day~0 and $D_k$ using the last FTC-TP measurement in $(t_{k-1}, t_{k}]$ for $k\geq 2$, rather than the first measurement between $[t_k,t_{k+1})$. The result, reported in Supplementary Material~\ref{app:sensitivity analysis for the case study}, is qualitatively similar to those presented above.

%conducted two sensitivity analyses. First, we defined $D_k$ using TFV-DP instead of FTC-TP. Second, we defined $D_1$ using the FTC-TP measurement on day~0 and $D_k$ using the last FTC-TP measurement in $(t_{k-1}, t_{k}]$ for $k\geq 2$, rather than the first measurement between $[t_k,t_{k+1})$. The results, reported in Supplementary Material~\ref{app:Sensitivity analysis for the case study}, are qualitatively similar to those presented above.

\subsection{Effect of VRC01 infusion and a per-protocol reanalysis}
\label{subsec: case study IV analysis}
We next studied the effect of VRC01 infusions administered every 8 weeks at 10 mg/kg or 30 mg/kg on the acquisition of VRC01-sensitive and VRC01-insensitive HIV-1 strains, as well as the counterfactual cause-specific cumulative incidence functions under full adherence to the VRC01 infusion schedule. After excluding participants with missing primary outcome data, unknown strain sensitivity, or no follow-up, the analysis included 2,642 participants who had been randomized in a 1:1:1 ratio to high-dose VRC01 at 30 mg/kg ($Z = 2$), low-dose VRC01 at 10 mg/kg ($Z = 1$), or placebo ($Z = 0$).

We estimated the effect of each VRC01 dose separately, comparing participants randomized to that dose (30 mg/kg or 10 mg/kg) with those randomized to placebo. We used the same $K = 10$ time intervals with administrative censoring at $\tau = 560$ days as in the previous analysis. These 10 intervals correspond to the 10 scheduled infusions. For each interval, we defined $D_k = 1$ if the participant received placebo or skipped the scheduled infusion, and $D_k = 0$ if the participant received the scheduled VRC01 infusion. Because only participants randomized to a VRC01 arm could receive VRC01, randomized assignment is a strong instrumental variable for subsequent VRC01 infusions.

\begin{figure}[ht]
    \centering
    \includegraphics[width=0.95\linewidth]{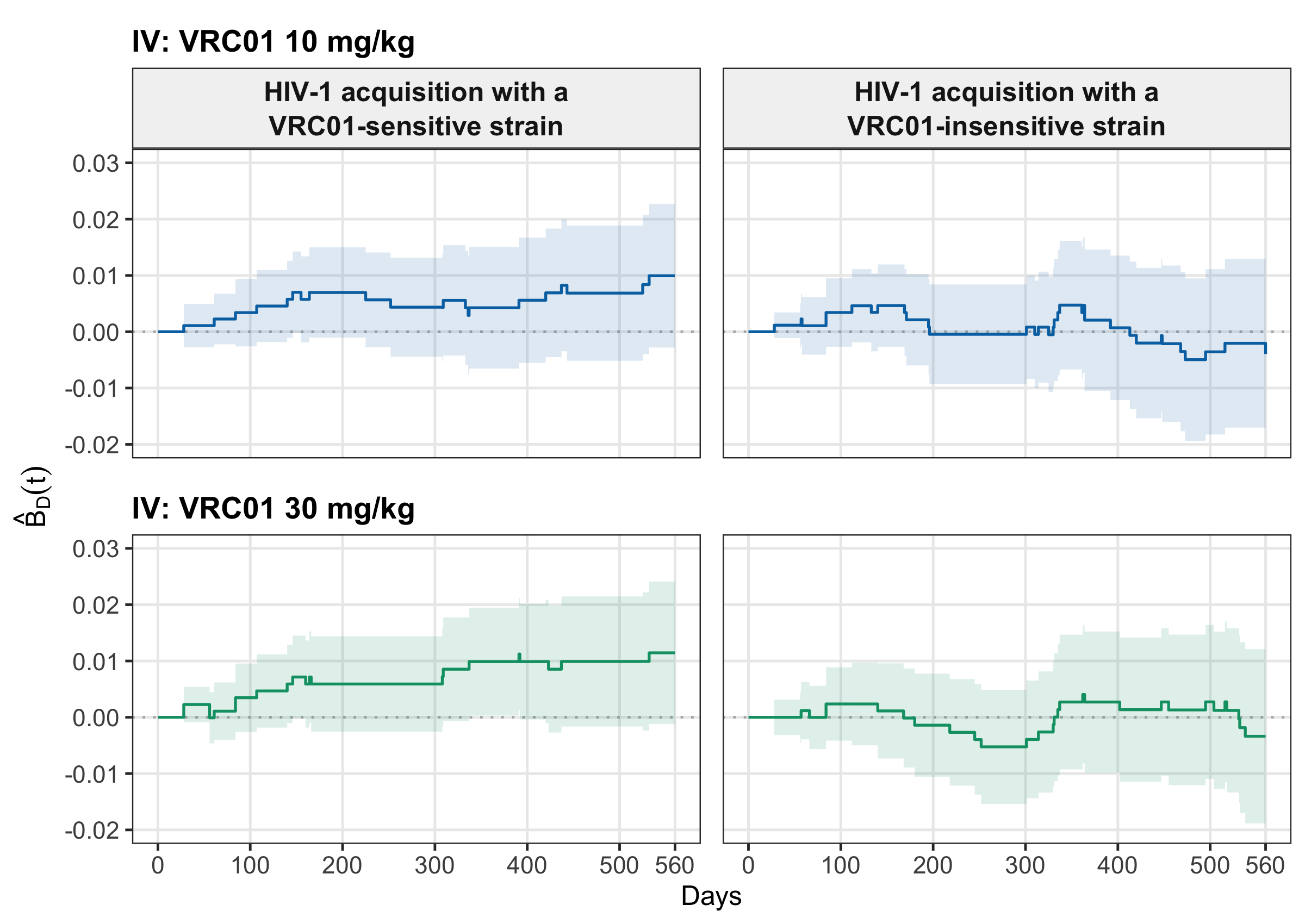}
    \caption{Estimated cumulative effects of skipping VRC01 infusions administered every 8 weeks at 10 mg/kg (top panels) and 30 mg/kg (bottom panels) on HIV-1 acquisition with a VRC01-sensitive strain (left panels) and a VRC01-insensitive strain (right panels). Shaded region indicates pointwise 95\% confidence intervals. A positive value of $\hat{B}_D(t)$ indicates that skipping VRC01 infusions increases the hazard of acquiring HIV-1 of the given strain type.}
    \label{fig:B_D_iv}
\end{figure}

Figure~\ref{fig:B_D_iv} shows the estimated cumulative effects, obtained using the IV-based methods described in Section~\ref{subsec: IV estimation and inference}, of omitting versus receiving VRC01 infusions every 8 weeks at 10 mg/kg (top panels) and 30 mg/kg (bottom panels). For both doses, omitting VRC01 appeared to increase the risk of acquiring VRC01-sensitive, but not VRC01-insensitive, HIV-1 strains.

If we further assume a constant additive effect on the hazard, that is, $B_D(t) = \beta t$, the estimated effects of omitting 10 mg/kg VRC01 infusions are $1.79 \times 10^{-5}$ per day (95\% CI: $-0.51 \times 10^{-5}$ to $4.08 \times 10^{-5}$) on the acquisition of VRC01-sensitive strains and $-0.57 \times 10^{-5}$ per day (95\% CI: $-3.30 \times 10^{-5}$ to $2.15 \times 10^{-5}$) on the acquisition of VRC01-insensitive strains. The corresponding estimates for the 30 mg/kg dose are $2.09 \times 10^{-5}$ per day (95\% CI: $-0.18 \times 10^{-5}$ to $4.36 \times 10^{-5}$) and $-0.54 \times 10^{-5}$ per day (95\% CI: $-3.29 \times 10^{-5}$ to $2.22 \times 10^{-5}$), respectively. Although all confidence intervals include zero, the estimates for VRC01-sensitive strains point toward increased risk without VRC01 at both doses, whereas those for VRC01-insensitive strains are close to zero. These findings are consistent with the expectation that the protective effect of VRC01 depends on the sensitivity of the infecting strain to the antibody \citep{wu2010rational,corey2021two}.

\begin{figure}[ht]
    \centering
    \includegraphics[width=0.95\linewidth]{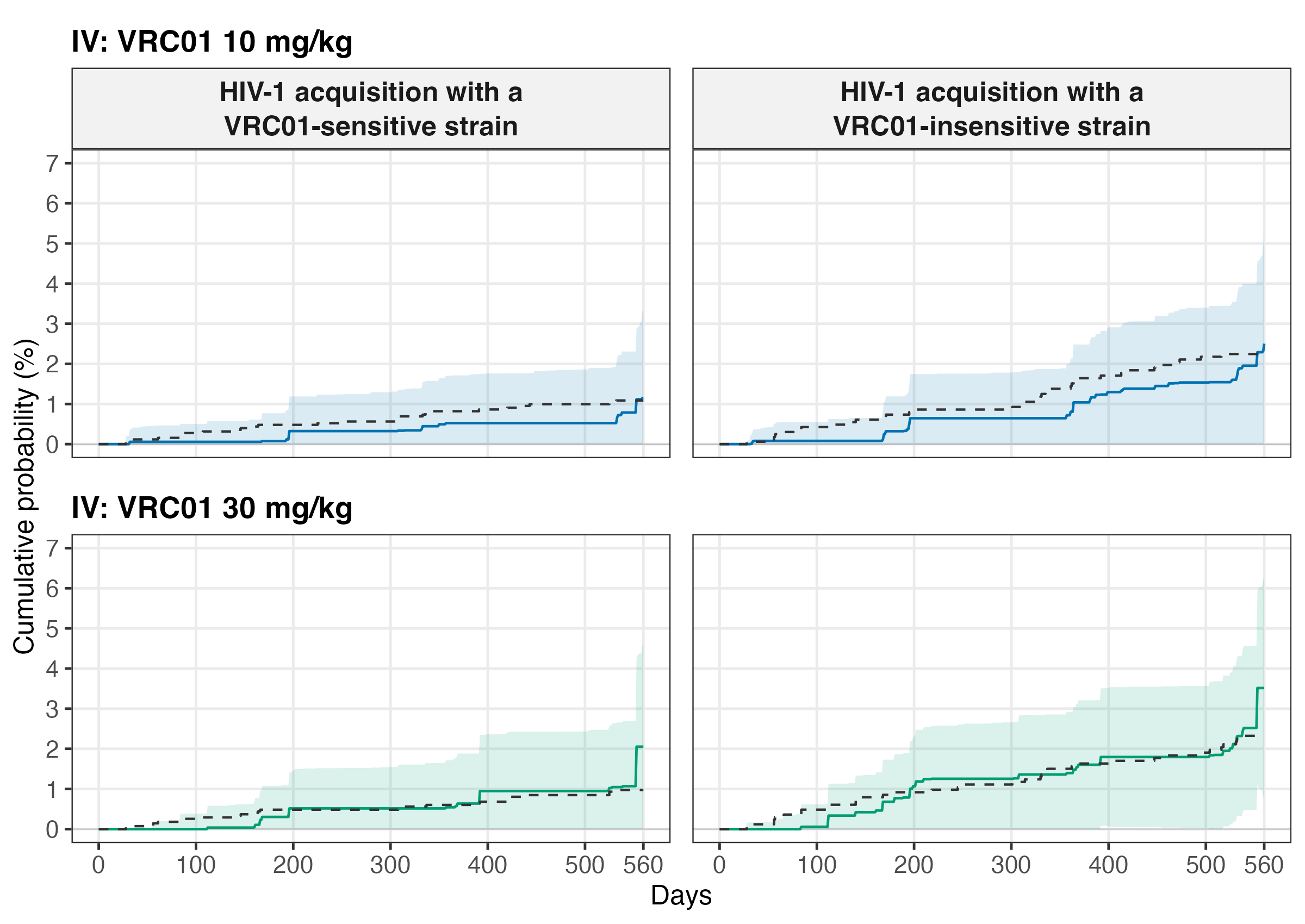}
    \caption{Solid lines: estimated counterfactual cause-specific cumulative incidence functions under full adherence to scheduled VRC01 infusions. Shaded region indicates pointwise 95\% confidence intervals. Black dashed lines: intention-to-treat estimates of cause-specific cumulative incidence functions using the Fine--Gray model.}
    \label{fig:cumulative_iv}
\end{figure}

Finally, Figure~\ref{fig:cumulative_iv} shows the estimated cause-specific CIFs under a counterfactual scenario in which participants received all scheduled VRC01 infusions (``per-protocol''), alongside the intention-to-treat estimates. For both doses and both causes, the per-protocol estimates were close to the intention-to-treat estimates, consistent with the high adherence to the infusion schedule in the trial: participants received a mean of 9.0 infusions, and 79\% received all 10.

\section{Discussion}
\label{sec: discussion}
We generalized the structural nested cumulative survival time model to a cause-specific version and studied estimation and inference under two complementary frameworks: sequential ignorability (SI) and instrumental variables (IV). The IV framework is suited to settings in which the randomized treatment assignment is associated with subsequent treatment uptake, for example, when uptake represents adherence to the assigned treatment. The SI framework applies when no such instrument is available, for example when uptake involves a treatment not influenced by the initial assignment, provided that all confounders of uptake and the outcome are measured. 

We applied both methods to HVTN 704/HPTN 085, one of the two Antibody Mediated Prevention (AMP) trials. In the first part of the case study, we estimated that oral PrEP uptake reduced the risk of HIV-1 acquisition for both VRC01-sensitive and VRC01-insensitive strains, consistent with oral PrEP's mechanism of HIV-1 prevention. Using the proposed method, we also recovered the cause-specific cumulative incidence functions in the absence of oral PrEP uptake. In the second part, we found that VRC01 infusions administered every 8 weeks tended to reduce the risk of acquiring VRC01-sensitive strains, with no apparent effect on VRC01-insensitive strains. This pattern agrees with the sieve analysis reported by \citet{corey2021two}. The proposed method also provides a principled way to estimate the per-protocol effect, that is, the effect under full adherence to the scheduled VRC01 infusions.

The proposed methods can also be applied to the other AMP trial, HVTN 703/HPTN 081. However, we did not pursue this analysis because oral PrEP uptake was very low among women participants in sub-Saharan Africa in HVTN 703/HPTN 081. During 3300 person-years of follow-up in HVTN 703/HPTN 081, oral PrEP was detectable in only 3.8\% (95\% CI: 2.8 to 4.9) person-years \citep{corey2021two}.

The oral PrEP analysis was constrained by the granularity of the available data. Because the oral PrEP biomarkers were measured approximately every 8 weeks, we used 8-week time intervals and treated oral PrEP uptake as constant within each interval. With more frequent measurements of oral PrEP use, the analysis could be adapted to a finer time grid, under which the assumption of constant uptake within intervals would be more plausible.

\section*{Acknowledgements}
We thank the participants, site staff, and investigators of the Antibody Mediated Prevention trials sponsored by the National Institute of Allergy and Infectious Diseases of the National Institutes of Health (NIAID NIH). We thank Dr. Peter Gilbert and Dr. Allan deCamp for helpful feedback and facilitating access to the data. This work is funded by the NIAID NIH (R01AI192632, R37AI054165, UM1AI068635 and UM1AI068614). The content is solely the responsibility of the authors and does not necessarily represent the official views of the National Institutes of Health.

\singlespacing
\bibliography{reference}

\clearpage
\appendix

\onehalfspacing
\begin{center}
    {\bf \Large Supplemental Materials to ``Cause-specific structural nested cumulative survival time models for competing risks, with application to the Antibody Mediated Prevention trials" by Yuxiao Li, Yuhao Deng, and Bo Zhang}
\end{center}

\section{Proofs }

We provide more detailed proofs of uniform consistency and weak convergence for the IV-based strategy. The corresponding proofs under sequential ignorability follow similar arguments, so we omit repetitive steps.

\subsection{Sequential ignorability-based Strategy}

\subsubsection{Proof of Theorem~\ref{thm:SI_moment_validity}}
\label{app:proof_SI_moment_validity}
\begin{proof}
In the following proof, we make no distinction between $L$ and $L^*$, and omit $Z$ in the potential outcome for simplicity.

For data without censoring, we establish the validity of its moment equation first.

\begin{lemma}
\label{lem:SI_moment_full_data}
Under the cause-specific SNCSTM \eqref{eq:cause_specific_SNCSTM},
consistency, and Assumption~\ref{assumption:SI}, for each
$k=1,\ldots,K$ and all $t\in[t_k,t_{k+1})$,
\begin{equation}
    \mathbb E\Bigl[
    \mathbf 1\{\tilde T\geq t_k\}
    \Bigl(
    D_k-
    \mathbb E[
    D_k
    \mid
    \bar D_{k-1},L,
    \tilde T\geq t_k,
    \tilde Y^1(t;\bar D_k,\bar 0)=1
    ]
    \Bigr)
    \tilde Y^1(t)
    \Bigl(
    d\tilde N^1(t)-D_k\,dB_D(t)
    \Bigr)
    \Bigr]
    =0.
\end{equation}
\end{lemma}

For notational simplicity, define the residual
$$
e_k(t\mid \bar D_k,L,\tilde T\geq t_k) := D_k - \mathbb{E}[D_k\mid \bar D_{k-1},L,\tilde T\geq t_k,\tilde Y^1(t;\bar D_{k},\bar 0)=1 ],
$$
so that $\mathbb{E}[e_k(t\mid \bar D_k,L,\tilde T\geq t_k) \mid \bar D_{k-1},L,\tilde T\geq t_k,\tilde Y^1(t;\bar D_k,\bar 0)=1]=0$. Then for $t\in[t_k,t_{k+1})$, we prove a stronger conclusion for future use.

$$
\begin{aligned}
&\mathbb E\Bigl[e_k(t\mid \bar D_k,L,\tilde T\geq t_k) \tilde Y^1(t)\Bigl(d\tilde N^1(t)- D_k\,dB_D(t) \Bigr) \mid L,\tilde T\geq t_k\Bigr]\\
=&\mathbb E\Bigl[e_k(t\mid \bar D_k,L,\tilde T\geq t_k) \tilde Y^1(t;\bar D_k,\bar 0)\Bigl(d\tilde N^1(t;\bar D_k,\bar 0)- D_k\,dB_D(t) \Bigr) \mid L,\tilde T\geq t_k\Bigr]\\
=& \mathbb E\Bigl[e_k(t\mid \bar D_k,L,\tilde T\geq t_k) \tilde Y^1(t;\bar D_k,\bar 0)\mathbb E [\Bigl(d\tilde N^1(t;\bar D_k,\bar 0)- D_k\,dB_D(t) \Bigr) \mid \bar D_k,L,\tilde T\geq t_k,\tilde Y^1(t;\bar D_k,\bar 0) ] \mid L,\tilde T\geq t_k\Bigr]\\
=& \mathbb E\Bigl[e_k(t\mid \bar D_k,L,\tilde T\geq t_k) \tilde Y^1(t;\bar D_k,\bar 0)\times \\
&\qquad \mathbb E [\Bigl(d\tilde N^1(t;\bar D_k,\bar 0)- D_k\,dB_D(t) \Bigr) \mid \bar D_k,L,\tilde T\geq t_k,\tilde Y^1(t;\bar D_k,\bar 0)=1 ] \mid L,\tilde T\geq t_k\Bigr]\\
=& \mathbb E\Bigl[e_k(t\mid \bar D_k,L,\tilde T\geq t_k) \tilde Y^1(t;\bar D_k,\bar 0)\mathbb E [d\tilde N^1(t;\bar D_{k-1},\bar 0)  \mid \bar D_{k-1},L,\tilde T\geq t_k,\tilde Y^1(t;\bar D_{k-1},\bar 0)=1 ] \mid L,\tilde T\geq t_k\Bigr]\\
=& \mathbb E\Bigl[e_k(t\mid \bar D_k,L,\tilde T\geq t_k) \tilde Y^1(t;\bar D_k,\bar 0)\delta(t\mid \bar D_{k-1},L ,\tilde T\geq t_k)\mid L,\tilde T\geq t_k\Bigr]\\
=& \mathbb E\Bigl[\mathbb{E}[ e_k(t\mid \bar D_k,L,\tilde T\geq t_k) \mid \bar D_{k-1},L,\tilde T\geq t_k,\tilde Y^1(t;\bar D_k,\bar 0)]\tilde Y^1(t;\bar D_k,\bar 0)\delta(t\mid \bar D_{k-1},L ,\tilde T\geq t_k)\mid L,\tilde T\geq t_k\Bigr]\\
=& \mathbb E\Bigl[\mathbb{E}[ e_k(t\mid \bar D_k,L,\tilde T\geq t_k) \mid \bar D_{k-1},L,\tilde T\geq t_k,\tilde Y^1(t;\bar D_k,\bar 0)=1]\tilde Y^1(t;\bar D_k,\bar 0)\delta(t\mid \bar D_{k-1},L ,\tilde T\geq t_k)\mid L,\tilde T\geq t_k\Bigr]\\
=& 0
\end{aligned}
$$

This proves the lemma.

Consider censoring $C$ and define $Y^{1,c}(t)=\mathbf1\{C\geq t \}\tilde Y^1(t)$. Assume that $Y^{1,c}(t)$ can be observed, like in censoring-complete data, $Y^1(t) = Y^{1,c}(t)$. We establish the validity of moment equation for this case.

\begin{lemma}
\label{lem:SI_moment_censoring_complete}
Under the conditions of Lemma~\ref{lem:SI_moment_full_data} and
Assumption~\ref{assumpion:conditional_independent_censoring_si},
for each $k=1,\ldots,K$ and all $t\in[t_k,t_{k+1})$,
\begin{equation}
\mathbb E\Bigl[
\mathbf 1\{T\geq t_k\}
\Bigl(
D_k-
\mathbb E[
D_k
\mid
\bar D_{k-1},L^*,
\tilde T\geq t_k,
\tilde Y^{1}(t;\bar D_k,\bar 0)=1
]
\Bigr)
Y^{1,c}(t)
\Bigl(
dN^1(t)-D_k\,dB_D(t)
\Bigr)
\Bigr]
=0.
\label{eq:SI_moment2}
\end{equation}
\end{lemma}
This result is derived from Lemma~\ref{lem:SI_moment_full_data}.
$$
\begin{aligned}
&\mathbb E\Bigl[ \mathbf 1\{T\geq t_k \}\Bigl(D_k - \mathbb{E}[D_k\mid \bar D_{k-1},L,\tilde T\geq t_k,\tilde Y^1(t;\bar D_{k},\bar 0)=1 ]\Bigr)Y^{1,c}(t)\Bigl(dN^1(t)- D_k\,dB_D(t) \Bigr) \mid L\Bigr] \\
=&\mathbb E \Bigl[\mathbf 1\{\tilde T\geq t_k \}\mathbf 1\{C\geq t_k \}
\Bigl(D_k - \mathbb{E}[D_k\mid \bar D_{k-1},L,\tilde T\geq t_k,\tilde Y^1(t;\bar D_{k},\bar 0)=1 ]\Bigr)\\
&\qquad {}\times \mathbf 1\{C\geq t \}\tilde Y^1(t)
\Bigl(\mathbf 1\{C\geq t\}d\tilde N^1(t)-D_k\,dB_D(t)\Bigr)\mid L\Bigr]\\
=&\mathbb E \Bigl[\mathbf 1\{C\geq t \}\mathbf 1\{\tilde T\geq t_k \} \Bigl(D_k - \mathbb{E}[D_k\mid \bar D_{k-1},L,\tilde T\geq t_k,\tilde Y^1(t;\bar D_{k},\bar 0)=1 ]\Bigr) \tilde Y^1(t)\Bigl(d\tilde N^1(t)- D_k\,dB_D(t) \Bigr)\mid L \Bigr]\\
=& P(C\geq t\mid L)\mathbb E \Bigl[ \mathbf 1\{\tilde T\geq t_k \} \Bigl(D_k - \mathbb{E}[D_k\mid \bar D_{k-1},L,\tilde T\geq t_k,\tilde Y^1(t;\bar D_{k},\bar 0)=1 ]\Bigr)\tilde Y^1(t)\Bigl(d\tilde N^1(t)- D_k\,dB_D(t) \Bigr) \mid L\Bigr] \\
=& 0
\end{aligned}
$$

This proves the lemma.

For incomplete data, where $Y^{1,c}(t)$ is unobservable, we need IPCW weights in \citet{fine1999proportional} to construct observable indicator $Y^{1,w}(t)$. Use $\mathcal O_t$ to denote observed history up to $t$. An important property of this weight is
$$
\mathbb E[ Y^{1,c}(t) \mid \mathcal O_t]=Y^{1,w}(t)
$$
Prove this first. Consider 4 cases. If $\mathcal O_t$ indicates $\{T\geq t\}$, then $Y^{1,c}(t)=1$. If $\mathcal O_t$ indicates $\{T< t,\Delta =1,V=1\}$ or $\{T<t,\Delta =0\}$, then $Y^{1,c}(t)=0$. If $\mathcal O_t$ indicates $\{T< t,\Delta =1,V\neq 1\}$, then
$$
\begin{aligned}
\mathbb{E}[Y^{1,c}(t)\mid \mathcal O_t]&=\mathbb{E}[\mathbf 1\{ C\geq t\}\mid \mathcal O_t]\\
&=P(C\geq t\mid C\geq T,L )\\
&=\frac{G(t\mid L)}{G(T\mid L)}\\
\end{aligned}
$$

This proves the stated property. So, within the expectation, $Y^{1,w}(t)$ can be viewed as $Y^{1,c}(t)$. Then we are able to establish the validity of the moment equation.
$$
\begin{aligned}
&\mathbb E [\mathbf 1\{T\geq t_k \} \Bigl( D_k - \mathbb{E}[D_k\mid \bar D_{k-1},L,\tilde T\geq t_k,\tilde Y^1(t;\bar D_{k},\bar 0)=1 ]\Bigr)Y^{1,w}(t)\Bigl(dN^1(t)- D_k\,dB_D(t) \Bigr) \mid L]\\
=&\mathbb E [\mathbf 1\{ T\geq t_k \} \Bigl( D_k - \mathbb{E}[D_k\mid \bar D_{k-1},L,\tilde T\geq t_k,\tilde Y^1(t;\bar D_{k},\bar 0)=1 ]\Bigr)\mathbb E[ Y^{1,c}(t) \mid \mathcal O_t]\Bigl(dN^1(t)- D_k\,dB_D(t) \Bigr) \mid L]\\
=&\mathbb E\Bigl[ \mathbb E [\mathbf 1\{ T\geq t_k \} \Bigl( D_k - \mathbb{E}[D_k\mid \bar D_{k-1},L,\tilde T\geq t_k,\tilde Y^1(t;\bar D_{k},\bar 0)=1 ]\Bigr) Y^{1,c}(t)\Bigl(dN^1(t)- D_k\,dB_D(t) \Bigr)  \mid \mathcal O_t] \mid L\Bigr]\\
=&\mathbb E\Bigl[ \mathbf 1\{ T\geq t_k \} \Bigl( D_k - \mathbb{E}[D_k\mid \bar D_{k-1},L,\tilde T\geq t_k,\tilde Y^1(t;\bar D_{k},\bar 0)=1 ]\Bigr)\mathbf1\{C\geq t \}\tilde Y^1(t)\Bigl(d\tilde N^1(t)- D_k\,dB_D(t) \Bigr) \mid L\Bigr]\\
=&P(C\geq t\mid L)\mathbb E\Bigl[ \mathbf 1\{\tilde T\geq t_k \} \Bigl( D_k - \mathbb{E}[D_k\mid \bar D_{k-1},L,\tilde T\geq t_k,\tilde Y^1(t;\bar D_{k},\bar 0)=1 ]\Bigr)\tilde Y^1(t)\Bigl(d\tilde N^1(t)- D_k\,dB_D(t) \Bigr) \mid L\Bigr]\\
=&0
\end{aligned}
$$
This proves the theorem.
\end{proof}

\subsubsection{Proof of Proposition~\ref{prop:conditional_treatment_mean}}
\label{app:conditional_treatment_mean}
\begin{proof}

Under the cause-specific SNCSTM and sequential ignorability, we have
$$
\begin{aligned}
\mathbb E[\tilde Y^1(t;\bar D_{k},\bar 0) \mid \bar D_k,L^*,\tilde T\geq t_k ] &= \mathbb E[\tilde Y^1(t;\bar D_{k-1},\bar 0) \mid \bar D_k,L^*,\tilde T\geq t_k ]\exp\{-\int_{t_k}^{t\wedge t_{k+1} } D_kdB_D(s) \}\\
&= \mathbb E[\tilde Y^1(t;\bar D_{k-1},\bar 0) \mid \bar D_{k-1},L^*,\tilde T\geq t_k ]\exp\{-\int_{t_k}^{t\wedge t_{k+1} } D_kdB_D(s) \}\\
\end{aligned}
$$

Then we prove this proposition using this result.

$$
\begin{aligned}
&\mathbb{E}[D_k\mid \bar D_{k-1},L,\tilde T\geq t_k,\tilde Y^1(t;\bar D_k,\bar 0)=1 ]\\
=& \frac{\mathbb E[\tilde Y^1(t)D_k\mid \bar D_{k-1},L,\tilde T\geq t_k] }{\mathbb E[\tilde Y^1(t)\mid \bar D_{k-1},L,\tilde T\geq t_k] }\\
=& \frac{\mathbb E[\mathbb E[\tilde Y^1(t;\bar D_k,\bar 0)\mid \bar D_k,L,\tilde T\geq t_k]D_k\mid \bar D_{k-1},L,\tilde T\geq t_k] }{\mathbb E[\mathbb E[\tilde Y^1(t;\bar D_k,\bar 0)\mid \bar D_k,L,\tilde T\geq t_k]\mid \bar D_{k-1},L,\tilde T\geq t_k] }\\
=& \frac{\mathbb E[\mathbb E[\tilde Y^1(t;\bar D_{k-1},\bar 0)\mid \bar D_k,L,\tilde T\geq t_k] \exp\{-\int_{t_k}^{t\wedge t_{k+1}}D_kdB_D(s) \}D_k\mid \bar D_{k-1},L,\tilde T\geq t_k] }{\mathbb E[\mathbb E[\tilde Y^1(t;\bar D_{k-1},\bar 0)\mid \bar D_k,L,\tilde T\geq t_k]\exp\{-\int_{t_k}^{t\wedge t_{k+1}}D_kdB_D(s) \}\mid \bar D_{k-1},L,\tilde T\geq t_k] }\\
=&  \frac{\mathbb E[\tilde Y^1(t;\bar D_{k-1},\bar 0)\mid \bar D_{k-1},L,\tilde T\geq t_k]\times \mathbb E[ \exp\{-\int_{t_k}^{t\wedge t_{k+1}}D_kdB_D(s) \}D_k\mid \bar D_{k-1},L,\tilde T\geq t_k] }{\mathbb E[\tilde Y^1(t;\bar D_{k-1},\bar 0)\mid \bar D_{k-1},L,\tilde T\geq t_k] \times \mathbb E[\exp\{-\int_{t_k}^{t\wedge t_{k+1}}D_kdB_D(s) \}\mid \bar D_{k-1},L,\tilde T\geq t_k] }\\
=&  \frac{\mathbb E[ \exp\{-\int_{t_k}^{t\wedge t_{k+1}}D_kdB_D(s) \}D_k\mid \bar D_{k-1},L,\tilde T\geq t_k] }{\mathbb E[\exp\{-\int_{t_k}^{t\wedge t_{k+1}}D_kdB_D(s) \}\mid \bar D_{k-1},L,\tilde T\geq t_k] }\\
\end{aligned}
$$

\end{proof}

\subsubsection{Proof of Proposition~\ref{prop:SI_uniformly_consistent}}
\label{app:SI_uniformly_consistent}

Fix one $k\in\{1,\ldots,K\}$, and the following analysis is on the interval $[t_k,t_{k+1}]$. For a real-valued function $g$ defined on this interval, let $\|g\|_\infty$ denote its supremum norm and $\mathcal V(g)$ its total variation. Let $\|A_k\|_\infty=M^o<\infty$. Define
\[
e_k(s,H(\cdot))
=
D_k-
\frac{
\mathbb E\left[
D_k\exp\{-D_kH(s^-)\}
\mid \bar D_{k-1},L^*,T\geq t_k
\right]
}{
\mathbb E\left[
\exp\{-D_kH(s^-)\}
\mid \bar D_{k-1},L^*,T\geq t_k
\right]
}.
\]

\begin{assumption}
\label{assumption:SI_uniform_assumption1}
$e_k$ is bounded by $M_{\max}$.
\end{assumption}

This holds when $D_k\in[0,1]$, which we assume throughout. We add subscript $i$ to denote the $i$-th individual. Define
\[
\begin{aligned}
E_n(s,H(\cdot))
&= \frac{1}{n}\sum_{i=1}^n
e_{i,k}(s,H(\cdot))\check Y_i^1(s)D_{i,k}\\
E(s,H(\cdot))
&= \mathbb E[
e_k(s,H(\cdot))\check Y^1(s)D_k]\\
\Upsilon_n(H(\cdot),t)
&= \int_{t_k}^t
\frac{
\sum_{i=1}^n e_{i,k}(s,H(\cdot))dN_i^1(s)
}{
nE_n(s,H(\cdot))
}\\
\bar\Upsilon_n(H(\cdot),t)
&= \int_{t_k}^t
\frac{
\sum_{i=1}^n e_{i,k}(s,H(\cdot))dN_i^1(s)
}{
nE(s,H(\cdot))
}\\
\Upsilon(H(\cdot),t)
&= \int_{t_k}^t
\frac{
\mathbb E[e_k(s,H(\cdot))dN^1(s)]
}{
E(s,H(\cdot))
}.
\end{aligned}
\]

\begin{assumption}
\label{assumption:SI_uniform_assumption2}
There exist constants $M_0>\|A_k\|_\infty$,
$M>\mathcal V(A_k)$, and $\nu>0$ such that
\[
\inf_{s\in[t_k,t_{k+1}],\,H\in\mathbb H_{M_0,M}}
E(s,H)>\nu,
\]
and
\[
M>\frac{M_{\max}}{\nu},
\]
where
\[
\mathbb H_{M_0,M}
=
\left\{
H\in D[t_k,t_{k+1}]:
H(t_k)=0,\,
\|H\|_\infty\le M_0,\,
\mathcal V(H)\le M
\right\}.
\]
\end{assumption}

\begin{assumption}
\label{assumption:SI_uniform_assumption3}
$A_k(t)$ is the unique continuous solution to
$B(t)=\Upsilon(B(\cdot),t)$ in $t\in[t_k,t_{k+1}]$.
\end{assumption}

\begin{assumption}
\label{assumption:SI_uniform_assumption4}
Assume that $N^1$ admits the Doob--Meyer decomposition
\[
dN^1(t)
=
Y(t)\lambda_1(t)\,dt
+
dM^1(t),
\]
where $\lambda_1(t)$ is the cause-specific hazard and
\[
\sup_{t\in[t_k,t_{k+1}]}\lambda_1(t)
\leq C_0<\infty.
\]
\end{assumption}

We use Helly's selection theorem as in the IV proof.

\begin{lemma}
\label{lemma:SI_uniform}
Under Assumptions~\ref{assumption:SI_uniform_assumption1}
and \ref{assumption:SI_uniform_assumption2},
\[
\sup_{s\in[t_k,t_{k+1}],\,H\in\mathbb H_{M_0,M}}
|E_n(s,H(\cdot))-E(s,H(\cdot))|
\xrightarrow{\mathrm{a.s.}}0.
\]
\end{lemma}

Define
\[
\phi_{H,s}(O)
=
e_k(s,H(\cdot))\check Y^1(s)D_k,
\qquad
\Phi_{M_0,M}
=
\left\{
\phi_{H,s}:H\in\mathbb H_{M_0,M},\
s\in[t_k,t_{k+1}]
\right\}.
\]

Since $H\in\mathbb H_{M_0,M}$ has uniformly bounded total variation, and $e_k(s,H(\cdot))$, $\check Y^1(\cdot)$, and $D_k$ are uniformly bounded, the class $\Phi_{M_0,M}$ is a uniformly bounded bounded-variation class. By \citet[Example~19.11]{vandervaart2000}, $\Phi_{M_0,M}$ is Glivenko--Cantelli. Therefore,
\[
\sup_{s\in[t_k,t_{k+1}],\,H\in\mathbb H_{M_0,M}}
\left|
(\mathbb P_n-\mathbb P)\phi_{H,s}
\right|
\xrightarrow{\mathrm{a.s.}}0,
\]
which proves the lemma.

\paragraph{Step 1}
We first work under the oracle nuisance parameters $\theta$ and write $\hat A_k(t)=\hat A_k(t;\theta)$. Define $\xi(y)=\operatorname{sgn}(y)\min\{|y|,M_0\}$. Let $\tilde A_n(t)$ be the solution to
$$
\tilde A_n(t)=\Upsilon_n(\xi(\tilde A_n)(\cdot),t).
$$
By Lemma~\ref{lemma:SI_uniform} and Assumption~\ref{assumption:SI_uniform_assumption2}, for sufficiently large $n$, there is
$$
\inf_{s\in[t_k,t_{k+1}],\,H\in\mathbb H_{M_0,M}}
E_n(s,H(\cdot))\geq\nu.
$$
As in the IV proof, since $\tilde A_n(s)$ is recursively constructed based on information in $[t_k,s)$ and $\tilde A_n(t_k)=0$, we can apply this denominator bound by induction. Thus,
$$
\begin{aligned}
\mathcal V(\tilde A_n)
&=
\int_{t_k}^{t_{k+1}}
\frac{
\left|\sum_{i=1}^n
e_{i,k}(s,\xi(\tilde A_n)(\cdot))
\,dN_i^1(s)\right|
}{
n|E_n(s,\xi(\tilde A_n)(\cdot))|
}
\\
&\leq
\frac{M_{\max}}{\nu}
\frac1n\sum_{i=1}^n N_i^1(t_{k+1})
\\
&\leq
\frac{M_{\max}}{\nu}<M.
\end{aligned}
$$
Since $\tilde A_n(t_k)=0$, we also have
$$
\|\tilde A_n\|_\infty
\leq\mathcal V(\tilde A_n)
\leq\frac{M_{\max}}{\nu}<M.
$$
Therefore, we can apply Helly's selection theorem.

\paragraph{Step 2} We prove
$$
\sup_{s\in[t_k,t_{k+1}],\,H\in\mathbb H_{M_0,M}}
|\Upsilon_n(H(\cdot),s)-\Upsilon(H(\cdot),s)|
\rightarrow 0,\qquad \text{a.s.}
$$

We separate it into two parts, $\|\Upsilon_n-\bar\Upsilon_n\|_\infty$ and $\|\bar\Upsilon_n-\Upsilon\|_\infty$. First, we notice that
$$
\bar\Upsilon_n(H(\cdot),t)
=\frac1n\sum_{i=1}^n\psi_{H,t}(O_i)
=\mathbb P_n(\psi_{H,t}),
$$
where
$$
\psi_{H,t}(O)
=
\frac{
e_k(T,H(\cdot))
\{N^1(t)-N^1(t_k^-)\}
}{
E(T,H(\cdot))
}.
$$

Since $\Upsilon(H(\cdot),t)=\mathbb P(\psi_{H,t})$, following the same Glivenko--Cantelli argument as in the IV proof, we obtain
$$
\sup_{H\in\mathbb H_{M_0,M},\,t\in[t_k,t_{k+1}]}
\left|
\bar\Upsilon_n(H,t)-\Upsilon(H,t)
\right|
=
\left\|
(\mathbb P_n-\mathbb P)\psi_{H,t}
\right\|_\infty
\overset{a.s.}{\longrightarrow}0.
$$

Besides, by Lemma~\ref{lemma:SI_uniform} and Assumption~\ref{assumption:SI_uniform_assumption2},
$$
\begin{aligned}
&\sup_{H\in\mathbb H_{M_0,M},\,t\in[t_k,t_{k+1}]}
|\Upsilon_n(H,t)-\bar\Upsilon_n(H,t)|\\
&\leq
\frac{M_{\max}}{\nu^2}
\sup_{s\in[t_k,t_{k+1}],\,H\in\mathbb H_{M_0,M}}
|E_n(s,H)-E(s,H)|
\xrightarrow{\mathrm{a.s.}}0.
\end{aligned}
$$
Together, these establish the almost sure uniform convergence of
$\Upsilon_n$ to $\Upsilon$.

\paragraph{Step 3}

We use contradiction to prove that
$$
\|\tilde A_n-A_k\|_\infty\rightarrow 0.
$$
If this does not hold, there is one sub-sequence $\tilde A_{n_j}$, s.t.
$$
\liminf_{j\rightarrow\infty}
\|\tilde A_{n_j}-A_k\|_\infty>0.
$$

Given that $\mathcal V(\tilde A_n)\leq M$, $\|\tilde A_n\|_\infty\leq M$, by Helly's selection theorem, there is one sequence $\tilde A_{n_j}(t)\rightarrow\tilde A(t)$ pointwise. As in the IV proof, for any $t_k\leq u_1<u_2\leq t_{k+1}$, we have
$$
|\tilde A_{n_j}(u_2)-\tilde A_{n_j}(u_1)|
\leq
C^*[F_{n_j}(u_2)-F_{n_j}(u_1)],
\qquad C^*=\frac{M_{\max}}{\nu},
$$
where
$$
F_{n_j}(t)=\frac1{n_j}\sum_{i=1}^{n_j}N_i^1(t),
\qquad
F(t)=\mathbb P(T\leq t,\Delta=1,V=1).
$$
Since $\|F_{n_j}-F\|_\infty\overset{a.s.}{\longrightarrow}0$ and, by Assumption~\ref{assumption:SI_uniform_assumption4},
$$
F(u_2)-F(u_1)\leq C_0(u_2-u_1),
$$
we obtain
$$
|\tilde A(u_2)-\tilde A(u_1)|
\leq C^*C_0(u_2-u_1).
$$
Thus $\tilde A$ is Lipschitz continuous. Following the same finite-grid argument as in the IV proof, the pointwise convergence can be strengthened to uniform convergence:
$$
\|\tilde A_{n_j}-\tilde A\|_\infty\rightarrow 0.
$$

Going further, we prove that $\tilde A(t)$ is one solution to
$$
B(t)=\Upsilon(B(\cdot),t).
$$

To see this, we focus on $|\tilde A(t)-\Upsilon(\xi(\tilde A)(\cdot),t)|$ and separate it into 4 terms.
$$
\begin{aligned}
&|\tilde A(t)-\Upsilon(\xi(\tilde A)(\cdot),t)|
\\
\leq{}&
|\tilde A(t)-\tilde A_{n_j}(t)|
\\
&+
|\tilde A_{n_j}(t)-\Upsilon_{n_j}(\xi(\tilde A_{n_j})(\cdot),t)|
\\
&+
|\Upsilon_{n_j}(\xi(\tilde A_{n_j})(\cdot),t)
-\Upsilon(\xi(\tilde A_{n_j})(\cdot),t)|
\\
&+
|\Upsilon(\xi(\tilde A_{n_j})(\cdot),t)
-\Upsilon(\xi(\tilde A)(\cdot),t)|.
\end{aligned}
$$

The first term uniformly converges to 0. The second term is exactly 0. For the third term, since
$$
\mathcal V(\xi(\tilde A_{n_j}))
\leq\mathcal V(\tilde A_{n_j})\leq M,
\qquad
\|\xi(\tilde A_{n_j})\|_\infty\leq M_0,
$$
by Step 2, we know it uniformly converges to 0.

For the fourth term, we prove a Lipschitz continuity first: For any two $H_1,H_2\in\mathbb H_{M_0,M}$, there is one $C_*$, s.t.
$$
|\Upsilon(H_1(\cdot),t)-\Upsilon(H_2(\cdot),t)|
\leq C_*\|H_1-H_2\|_\infty.
$$

Define
$$
m_k(s,h)=
\frac{
\mathbb E[D_ke^{-D_k h}\mid\bar D_{k-1},L^*,T\geq t_k]
}{
\mathbb E[e^{-D_k h}\mid\bar D_{k-1},L^*,T\geq t_k]
}.
$$
By Cauchy--Schwarz inequality and $D_k\in[0,1]$, we know
$$
\left|\frac{\partial}{\partial h}m_k(s,h)\right|
\leq
\frac{
\mathbb E[D_k^2e^{-D_kh}\mid\bar D_{k-1},L^*,T\geq t_k]
}{
\mathbb E[e^{-D_kh}\mid\bar D_{k-1},L^*,T\geq t_k]
}
\leq 1.
$$
By the mean value theorem,
$$
\begin{aligned}
|e_k(s,H_1(\cdot))-e_k(s,H_2(\cdot))|
&=
|m_k(s,H_1(s^-))-m_k(s,H_2(s^-))|
\\
&\leq
|H_1(s^-)-H_2(s^-)|
\\
&\leq
\|H_1-H_2\|_\infty.
\end{aligned}
$$
Since $D_k\in[0,1]$ and $\check Y^1(s)\in[0,1]$, we also have
$$
|E(s,H_1)-E(s,H_2)|
\leq\|H_1-H_2\|_\infty.
$$
Therefore, following the same numerator and denominator decomposition, we established the Lipschitz continuity, with
$$
C_*=\frac1\nu+\frac{M_{\max}}{\nu^2}<\infty.
$$

Since $\xi(\tilde A_{n_j}),\xi(\tilde A)\in\mathbb H_{M_0,M}$ and truncation $\xi$ is 1-Lipschitz, we have
$$
\begin{aligned}
|\Upsilon(\xi(\tilde A_{n_j})(\cdot),t)
-\Upsilon(\xi(\tilde A)(\cdot),t)|
&\leq
C_*\|\xi(\tilde A_{n_j})-\xi(\tilde A)\|_\infty
\\
&\leq
C_*\|\tilde A_{n_j}-\tilde A\|_\infty
\rightarrow 0.
\end{aligned}
$$
Let $n_j\rightarrow\infty$, all four terms vanish, and we get $\tilde A(t)=\Upsilon(\xi(\tilde A)(\cdot),t)$.

To eliminate $\xi$, define
$$
\tau^*=\inf\{t\in[t_k,t_{k+1}]:|\tilde A(t)|\geq M_0\}.
$$
If $\tau^*\leq t_{k+1}$, then in $t\in[t_k,\tau^*]$, there is
$$
\tilde A(t)=\Upsilon(\tilde A(\cdot),t).
$$
In the discussion above, we enlarge $|e_k(s,H_1(\cdot))-e_k(s,H_2(\cdot))|$ to $\|H_1-H_2\|_\infty$. However, we can get another bound
$$
|e_k(s,H_1(\cdot))-e_k(s,H_2(\cdot))|
\leq
\sup_{u\in[t_k,s]}|H_1(u)-H_2(u)|.
$$
Using this bound and repeating the above process, we know that there exists a constant $C<\infty$ such that
$$
|\Upsilon(H_1,t)-\Upsilon(H_2,t)|
\leq C\int_{[t_k,t]}
\sup_{u\in[t_k,s]}|H_1(u)-H_2(u)|\,dF(s).
$$
Define $h(t)=\sup_{u\in[t_k,t]}|\tilde A(u)-A_k(u)|$. As in the IV proof, for $t\in[t_k,\tau^*]$, we have
$$
h(t)\leq C\int_{[t_k,t]}h(s)\,dF(s)
\leq CC_0\int_{t_k}^t h(s)\,ds.
$$
By Gronwall's inequality, $h(t)=0$ on $[t_k,\tau^*]$. Therefore,
$$
\tilde A(t)=A_k(t),\qquad t\in[t_k,\tau^*].
$$

But this means $|A_k(\tau^*)|=|\tilde A(\tau^*)|=M_0$. Recall that $\|A_k\|_\infty\leq M^o<M_0$. So this leads to contradiction. This means $\|\tilde A\|_\infty\leq M_0$. So we can eliminate $\xi$:
$$
\tilde A(t)=\Upsilon(\tilde A(\cdot),t),
\qquad t\in[t_k,t_{k+1}].
$$
By Assumption~\ref{assumption:SI_uniform_assumption3}, we know $\tilde A(t)=A_k(t)$, $t\in[t_k,t_{k+1}]$. This means $\|\tilde A_{n_j}-A_k\|_\infty\rightarrow 0$ and contradicts the assumption that $\liminf_{j\rightarrow\infty}\|\tilde A_{n_j}-A_k\|_\infty>0$. Therefore, Step 3 is established.

\paragraph{Step 4}

By step 3, $\|\tilde A_n-A_k\|_\infty\rightarrow 0$. Given that $||A_k||_{\infty}=M^o<M_0$, for sufficiently large $n$, $\|\tilde A_n\|_\infty\leq M_0$. This means when $n$ is large,
$$
\xi(\tilde A_n)=\tilde A_n.
$$

In other words, $\hat A_k=\tilde A_n$. Therefore,
\[
\sup_{t\in[t_k,t_{k+1}]}
|\hat A_k(t)-A_k(t)|
\xrightarrow{\mathrm{a.s.}}0.
\]
 Now take $\hat B_D(t)$ into consideration. 
 For $t\in[t_k,t_{k+1})$, recall that
\[
\hat B_D(t)
=
\sum_{j=1}^k
\hat A_j(t\wedge t_{j+1}^-),
\qquad
B_D(t)
=
\sum_{j=1}^k
A_j(t\wedge t_{j+1}^-).
\]
Therefore,
\[
\begin{aligned}
|\hat B_D(t)-B_D(t)|
&\leq
\sum_{j=1}^k
\left|
\hat A_j(t\wedge t_{j+1}^-)
-
A_j(t\wedge t_{j+1}^-)
\right|
\\
&\leq
\sum_{j=1}^k
\sup_{s\in[t_j,t_{j+1}]}
|\hat A_j(s)-A_j(s)|.
\end{aligned}
\]
Hence, if there are $K<\infty$ treatment intervals,
\[
\sup_{t\in[0,\tau]}
|\hat B_D(t)-B_D(t)|
\leq
\sum_{j=1}^K
\sup_{s\in[t_j,t_{j+1}]}
|\hat A_j(s)-A_j(s)|.
\]
By the stagewise consistency, each term on the
right side converges almost surely to zero. Since $K$ is finite,
\[
\sup_{t\in[0,\tau]}
|\hat B_D(t)-B_D(t)|
\xrightarrow{\mathrm{a.s.}}0.
\]

Finally, under suitable smoothness conditions, since
$\hat\theta\xrightarrow{P}\theta$, a Taylor expansion gives
\[
\sup_{t\in[0,\tau]}
|\hat B_D(t;\hat\theta)-\hat B_D(t;\theta)|
=o_P(1).
\]
Combining this with the oracle consistency establishes
\[
\sup_{t\in[0,\tau]}
|\hat B_D(t;\hat\theta)-B_D(t)|
\xrightarrow{P}0.
\]

\subsubsection{Proof of Proposition~\ref{prop:SI_weakly_convergence}}
\label{app:SI_weakly_convergence}
We aim to provide a sum of i.i.d representation of $\sqrt{n}(\hat A_k(t)-A_k(t) )$ heuristically. Start from $\hat A_k(t;\theta)-A_k(t;\theta)$, where the nuisance parameters are fixed at the oracle values. The following $\hat A_k(t)$ denotes $\hat A_k(t;\theta)$.
$$
d\hat A_k(t)-dA_k(t)
=
\frac{
\sum_i e_{i,k}(t,\hat A_k(\cdot))
[dN_i^1(t)-\check Y_i^1(t)D_{i,k}\,dA_k(t)]
}{
nE_n(t,\hat A_k(\cdot))
}.
$$
Recall the proof in step 3 of \ref{app:SI_uniformly_consistent}, we have
$$
\sup_{s\in[t_k,t_{k+1}]}
\left|
E(s,H_1)-E(s,H_2)
\right|
\leq
\|H_1-H_2\|_\infty.
$$

Then by Lemma~\ref{lemma:SI_uniform}, we have
$$
\begin{aligned}
&\sup_{s\in[t_k,t_{k+1}]}
|E_n(s,\hat A_k(\cdot))-E(s,A_k(\cdot))|
\\
&\leq
\sup_{s\in[t_k,t_{k+1}]}
|E_n(s,\hat A_k(\cdot))-E(s,\hat A_k(\cdot))|
\\
&\quad+
\sup_{s\in[t_k,t_{k+1}]}
|E(s,\hat A_k(\cdot))-E(s,A_k(\cdot))|
\overset{a.s.}{\longrightarrow}0.
\end{aligned}
$$

We focus on the standardized error.
By the uniform consistency of $\hat A_k$
and an application of Slutsky's theorem, under suitable
regularity conditions ensuring that the integrated
denominator-replacement error is uniformly $o_p(1)$
on the standardized scale, we have

$$
\begin{aligned}
V_n(t)
&:=\sqrt n(\hat A_k(t)-A_k(t))
\\
&=
\frac1{\sqrt n}\sum_i
\int_{[t_k,t]}
\frac{
e_{i,k}(s,\hat A_k(\cdot))
[dN_i^1(s)-\check Y_i^1(s)D_{i,k}\,dA_k(s)]
}{
E(s,A_k(\cdot))
}
+o_p(1)
\\
&=
\frac1{\sqrt n}\sum_i
\int_{[t_k,t]}
H_i(s,\hat A_k)\,dR_i(s)
+o_p(1),
\end{aligned}
$$
where
$$
dR_i(t)
=
dN_i^1(t)-\check Y_i^1(t)D_{i,k}\,dA_k(t),
$$
and
$$
H_i(s,\hat A_k)
=
\frac{
e_{i,k}(s,\hat A_k(\cdot))
}{
E(s,A_k(\cdot))
},
\qquad
\dot H_i(s,A_k)
=
-\frac{
\left.
\displaystyle\frac{\partial}{\partial h}m_{i,k}(s,h)
\right|_{h=A_k(s^-)}
}{
E(s,A_k(\cdot))
}.
$$

Define
$$
W_n(t)
=
\frac1{\sqrt n}\sum_i
\int_{[t_k,t]}
H_i(s,A_k)\,dR_i(s),
$$
and
$$
v_k(s)
=
\frac{
\mathbb E[
D_k^2e^{-D_kA_k(s^-)}
\mid\bar D_{k-1},L^*,T\geq t_k]
}{
\mathbb E[
e^{-D_kA_k(s^-)}
\mid\bar D_{k-1},L^*,T\geq t_k]
}
-
m_k(s,A_k(s^-))^2.
$$
Let $v_{i,k}(s)$ denote its value for individual $i$. Since
$e_k(s,H(\cdot))=D_k-m_k(s,H(s^-))$, we have
$$
-\left.
\frac{\partial}{\partial h}m_{i,k}(s,h)
\right|_{h=A_k(s^-)}
=
v_{i,k}(s),
\qquad
\dot H_i(s,A_k)
=
\frac{v_{i,k}(s)}{E(s,A_k(\cdot))}.
$$

Under suitable regularity conditions ensuring that
the integrated Taylor remainder is uniformly $o_p(1)$
on the standardized scale, a Taylor expansion yields

$$
\begin{aligned}
V_n(t)
&:=\sqrt n\{\hat A_k(t)-A_k(t)\}
\\
&=
\frac1{\sqrt n}\sum_{i=1}^n
\int_{[t_k,t]}
H_i(s,\hat A_k)\,dR_i(s)
+o_p(1)
\\
&=
\frac1{\sqrt n}\sum_{i=1}^n
\int_{[t_k,t]}
\left[
H_i(s,A_k)
+
\frac{v_{i,k}(s)}{E(s,A_k(\cdot))}
\{\hat A_k(s^-)-A_k(s^-)\}
\right]dR_i(s)
+o_p(1)
\\
&=
W_n(t)
+
\int_{[t_k,t]}
V_n(s^-)\frac1n\sum_{i=1}^n dK_i(s)
+o_p(1),
\end{aligned}
$$
where
$$
dK_i(s)
=
\frac{v_{i,k}(s)}{E(s,A_k(\cdot))}\,dR_i(s),
\qquad
W_n(t)
=
\frac1{\sqrt n}\sum_{i=1}^n
\int_{[t_k,t]}
H_i(s,A_k)\,dR_i(s).
$$

Define
$$
d\bar K_n(s)=\frac1n\sum_{i=1}^n dK_i(s),
\qquad
\bar K_n(t_k-)=0.
$$
By the Volterra equation, the product-integral representation is
$$
V_n(t)
=
\int_{[t_k,t]}
\mathcal F_n(s,t)\,dW_n(s)
+o_p(1),
$$
where
$$
\mathcal F_n(s,t)
=
\prod_{(s,t]}\{1+d\bar K_n(u)\}.
$$
Here $\prod$ denotes the product integral, and $dK_i(s)$ does not depend on $t$.

Define
$$
\begin{aligned}
dK_0(s)
&=
\mathbb E[dK_i(s)]
\\
&=
\frac{
\mathbb E\left[
v_k(s)\{dN^1(s)-\check Y^1(s)D_k\,dA_k(s)\}
\right]
}{
E(s,A_k(\cdot))
},
\qquad K_0(t_k-)=0,
\end{aligned}
$$
and
$$
\mathcal F_0(s,t)
=
\prod_{(s,t]}\{1+dK_0(u)\}.
$$

Assume
$$
\sup_{t\in[t_k,t_{k+1}]}
|\bar K_n(t)-K_0(t)|=o_p(1)
$$
and uniformly bounded total variation. Then
$$
\sup_{t_k\leq s\leq t\leq t_{k+1}}
|\mathcal F_n(s,t)-\mathcal F_0(s,t)|
=o_p(1).
$$

Under suitable regularity conditions ensuring that
\[
\sup_{t\in[t_k,t_{k+1}]}
\left|
\int_{[t_k,t]}
\{\mathcal F_n(s,t)-\mathcal F_0(s,t)\}
dW_n(s)
\right|
=o_p(1),
\]
we obtain the following asymptotic linear
representation under oracle nuisance:
$$
\begin{aligned}
V_n(t)
&=
\int_{[t_k,t]}
\mathcal F_0(s,t)\,dW_n(s)
+o_p(1)
\\
&=
\frac1{\sqrt n}\sum_{i=1}^n
\int_{[t_k,t]}
\mathcal F_0(s,t)H_i(s,A_k)\,dR_i(s)
+o_p(1)
\\
&=
\frac1{\sqrt n}\sum_{i=1}^n
\phi_{i,k}^{A}(t)
+o_p(1),
\end{aligned}
$$
where
$$
\phi_{i,k}^{A}(t)
=
\int_{[t_k,t]}
\mathcal F_0(s,t)H_i(s,A_k)\,dR_i(s).
$$

Now calculate the full influence function of $\hat A_k(t;\hat\theta)$.
Assume
$$
\sqrt n(\hat\theta-\theta)
=
\frac1{\sqrt n}\sum_{i=1}^n\phi_i^\theta
+o_p(1).
$$
Then
$$
\begin{aligned}
&\sqrt n\{\hat A_k(t;\hat\theta)-A_k(t;\theta)\}
\\
={}&
\sqrt n\{\hat A_k(t;\hat\theta)-\hat A_k(t;\theta)\}
+
\sqrt n\{\hat A_k(t;\theta)-A_k(t;\theta)\}
\\
={}&
\sqrt n\{\hat A_k(t;\hat\theta)-\hat A_k(t;\theta)\}
+
\frac1{\sqrt n}\sum_{i=1}^n\phi_{i,k}^{A}(t)
+o_p(1)
\\
={}&
\left\{
\frac{\partial A_k(t;\theta)}{\partial\theta}
+o_p(1)
\right\}
\sqrt n(\hat\theta-\theta)
+
\frac1{\sqrt n}\sum_{i=1}^n\phi_{i,k}^{A}(t)
+o_p(1)
\\
={}&
\frac{\partial A_k(t;\theta)}{\partial\theta}
\sqrt n(\hat\theta-\theta)
+
\frac1{\sqrt n}\sum_{i=1}^n\phi_{i,k}^{A}(t)
+o_p(1)
\\
={}&
\frac1{\sqrt n}\sum_{i=1}^n\phi_{i,k}(t)
+o_p(1),
\end{aligned}
$$
where
$$
\phi_{i,k}(t)
:=
\frac{\partial A_k(t;\theta)}{\partial\theta}
\phi_i^\theta
+
\phi_{i,k}^{A}(t).
$$

In practice, we recursively update
$\frac{\partial A_k(t;\theta)}{\partial\theta}\phi_i^\theta$
and $\phi_{i,k}^{A}(t)$ to calculate the influence function.

For $t\in[t_k,t_{k+1})$, define
$$
\begin{aligned}
\psi_i(t;\theta)
&=
\sum_{j=1}^{k-1}\phi_{i,j}(t_{j+1}^-)
+\phi_{i,k}(t)
\\
&=
\sum_{j=1}^{k}
\left\{
\frac{\partial A_j(t\wedge t_{j+1}^-;\theta)}
{\partial\theta}\phi_i^\theta
+
\phi_{i,j}^{A}(t\wedge t_{j+1}^-)
\right\}.
\end{aligned}
$$
Then
$$
\sqrt n\{\hat B_D(t)-B_D(t)\}
=
\frac1{\sqrt n}\sum_{i=1}^n\psi_i(t;\theta)
+o_p(1).
$$
Let $\hat\psi_i(t;\hat\theta)$ denote its plug-in estimate. 
With this i.i.d. representation, similar arguments
as in \citet{Lin2000} and \citet{Martinussen2017}
can be adopted to establish the convergence of
$\sqrt n(\hat B_D(t)-B_D(t))$ in distribution
to a mean-zero Gaussian process. The variance function of the limiting Gaussian process is consistently estimated by
$$
\frac1n\sum_{i=1}^n\hat\psi_i^2(t;\hat\theta).
$$

\subsubsection{Proof of Theorem~\ref{thm:observed_blip_down}}
\label{app:observed_blip_down}
\begin{proof}

Throughout the proof, we use $L$ to denote $L^*=(Z,L)$ under
sequential ignorability and $L$ under the IV approach,
and omit $Z$ in the potential outcome notation for simplicity.
Under either approach, the conditional independent censoring
assumption implies
\[
C\perp
(\{d\tilde N^j(t):t\in[0,\tau],j=1,\ldots,J\},\bar D_K)
\mid L.
\]

\paragraph{Step 1 }Consider the case without censoring and prove its blip-down equation first.

\begin{lemma}
\label{lem:blip_down_representation}

Define
\[
\mathcal A_k(t)=\int_{t_k}^{t\wedge t_{k+1}}D_k\,dB_D(s),\quad
t\ge t_k,
\qquad
\Gamma_j(t)=\sum_{k=1}^j \mathcal A_k(t), \quad t\ge t_j.
\]

Under the cause-specific SNCSTM model
\eqref{eq:cause_specific_SNCSTM}, for
$t\in[t_k,t_{k+1})$,
\begin{equation}
\begin{aligned}
\mathbb E\Bigl[
\tilde Y^1(t;\bar 0)
\mid L
\Bigr]
&=
\begin{cases}
\mathbb E\Bigl[
\tilde Y^1(t)\exp\{\Gamma_1(t)\}
\mid L
\Bigr],
& k=1,
\\[1em]
\mathbb E\Bigl[
\displaystyle
\sum_{j=1}^{k-1}
\mathbf 1\{t_j\leq \tilde T<t_{j+1},\,V\neq 1\}
\exp\{\Gamma_j(t)\}
\\
\qquad\qquad
+\mathbf 1\{\tilde T\geq t_k\}
\tilde Y^1(t)\exp\{\Gamma_k(t)\}
\mid L
\Bigr],
& k\geq 2.
\end{cases}
\end{aligned}
\end{equation}
\end{lemma}

From cause-specific SNCSTM \eqref{eq:cause_specific_SNCSTM}, we know that for $t\geq t_k$,
$$
\frac{\mathbb E[\tilde Y^1(t;\bar D_k,\bar 0)\mid \bar D_k,L,\tilde T\geq t_k] }{\mathbb E[\tilde Y^1(t;\bar D_{k-1},\bar 0)\mid \bar D_k,L,\tilde T\geq t_k] }
=\exp\{-\int_{t_k}^{t\wedge t_{k+1}} D_kdB_D(s) \}.
$$

For $t\in[t_1,t_{2}),t_1=0$, we have
$$
\begin{aligned}
\mathbb E\Bigl[\tilde Y^1(t;\bar 0)\mid L  \Bigr]
=& \mathbb E\Bigl[\mathbb E\Bigl[\tilde Y^1(t;\bar 0)\mid \bar D_1,L,\tilde T\geq 0   \Bigr]   \mid L\Bigr]\\
=& \mathbb E\Bigl[\exp\{\int_{t_1}^{t}D_1dB_D(s) \}\mathbb E\Bigl[\tilde Y^1(t;\bar D_1,\bar 0)\mid \bar D_1,L,\tilde T\geq 0   \Bigr]   \mid L\Bigr]\\
=& \mathbb E\Bigl[\exp\{\mathcal A_1(t) \} \tilde Y^1(t;\bar D_1,\bar 0)  \mid L\Bigr]\\
=& \mathbb E\Bigl[\exp\{\mathcal A_1(t) \} \tilde Y^1(t)  \mid L\Bigr]\\
\end{aligned}
$$

For $t\in[t_2,t_3)$, things become different. Define $H_k = \{\bar D_k,L,\tilde T\geq t_k \}$. We still have
$$
\begin{aligned}
\mathbb E\Bigl[\tilde Y^1(t;\bar 0)\mid L  \Bigr]
=& \mathbb E\Bigl[\mathbb E\Bigl[\tilde Y^1(t;\bar 0)\mid \bar D_1,L,\tilde T\geq 0   \Bigr]   \mid L\Bigr]\\
=& \mathbb E\Bigl[\exp\{\int_{t_1}^{t\wedge t_2 }D_1dB_D(s) \}\mathbb E\Bigl[\tilde Y^1(t;\bar D_1,\bar 0)\mid \bar D_1,L,\tilde T\geq 0   \Bigr]   \mid L\Bigr]\\
=& \mathbb E\Bigl[\exp\{\mathcal A_1(t)\}\mathbb E\Bigl[\tilde Y^1(t;\bar D_1,\bar 0)\mid H_1   \Bigr]   \mid L\Bigr]\\
\end{aligned}
$$
But now we need to consider 3 cases, $\{\tilde T\geq t_2\},\{\tilde T<t_2\text{ and }V=1$\}, $\{\tilde T<t_2\text{ and }V\neq 1\}$. Using total probability equation, we obtain
$$
\begin{aligned}
&\mathbb E\Bigl[\tilde Y^1(t;\bar D_1,\bar 0)\mid H_1 \Bigr]\\
=&\mathbb E\Bigl[\tilde Y^1(t;\bar D_1,\bar 0)\mid H_1,\tilde T\geq t_2 \Bigr]P(\tilde T\geq t_2\mid H_1)
+ \mathbb E\Bigl[\tilde Y^1(t;\bar D_1,\bar 0)\mid H_1,\tilde T<t_2,V= 1 \Bigr]P(\tilde T<t_2,V= 1\mid H_1 )\\
&\qquad \qquad
+ \mathbb E\Bigl[\tilde Y^1(t;\bar D_1,\bar 0)\mid H_1,\tilde T<t_2,V\neq 1 \Bigr]P(\tilde T<t_2,V\neq 1\mid H_1 ) \\
=& \mathbb E\Bigl[\tilde Y^1(t;\bar D_1,\bar 0)\mid H_1,\tilde T\geq t_2 \Bigr]P(\tilde T\geq t_2\mid H_1)
+ 0 + 1\times P(\tilde T<t_2,V\neq 1\mid H_1 )  \\
=& \mathbb E\Bigl[\mathbf1\{\tilde T<t_2,V\neq 1\} +  \mathbb E\Bigl[\tilde Y^1(t;\bar D_1,\bar 0)\mid H_1,\tilde T\geq t_2 \Bigr]\mathbf1\{ \tilde T\geq t_2\}\mid H_1\Bigr] \\
\end{aligned}
$$
Then we plug in this equality, and use tower property to apply SNCSTM. 
$$
\begin{aligned}
&\mathbb E\Bigl[\tilde Y^1(t;\bar 0)\mid L  \Bigr]\\
=& \mathbb E\Bigl[\exp\{\mathcal A_1(t)\}\mathbb E\Bigl[\tilde Y^1(t;\bar D_1,\bar 0)\mid H_1   \Bigr]   \mid L\Bigr]\\
=& \mathbb E\Bigl[\exp\{\mathcal A_1(t)\}\Bigl(\mathbb E\Bigl[\mathbf1\{\tilde T<t_2,V\neq 1\} +  \mathbb E\Bigl[\tilde Y^1(t;\bar D_1,\bar 0)\mid H_1,\tilde T\geq t_2 \Bigr]\mathbf1\{ \tilde T\geq t_2\}\mid H_1\Bigr]  \Bigr)  \mid L\Bigr]\\
=& \mathbb E\Bigl[\exp\{\mathcal A_1(t)\}\Bigl(\mathbb E\Bigl[\mathbf1\{\tilde T<t_2,V\neq 1\} +  \mathbb E\Bigl[\mathbb E[\tilde Y^1(t;\bar D_1,\bar 0)\mid H_2]\mid H_1,\tilde T\geq t_2 \Bigr]\mathbf1\{ \tilde T\geq t_2\}\mid H_1\Bigr]  \Bigr)  \mid L\Bigr]\\
=& \mathbb E\Bigl[\exp\{\mathcal A_1(t)\}\Bigl(\mathbb E\Bigl[\mathbf1\{\tilde T<t_2,V\neq 1\} +  \mathbb E\Bigl[\exp\{\mathcal A_2(t)\}\mathbb E[\tilde Y^1(t;\bar D_2,\bar 0)\mid H_2]\mid H_1,\tilde T\geq t_2 \Bigr]\mathbf1\{ \tilde T\geq t_2\}\mid H_1\Bigr]  \Bigr)  \mid L\Bigr]\\
=& \mathbb E\Bigl[\exp\{\mathcal A_1(t)\}\Bigl(\mathbb E\Bigl[\mathbf1\{\tilde T<t_2,V\neq 1\} +  \mathbb E\Bigl[\exp\{\mathcal A_2(t)\}\tilde Y^1(t;\bar D_2,\bar 0)\mid H_1,\tilde T\geq t_2 \Bigr]\mathbf1\{ \tilde T\geq t_2\}\mid H_1\Bigr]  \Bigr)  \mid L\Bigr]\\
=& \mathbb E\Bigl[\exp\{\mathcal A_1(t)\}\mathbf1\{\tilde T<t_2,V\neq 1\} 
+ \mathbb E\Bigl[\mathbb E\Bigl[\exp\{\mathcal A_1(t)+\mathcal A_2(t)\}\tilde Y^1(t;\bar D_2,\bar 0)\mid H_1,\tilde T\geq t_2 \Bigr]\mathbf1\{ \tilde T\geq t_2\}\mid H_1\Bigr]   \mid L\Bigr]\\
=& \mathbb E\Bigl[\exp\{\mathcal A_1(t)\}\mathbf1\{\tilde T<t_2,V\neq 1\} 
+ \exp\{\Gamma_2(t)\}\tilde Y^1(t)\mathbf1\{ \tilde T\geq t_2\}   \mid L\Bigr]\\
\end{aligned}
$$
The final equality is because
$$
\begin{aligned}
&\mathbb E \Bigl[ \mathbb{E}[ \exp\{\Gamma_2(t) \}\tilde Y^1(t;\bar D_2,\bar 0) \mid H_1,\mathbf{1}\{\tilde T\geq t_2 \}=1] \mathbf 1\{ \tilde T\geq t_2\}\mid H_1\Bigr]\\
=& \mathbb E \Bigl[ \mathbb{E}[ \exp\{\Gamma_2(t) \}\tilde Y^1(t;\bar D_2,\bar 0)  \mid H_1,\mathbf{1}\{\tilde T\geq t_2 \}] \mathbf 1\{ \tilde T\geq t_2\}\mid H_1\Bigr]\\
=& \mathbb E \Bigl[ \exp\{\Gamma_2(t) \}\tilde Y^1 (t;\bar D_2,\bar 0)  \mathbf 1\{ \tilde T\geq t_2\}\mid H_1\Bigr]\\
\end{aligned}
$$

So iteratively, we can prove that for $t\in[t_k,t_{k+1})$, there exists
$$
\begin{aligned}
&\mathbb E\Bigl[\tilde Y^1(t;\bar 0)\mid L  \Bigr]\\
=&
\begin{cases}
\mathbb E\Bigl[\tilde Y^1(t)\exp\{\Gamma_1(t)\} \mid L \Bigr], & k=1,\\[1em]
\mathbb E\Bigl[\sum_{j=1}^{k-1}\mathbf 1\{t_j\leq \tilde T <t_{j+1}, V\neq 1\}\exp\{\Gamma_j(t) \}+\mathbf 1\{\tilde T\geq t_k \}\tilde Y^1(t)\exp\{\Gamma_k(t)\} \mid L  \Bigr], & k\geq 2.
\end{cases}
\end{aligned}
$$

This proves the lemma.

For data without censoring, we can directly use this result to obtain 
\begin{equation}
P(\tilde T(\bar 0)\leq t,V(\bar 0) = 1\mid L)= 1- \mathbb{E}[\tilde Y^1(t;\bar 0)\mid L]
\end{equation}

\paragraph{Step 2} Consider censoring $C$ and observed survival time $T =\min\{\tilde T,C \}$. We can observe
$$
\begin{aligned}
N^1(t) &:= \mathbf 1\{ T\leq t,\Delta =1,V = 1 \}\\
Y^{1}(t) &:=  \mathbf 1\{T\geq t \} + \mathbf1\{T<t,\Delta =1, V\neq 1 \}
\end{aligned}
$$

Define $Y^{1,c}(t)=\mathbf1\{C\geq t \}\tilde Y^1(t)$. Consider $Y^{1,c}(t)$ observable first, like in censoring-complete data, $Y^1(t) = Y^{1,c}(t)$. For $k=1$, we have
$$
\begin{aligned}
\mathbb{E}\Bigl[ Y^{1,c}(t)\exp\{ \Gamma_1(t)\} \mid L\Bigr]& = \mathbb{E}\Bigl[ \mathbf 1\{ C\geq t\}\tilde Y^1(t)\exp\{ \Gamma_1(t)\} \mid L\Bigr]\\
& = P(C\geq t\mid L)\mathbb{E}\Bigl[\tilde Y^1(t)\exp\{ \Gamma_1(t)\} \mid L\Bigr]\\
\end{aligned}
$$

So $\mathbb{E}\Bigl[\tilde Y^1(t)\exp\{ \Gamma_1(t)\} \mid L\Bigr] = \frac{\mathbb{E}\Bigl[ Y^{1,c}(t)\exp\{ \Gamma_1(t)\} \mid L\Bigr] }{P(C\geq t\mid L)}$

For $k\geq 2$, let us focus on the right term first.
$$
\begin{aligned}
\mathbb{E}\Bigl[ \mathbf{1}\{T\geq t_k \}Y^{1,c}(t)\exp\{ \Gamma_k(t)\} \mid L\Bigr]& = \mathbb{E}\Bigl[ \mathbf 1\{ C\geq t_k\}\mathbf 1\{\tilde T\geq t_k \}\mathbf 1\{C\geq t \} \tilde Y^1(t)\exp\{ \Gamma_k(t)\} \mid L\Bigr]\\
& = \mathbb{E}\Bigl[ \mathbf 1\{C\geq t \} \mathbf 1\{ \tilde T\geq t_k\} \tilde Y^1(t)\exp\{ \Gamma_k(t)\} \mid L\Bigr]\\
& = P(C\geq t\mid L)\mathbb{E}\Bigl[\mathbf 1\{\tilde T\geq t_k \} \tilde Y^1(t)\exp\{ \Gamma_k(t)\} \mid L\Bigr]\\
\end{aligned}
$$

So we can use the observed variables to do expectations.

$$
\mathbb{E}\Bigl[\mathbf 1\{\tilde T\geq t_k \} \tilde Y^1(t)\exp\{ \Gamma_k(t)\} \mid L\Bigr] = \frac{\mathbb{E}\Bigl[ \mathbf{1}\{T\geq t_k \}Y^{1,c}(t)\exp\{ \Gamma_k(t)\} \mid L\Bigr]}{P(C\geq t\mid L)}
$$

For the left term, we have
$$
\begin{aligned}
&\mathbb E\Bigl[\sum_{j=1}^{k-1}\mathbf 1\{t_j\leq T <t_{j+1}, \Delta=1, V\neq 1\}\frac{1}{G(T\mid L)}\exp\{\Gamma_j(t) \}\mid L \Bigr]\\
=&  \sum_{j=1}^{k-1}\mathbb E\Bigl[\mathbf 1\{t_j\leq \tilde T <t_{j+1}, V\neq 1\}\frac{\mathbf{1} \{C\geq \tilde T \}}{G(T\mid L)} \exp\{\Gamma_j(t) \}\mid L\Bigr]\\
=&  \sum_{j=1}^{k-1}\mathbb E\Bigl[\mathbf 1\{t_j\leq \tilde T <t_{j+1}, V\neq 1\}\frac{\mathbb E[\mathbf{1} \{C\geq \tilde T \} \mid \tilde T,V,\bar D,L]}{G(\tilde T\mid L)} \exp\{\Gamma_j(t) \}\mid L\Bigr]\\
=&  \sum_{j=1}^{k-1}\mathbb E\Bigl[\mathbf 1\{t_j\leq \tilde T <t_{j+1}, V\neq 1\} \exp\{\Gamma_j(t) \}\mid L\Bigr]\\
\end{aligned}
$$

As a result, we can use observed variables to blip down and calculate the expectation.

\begin{equation}
\begin{aligned}
&\mathbb E\Bigl[\tilde Y^1(t;\bar 0)\mid L\Bigr]\\
=&
\begin{cases}
\mathbb E\Bigl[
\frac{Y^{1,c}(t)\exp\{\Gamma_1(t)\}}{G(t\mid L)}
\mid L\Bigr],
& k=1,\\[1em]
\mathbb E\Bigl[
\displaystyle\sum_{j=1}^{k-1}
\mathbf 1\{t_j\leq T<t_{j+1},\Delta=1,V\neq 1\}
\frac{\exp\{\Gamma_j(t)\}}{G(T\mid L)}
\\
\qquad\qquad
+\mathbf1\{T\geq t_k\}
\frac{Y^{1,c}(t)\exp\{\Gamma_k(t)\}}{G(t\mid L)}
\mid L\Bigr],
& k\geq 2.
\end{cases}
\end{aligned}
\end{equation}

Recall that for incomplete-censoring data, $Y^{1,w}(t)$ can replace unobservable $Y^{1,c}(t)$ within expectations. So this equation becomes
\begin{equation}
\begin{aligned}
&\mathbb E\Bigl[\tilde Y^1(t;\bar 0)\mid L\Bigr]\\
=&
\begin{cases}
\mathbb E\Bigl[
\frac{Y^{1,w}(t)\exp\{\Gamma_1(t)\}}{G(t\mid L)}
\mid L\Bigr],
& k=1,\\[1em]
\mathbb E\Bigl[
\displaystyle\sum_{j=1}^{k-1}
\mathbf 1\{t_j\leq T<t_{j+1},\Delta=1,V\neq 1\}
\frac{\exp\{\Gamma_j(t)\}}{G(T\mid L)}
\\
\qquad\qquad
+\mathbf1\{T\geq t_k\}
\frac{Y^{1,w}(t)\exp\{\Gamma_k(t)\}}{G(t\mid L)}
\mid L\Bigr],
& k\geq 2.
\end{cases}
\end{aligned}
\end{equation}

Finally, marginalizing over $L$ yields the desired result.

\end{proof}

\subsection{IV-based Strategy}

\subsubsection{Proof of Theorem~\ref{thm:IV_moment}}
\label{app:proof_IV_moment_validity}

We view $Z$ as a baseline covariate first. According to Lemma~\ref{lem:blip_down_representation}, there exists
\begin{equation}
\begin{aligned}
&\mathbb E\Bigl[\tilde Y^1(t;\bar 0)\mid Z,L  \Bigr]\\
=&
\begin{cases}
\mathbb E\Bigl[\tilde Y^1(t)\exp\{\Gamma_1(t)\} \mid Z,L \Bigr], & k=1,\\[1em]
\mathbb E\Bigl[\sum_{j=1}^{k-1}\mathbf 1\{t_j\leq \tilde T <t_{j+1}, V\neq 1\}\exp\{\Gamma_j(t) \}+\mathbf 1\{\tilde T\geq t_k \}\tilde Y^1(t)\exp\{\Gamma_k(t)\} \mid Z,L  \Bigr], & k\geq 2.
\end{cases}
\end{aligned}
\label{eq:Z_blip}
\end{equation}

Define the subdistribution hazard function as follows.
$$
\begin{aligned}
\lambda_0^1(t\mid L):=&\frac{
\frac{d}{dt}P(\tilde T{(\bar 0)}\leq t,V(\bar 0)=1\mid L)
}{
1-P(\tilde T{(\bar 0)}\leq t,V(\bar 0)=1\mid L)
} \\
=&\frac{
\frac{d}{dt}P(\tilde T{(\bar 0)}\leq t,V(\bar 0)=1\mid Z,L)
}{
1-P(\tilde T{(\bar 0)}\leq t,V(\bar 0)=1\mid Z,L)
}\\
=&-\frac{d}{dt}\log(1- P(\tilde T(\bar 0)\leq t,V(\bar 0)=1\mid Z,L))\\
=&-\frac{d}{dt}\log(\mathbb E\Bigl[\tilde Y^1(t;\bar 0)\mid Z,L  \Bigr])
\end{aligned}
$$

The first equality is due to IV independence. Then based on \eqref{eq:Z_blip}, we keep on calculating. For simplicity, define 
$$
Q_k(t)
=
\sum_{j=1}^{k-1}
\mathbf 1\{t_j\leq \tilde T<t_{j+1},V\neq1\}
\exp\{\Gamma_j(t)\}
+
\mathbf 1\{\tilde T\geq t_k\}
\tilde Y^1(t)
\exp\{\Gamma_k(t)\}.
$$
Then we have
$$
\begin{aligned}
\lambda_0^1(t\mid L) &= -\frac{d}{dt}\log \Bigl(\mathbb E\Bigl[Q_k(t)\mid Z,L \Bigr] \Bigr)\\
\lambda_0^1(t\mid L)\mathbb E\Bigl[Q_k(t)\mid Z,L \Bigr]dt &= -d\mathbb E\Bigl[Q_k(t)\mid Z,L \Bigr] 
\end{aligned}
$$

Now focus on $d\mathbb{E}[Q_k(t)\mid Z,L]$. For $t\in [t_k,t_{k+1})$, the first part of $Q_k(t)$ is fixed. So we only need to differentiate on the second part. So the right term is
$$
\begin{aligned}
&d \mathbb E\Bigl[Q_k(t)\mid Z,L \Bigr]\\
=& \mathbb E\Bigl[d\Bigl(\mathbf 1\{\tilde T\geq t_k \}\tilde Y^1(t)\exp\{\Gamma_k(t) \} \Bigr)\mid Z,L \Bigr]\\
=& \mathbb E\Bigl[ \mathbf 1\{\tilde T\geq t_k \}\Bigl( d\tilde Y^1(t)\exp\{\Gamma_k(t) \} +\tilde Y^1(t)\frac{d}{dt}\exp\{ \Gamma_k(t)\}dt  \Bigr)\mid Z,L \Bigr]\\
=& \mathbb E\Bigl[ \mathbf 1\{\tilde T\geq t_k \}\Bigl( -d\tilde N^1(t)\exp\{\Gamma_k(t) \} +\tilde Y^1(t)\exp\{ \Gamma_k(t)\}D_kdB_D(t)  \Bigr)\mid Z,L \Bigr]\\
\end{aligned}
$$
The left term is
$$
\begin{aligned}
\lambda_0^1(t\mid L)\mathbb E\Bigl[Q_k(t)\mid Z,L \Bigr]dt &=\lambda_0^1(t\mid L)\mathbb E\Bigl[\tilde Y^1(t;\bar 0) \mid Z,L\Bigr]dt\\
&=\lambda_0^1(t\mid L)\mathbb E\Bigl[\tilde Y^1(t;\bar 0) \mid L\Bigr]dt\\
\end{aligned}
$$

Rearranging yields
\begin{equation}
\mathbb E\Bigg[
\mathbf 1\{\tilde T\geq t_k\}
\tilde Y^1(t)
\exp\{\Gamma_k(t)\}
\Bigl\{
d\tilde N^1(t)
- D_kdB_D(t)
\Bigr\}
-
\mathbb E\Bigl[\tilde Y^1(t;\bar 0) \mid L\Bigr]\lambda_0^1(t\mid L)dt
\;\Bigm|\; Z,L
\Bigg]=0
\end{equation}

Add the residual term $Z-\mathbb E[Z\mid L]$ in the above equation, then we get

\begin{equation}
\mathbb E\Bigg[
\Bigl(
Z-\mathbb E[Z\mid L]
\Bigr)
\mathbf 1\{ \tilde T\geq t_k\}
\tilde Y^1(t)
\exp\{\Gamma_k(t)\}
\Bigl(
d\tilde N^1(t)
-
D_kdB_D(t)
\Bigr)
\mid L\Bigg]
=0.
\end{equation}

For data without censoring, this can be the moment equation. Consider censoring $C$. We assume $Y^{1,c}(t)=\mathbf 1\{C\geq t \}\tilde Y^1(t)$ can be observed first, like censoring-complete data. Consider the following equation.

$$
\begin{aligned}
&\mathbb E\Bigg[
\Bigl(
Z-\mathbb E[Z\mid L]
\Bigr)
\mathbf 1\{ T\geq t_k\}
Y^{1,c}(t)
\exp\{\Gamma_k(t)\}
\Bigl(
d N^1(t)
-
D_kdB_D(t)
\Bigr)
\mid L\Bigg]\\
=& \mathbb E\Bigg[
\Bigl(
Z-\mathbb E[Z\mid L]
\Bigr)
\mathbf 1\{ \tilde T\geq t_k\}\mathbf{1}\{C\geq t_k \}
\mathbf 1\{C\geq t \}\tilde Y^1(t)
\exp\{\Gamma_k(t)\}
\mathbf{1}\{C\geq t \}\Bigl(
d \tilde N^1(t)
-
D_kdB_D(t)
\Bigr)
\mid L\Bigg]\\
=& \mathbb E\Bigg[
\Bigl(
Z-\mathbb E[Z\mid L]
\Bigr)
\mathbf 1\{ \tilde T\geq t_k\}
\mathbf 1\{C\geq t \}\tilde Y^1(t)
\exp\{\Gamma_k(t)\}\Bigl(
d \tilde N^1(t)
-
D_kdB_D(t)
\Bigr)
\mid L\Bigg]\\
=& P(C\geq t\mid L)\mathbb E\Bigg[
\Bigl(
Z-\mathbb E[Z\mid L]
\Bigr)
\mathbf 1\{ \tilde T\geq t_k\}\tilde Y^1(t)
\exp\{\Gamma_k(t)\}\Bigl(
d \tilde N^1(t)
-
D_kdB_D(t)
\Bigr)
\mid L\Bigg]\\
=& 0
\end{aligned}
$$

So this is a valid moment equation. For incomplete-censoring data, as we discussed before, observable $Y^{1,w}$ can replace unobservable $Y^{1,c}(t)$ within expectations. So the moment equation turns out to be

\begin{equation}
    \mathbb E\Bigg[
\Bigl(
Z-\mathbb E[Z\mid L]
\Bigr)
\mathbf 1\{ T\geq t_k\}
Y^{1,w}(t)
\exp\{\Gamma_k(t)\}
\Bigl(
d N^1(t)
-
D_kdB_D(t)
\Bigr)
\mid L\Bigg]=0
\end{equation}

\subsubsection{Proof of Proposition~\ref{prop:IV_uniformly_consistent}}
\label{app:IV_uniformly_consistent}

For a real-valued function $g$ defined on $[0,\tau]$, write $\|g\|_{\infty}$
for its supremum norm on $[0,\tau]$, and $\mathcal V(g)$ denote
its total variation over $[0,\tau]$. Let $\|B_D\|_{\infty}=M^o<\infty$.

\begin{assumption}
\label{assumption:IV_uniform_assumption1}
$Z$ is bounded by $M_{\max}$ and $\mathcal V(D_i)\le K_D$ uniformly.
\end{assumption}

Bounded $Z$ is common in applications. For bounded $D_i(t)$ and fixed time points $K$, it's also reasonable to assume such an upper bound $K_D$ for $D_i(t)$'s total variation.

Define 5 terms at first.
\[
\begin{aligned}
A_n(s,H(\cdot))
&= \frac{1}{n}\sum_{i=1}^n Z_i^c \check Y_i^1(s)
\exp\{\int_0^{s^-}D_i(u)dH(u)\}D_i(s)\\
a(s,H(\cdot))
&= \mathbb E[Z_i^c \check Y_i^1(s)
\exp\{\int_0^{s^-}D_i(u)dH(u)\}D_i(s)]\\
\Upsilon_n(H(\cdot),t)
&=\int_0^t
\frac{\sum_{i=1}^n Z_i^c
\exp\{\int_0^{s^-}D_i(u)dH(u)\}dN_i^1(s)}
{nA_n(s,H(\cdot))}\\
\bar\Upsilon_n(H(\cdot),t)
&=\int_0^t
\frac{\sum_{i=1}^n Z_i^c
\exp\{\int_0^{s^-}D_i(u)dH(u)\}dN_i^1(s)}
{na(s,H(\cdot))}\\
\Upsilon(H(\cdot),t)
&=\int_0^t
\frac{\mathbb E[Z_i^c
\exp\{\int_0^{s^-}D_i(u)dH(u)\}dN_i^1(s)]}
{a(s,H(\cdot))}
\end{aligned}
\]

\begin{assumption}
\label{assumption:IV_uniform_assumption2}
There exist constants $M_0>\|B_D\|_\infty$,
$M>\mathcal V(B_D)$, and $\nu>0$ such that
\[
\inf_{s\in[0,\tau],\,H\in\mathbb H_{M_0,M}}
a(s,H)>\nu,
\]
and
\[
M>
\frac{2M_{\max}e^{(1+K_D)M_0}}{\nu},
\]
where
\[
\mathbb H_{M_0,M}
=
\left\{
H\in D[0,\tau]:
H(0)=0,\,
\|H\|_\infty\le M_0,\,
\mathcal V(H)\le M
\right\}.
\]
\end{assumption}

This is a sufficient condition. We can choose $M_0$ close to
$\|B_D\|_\infty$ and a sufficiently large $M$. However, increasing
$M$ enlarges the function class and may reduce the
lower bound $\nu$. In particular, there should exist
an $M$ such that
\[
M\nu>2M_{\max}e^{(1+K_D)M_0}.
\]
We note that this condition can be further relaxed
by accounting for the cause-1 event rate in the
total variation bound. We further assume that

\begin{assumption}
\label{assumption:IV_uniform_assumption3}
    $B_D(t)$ is the unique continuous solution to $B(t)=\Upsilon(B(\cdot) ,t)$.
\end{assumption}

\begin{assumption}
\label{assumption:IV_uniform_assumption4}
Assume that \(N^1\) admits the Doob--Meyer decomposition
\[
dN^1(t)
=
Y(t)\lambda_1(t)\,dt
+
dM^1(t),
\]
where \(\lambda_1(t)\) is the cause-specific hazard and $\sup_{t\in[0,\tau]}
\lambda_1(t)
\leq
C_0
<
\infty$.
\end{assumption}

We shall use the generalized Helly’s selection theorem on class of functions with finite total
variations to prove this proposition.

\begin{theorem}[Helly's Selection Theorem]
Let $\{f_n\}$ be a sequence of real-valued functions on $[0,\tau]$ such that
\[
\sup_n \|f_n\|_\infty \le A_1,
\qquad
\sup_n \mathcal V(f_n) \le A_2,
\]
where $A_1$ and $A_2$ are finite constants. Then there exists a subsequence
$\{f_{n_j}\}$ that converges pointwise on $[0,\tau]$ to a function $f$.
\end{theorem}

\begin{lemma}
\label{lemma:IV_uniform}
Under Assumptions~\ref{assumption:IV_uniform_assumption1}
and \ref{assumption:IV_uniform_assumption2},
\[
\sup_{s\in[0,\tau],\,H\in\mathbb H_{M_0,M}}
|A_n(s,H(\cdot))-a(s,H(\cdot))|
\xrightarrow{\mathrm{a.s.}}0.
\]
\end{lemma}

Define 
$$
\phi_{H,s}(O) = Z^c \check Y^1(s)\exp\{\int_0^{s^-}D(u)dH(u) \}D(s),
$$
and
\[
\Phi_{M_0,M}
=
\left\{
\phi_{H,s}:H\in\mathbb H_{M_0,M},
\ s\in[0,\tau]
\right\}.
\]

Since \(H\in\mathbb H_{M_0,M}\) has uniformly bounded total variation, and
\(Z^c\), \(\check Y^1(\cdot)\), and \(D(\cdot)\) are uniformly bounded, the class
\(\Phi_{M_0,M}\) is a uniformly bounded bounded-variation class. By
\citet[Example~19.11]{vandervaart2000}, \(\Phi_{M_0,M}\) is Glivenko--Cantelli.
Therefore,
\[
\sup_{s\in[0,\tau],\,H\in\mathbb H_{M_0,M}}
\left|
(\mathbb P_n-\mathbb P)\phi_{H,s}
\right|
\xrightarrow{\mathrm{a.s.}}0,
\]
which proves the lemma.

\paragraph{Step 1} The following analysis is under the oracle nuisance parameters $\theta$. The consistency of $\hat B_D(t,\hat \theta)$ follows immediately by a Taylor series expansion since
$\hat \theta$ is consistent for $\theta$. We use $\hat B_D(t)$ to denote $\hat B_D(t;\theta)$. Define $\xi(y)=\operatorname{sgn}(y)\min\{|y|,M_0\}$. Let $\tilde B_n(t)$ be the solution to
$$
B(t)=\Upsilon_n(\xi(B)(\cdot),t).
$$
Notice that for sufficiently large $n$, there is
$$
\inf_{s\in[0,\tau],\,H\in\mathbb H_{M_0,M}} A_n(s,H(\cdot))\geq\nu
$$
Then we construct an upper bound for $\mathcal V(\tilde B_n)$. Since $\tilde B_n(s)$ is recursively constructed based on information in $[0,s)$ and $\tilde B_n(0)=0$, we can consider $\xi(\tilde B_n)\in \mathbb H_{M_0,M}$ by induction. So the denominator has a lower bound $\nu$.

$$
\begin{aligned}
\mathcal V(\tilde B_n)
= {} & \int_0^\tau
\frac{
\left|\sum_{i=1}^n Z_i^c
\exp\left\{\int_0^{s^-}D_i(u)\,d\xi(\tilde B_n)(u)\right\}
\,dN_i^1(s)\right|
}{
n\left|A_n(s,\xi(\tilde B_n)(\cdot))\right|
}
\\
\leq {} & \int_0^\tau
\frac{
\sum_{i=1}^n
\left|Z_i^c
\exp\left\{\int_0^{s^-}D_i(u)\,d\xi(\tilde B_n)(u)\right\}\right|
\,|dN_i^1|(s)
}{
n\left|A_n(s,\xi(\tilde B_n)(\cdot))\right|
}
\\
\leq {} & \frac1n\sum_{i=1}^n\int_0^\tau
\frac{
2M_{\max}
\left|\exp\left\{\int_0^{s^-}D_i(u)\,d\xi(\tilde B_n)(u)\right\}\right|
\,dN_i^1(s)
}{\nu}.
\end{aligned}
$$

Without loss of generality, assume $||D_i||_{\infty}\leq 1$. Then we have

$$
\begin{aligned}
\left|\int_0^{s^-}D_i(u)\,d\xi(\tilde B_n)(u)\right|
&\leq
\|D_i\|_\infty\|\xi(\tilde B_n)\|_\infty
+
\left|\int_0^{s^-}\xi(\tilde B_n)(u^-)\,dD_i(u)\right|
\\
&\leq M_0(1+K_D)
\end{aligned}
$$

So $\mathcal V(\tilde B_n)$ has a uniform upper bound:

$$
\mathcal V(\tilde B_n)
\leq
\frac{2M_{\max}e^{(1+K_D)M_0}}{\nu}
\frac1n\sum_{i=1}^n N_i^1(\tau)
\leq
\frac{2M_{\max}e^{(1+K_D)M_0}}{\nu}.
$$

Therefore, $\tilde B_n$ has uniformly bounded total variation over $n$. By $\tilde B_n(0)=0$, $\|\tilde B_n\|_\infty$ has the same upper bound as follows. So we can apply Helly's selection theorem by the following statement.

$$
|\tilde B_n(t)-\tilde B_n(0)|
\leq \mathcal V(\tilde B_n)
\leq
\frac{2M_{\max}e^{(1+K_D)M_0}}{\nu}
$$

\paragraph{Step 2} We prove
$$
\sup_{s\in[0,\tau],\,H\in\mathbb H_{M_0,M}}
|\Upsilon_n(H(\cdot),s)-\Upsilon(H(\cdot),s)|
\rightarrow 0,\qquad \text{a.s.}
$$
We separate it into two parts, $\|\Upsilon_n-\bar\Upsilon_n\|_\infty$ and $\|\bar\Upsilon_n-\Upsilon\|_\infty$. First, we notice that
$$
\begin{aligned}
\bar\Upsilon_n(H(\cdot),t)
&=\sum_{i=1}^n\int_0^t
\frac{
Z_i^c\exp\left\{\int_0^{s^-}D_i(u)\,dH(u)\right\}
\,dN_i^1(s)
}{
n\,a(s,H(\cdot))
}
\\
&=\sum_{i=1}^n
\frac{
Z_i^c\exp\left\{\int_0^{T_i^-}D_i(u)\,dH(u)\right\}
N_i^1(t)
}{
n\,a(T_i,H(\cdot))
}
\\
&=\frac1n\sum_{i=1}^n\psi_{H,t}(O_i)
=\mathbb P_n(\psi_{H,t}),
\end{aligned}
$$
where
$$
\psi_{H,t}(O)
=
\frac{
Z^c\exp\left\{\int_0^{T^-}D(s)\,dH(s)\right\}N^1(t)
}{
a(T,H(\cdot))
}.
$$

Known that $\Upsilon(H(\cdot),t)=\mathbb P(\psi_{H,t})$,
it suffices to show that $\left\{
\psi_{H,t}:H\in\mathbb H_{M_0,M},\ t\in[0,\tau]
\right\}$ is Glivenko--Cantelli. Since \(H\in\mathbb H_{M_0,M}\) has uniformly bounded total variation and \(D(\cdot)\) is bounded, the class
\[
\left\{
Z^c
\exp\left\{
\int_0^{T^-}D(s)\,dH(s)
\right\}
N^1(t)
:
H\in\mathbb H_{M_0,M},\ t\in[0,\tau]
\right\}
\]
has uniformly bounded total variation and envelope. Hence, by \citet[Example~19.11]{vandervaart2000}, it is Glivenko--Cantelli.

Moreover, since
\[
\inf_{s\in[0,\tau],\,H\in\mathbb H_{M_0,M}}
|a(s,H)|
> \nu>0,
\]
we have
\[
\frac{1}{|a(T,H)|}
\leq
\frac{1}{\nu}.
\]
Thus \(1/a(T,H)\) is uniformly bounded and measurable. By
\citet[Example~2.10.10]{vandervaartwellner1996}, a
Glivenko--Cantelli class multiplied by a uniformly bounded
measurable function remains Glivenko--Cantelli. Therefore,
\[
\left\{
\psi_{H,t}
:
H\in\mathbb H_{M_0,M},\,
t\in[0,\tau]
\right\}
\]
is Glivenko--Cantelli. Consequently,
\[
\sup_{H\in\mathbb H_{M_0,M},\,t\in[0,\tau]}
\left|
\bar\Upsilon_n(H,t)-\Upsilon(H,t)
\right|
=
\left\|
(\mathbb P_n-\mathbb P)\psi_{H,t}
\right\|_\infty
\overset{a.s.}{\longrightarrow} 0,
\quad \text{as }n\rightarrow \infty
\]

Moreover, by Lemma~\ref{lemma:IV_uniform} and Assumption~\ref{assumption:IV_uniform_assumption2},
\[
\begin{aligned}
&\sup_{H\in\mathbb H_{M_0,M},\,t\in[0,\tau]}
|\Upsilon_n(H,t)-\bar\Upsilon_n(H,t)|\\
&\leq
\frac{2M_{\max}e^M}{\nu^2}
\sup_{s,H\in\mathbb H_{M_0,M}}
|A_n(s,H)-a(s,H)|
\xrightarrow{\mathrm{a.s.}}0.
\end{aligned}
\]
Together, these establish the almost sure uniform convergence of
$\Upsilon_n$ to $\Upsilon$.

\paragraph{Step 3}

We use contradiction to prove that
$$
\|\tilde B_n-B_D\|_\infty\rightarrow 0.
$$
If this does not hold, there is one sub-sequence $\tilde B_{n_j}$, s.t.
$$
\liminf_{j\rightarrow\infty}
\|\tilde B_{n_j}-B_D\|_\infty>0.
$$
We view this sub-sequence as $\tilde B_n$ in the following analysis. So now $\liminf_{j\rightarrow\infty}
\|\tilde B_{n_j}-B_D\|_\infty>0$.

Given that $\mathcal V(\tilde B_n)\leq M$, $\|\tilde B_n\|_\infty\leq M$, by Helly's selection theorem, there is one sequence $\tilde B_{n_j}(t)\rightarrow\tilde B(t)$ pointwise. We prove $\tilde B(t)$ is continuous first.

Fix any $0\leq t_1<t_2\leq\tau$, for sufficiently large $n_j$,
$$
\begin{aligned}
&|\tilde B(t_2)-\tilde B(t_1)|
\\
\leq{}&
|\tilde B(t_2)-\tilde B_{n_j}(t_2)|
+
|\tilde B_{n_j}(t_2)-\tilde B_{n_j}(t_1)|
+
|\tilde B_{n_j}(t_1)-\tilde B(t_1)|
\\
\leq{}&
o(1)+|\tilde B_{n_j}(t_2)-\tilde B_{n_j}(t_1)|.
\end{aligned}
$$
To obtain further upper bound, we have
$$
\begin{aligned}
&|\tilde B_{n_j}(t_2)-\tilde B_{n_j}(t_1)|
\\
\leq{}&
\left|
\int_{(t_1,t_2]}
\frac{
\sum_{i=1}^{n_j}Z_i^c
\exp\left\{\int_0^{s^-}D_i(u)\,d\xi(\tilde B_{n_j})(u)\right\}
\,dN_i^1(s)
}{
n_j A_{n_j}(s,\xi(\tilde B_{n_j}))
}
\right|
\\
\leq{}&
\frac1{n_j}\sum_{i=1}^{n_j}
\int_{(t_1,t_2]}
\frac{
\left|Z_i^c
\exp\left\{\int_0^{s^-}D_i(u)\,d\xi(\tilde B_{n_j})(u)\right\}\right|
\,|dN_i^1|(s)
}{
|A_{n_j}(s,\xi(\tilde B_{n_j}))|
}.
\end{aligned}
$$
$|Z_i^c|\leq 2M_{\max}$. For sufficiently large $n_j$, $\left|\frac1{A_{n_j}(s,\xi(\tilde B_{n_j}))}\right|\leq\frac1\nu$, and
$$
\begin{aligned}
\left|\int_0^{s^-}D_i(u)\,d\xi(\tilde B_{n_j})(u)\right|
&\leq
\|D_i\|_\infty\|\xi(\tilde B_{n_j})\|_\infty
+
\left|\int_0^{s^-}\xi(\tilde B_{n_j})(u^-)\,dD_i(u)\right|
\\
&\leq M_0+M_0K_D.
\end{aligned}
$$
Then the above term is
$$
\begin{aligned}
\leq{}&
\frac{2M_{\max}\exp\{(1+K_D)M_0\}}{\nu}
\frac1{n_j}\sum_{i=1}^{n_j}
\int_{(t_1,t_2]}dN_i^1(s)
\\
={}&
C^*\frac1{n_j}\sum_{i=1}^{n_j}
[N_i^1(t_2)-N_i^1(t_1)].
\end{aligned}
$$
Define $F(t)=\mathbb P(T\leq t,\Delta=1,V=1)$. Define $F_{n_j}(t)=\frac1{n_j}\sum_{i=1}^{n_j}N_i^1(t)$ and this is Glivenko--Cantelli. Then we have
$$
\|F_{n_j}-F\|_\infty\overset{a.s.}{\longrightarrow} 0,\quad \text{as }j\rightarrow \infty
$$
Thus,
$$
\begin{aligned}
&\left|
\frac1{n_j}\sum_{i=1}^{n_j}
[N_i^1(t_2)-N_i^1(t_1)]
\right|
\\
\leq{}&
\left|
\frac1{n_j}\sum_{i=1}^{n_j}N_i^1(t_2)-F(t_2)
\right|
+
\left|
\frac1{n_j}\sum_{i=1}^{n_j}N_i^1(t_1)-F(t_1)
\right|
+
|F(t_2)-F(t_1)|
\\
\leq{}&
2\|F_{n_j}-F\|_\infty+|F(t_2)-F(t_1)|.
\end{aligned}
$$

By Assumption~\ref{assumption:IV_uniform_assumption4}, we further notice that
$$
\begin{aligned}
F(t_2)-F(t_1)
&=\mathbb E[N^1(t_2)-N^1(t_1)]
\\
&=\mathbb E\left[\int_{(t_1,t_2]}dN^1(s)\right]
\\
&=\int_{t_1}^{t_2}\mathbb E[Y(s)\lambda_1(s)]\,ds
\\
&\leq C_0(t_2-t_1).
\end{aligned}
$$

Therefore, we know that
$$
\begin{aligned}
|\tilde B_{n_j}(t_2)-\tilde B_{n_j}(t_1)|
&\leq
C^*C_0(t_2-t_1)+2C^*\|F_{n_j}-F\|_\infty
\\
|\tilde B(t_2)-\tilde B(t_1)|
&\leq
o(1)+C^*C_0(t_2-t_1)+2C^*\|F_{n_j}-F\|_\infty.
\end{aligned}
$$
Let $n_j\rightarrow\infty$, we obtain the Lipschitz continuity
$$
|\tilde B(t_2)-\tilde B(t_1)|
\leq C^*C_0(t_2-t_1).
$$
Since Helly's selection theorem only gives pointwise convergence, we further prove that the following convergence is uniform.
$$
\tilde B_{n_j}(t)\rightarrow\tilde B(t)
$$
For any given $\epsilon>0$, $\exists\,\delta_1>0$, s.t if $|t_1-t_2|\leq\delta_1$, there is
$$
|\tilde B(t_2)-\tilde B(t_1)|\leq\frac\epsilon3
$$
Also, for sufficiently large $n_j,j>N_1$ and $\delta_2>0$,
$$
\begin{aligned}
2C^*\|F_{n_j}-F\|_\infty&\leq\frac\epsilon6,
\\
C^*C_0|t_2-t_1|&\leq C^*C_0\delta_2\leq\frac\epsilon6,
\qquad |t_2-t_1|\leq\delta_2,
\end{aligned}
$$
so that
$$
|\tilde B_{n_j}(t_2)-\tilde B_{n_j}(t_1)|
\leq\frac\epsilon3
$$
Define $\delta=\min\{\delta_1,\delta_2\}$. We make a grid $0=s_1<s_2<\cdots<s_m=\tau$, so that $s_{i+1}-s_i\leq\delta$.

For these fixed $m$ points, we can obtain a uniform $N_2$, if $j>N_2$, there is
$$
|\tilde B_{n_j}(s_i)-\tilde B(s_i)|
\leq\frac\epsilon3,\qquad i=1,\ldots,m
$$
For any $t\in[0,\tau]$, there is an $i\in\{1,\ldots,m-1\}$, s.t $t\in[s_i,s_{i+1}]$. Define $N=\max\{N_1,N_2\}$. For $n_j,j>N$, we have
$$
\begin{aligned}
|\tilde B_{n_j}(t)-\tilde B(t)|
&\leq
|\tilde B_{n_j}(t)-\tilde B_{n_j}(s_i)|
+
|\tilde B_{n_j}(s_i)-\tilde B(s_i)|
+
|\tilde B(s_i)-\tilde B(t)|
\\
&\leq\epsilon.
\end{aligned}
$$
Thus we proved the uniform convergence of $\tilde B_{n_j}\rightarrow\tilde B$.

We prove that $\tilde B(t)$ is one solution to
$$
B(t)=\Upsilon(B(\cdot),t).
$$

To see this, we focus on $|\tilde B(t)-\Upsilon(\xi(\tilde B)(\cdot),t)|$ and separate it into 4 terms.
$$
\begin{aligned}
&|\tilde B(t)-\Upsilon(\xi(\tilde B)(\cdot),t)|
\\
\leq{}&
|\tilde B(t)-\tilde B_{n_j}(t)|
\\
&+
|\tilde B_{n_j}(t)-\Upsilon_{n_j}(\xi(\tilde B_{n_j})(\cdot),t)|
\\
&+
|\Upsilon_{n_j}(\xi(\tilde B_{n_j})(\cdot),t)
-\Upsilon(\xi(\tilde B_{n_j})(\cdot),t)|
\\
&+
|\Upsilon(\xi(\tilde B_{n_j})(\cdot),t)
-\Upsilon(\xi(\tilde B)(\cdot),t)|.
\end{aligned}
$$
The first term uniformly converges to 0. The second term is exactly 0. For the third term,
$$
\mathcal V(\xi(\tilde B_{n_j}))
\leq\mathcal V(\tilde B_{n_j})\leq M,
\qquad
\|\xi(\tilde B_{n_j})\|_\infty\leq M_0.
$$
So by step 2, we know it uniformly converges to 0.

For the fourth term, we prove a Lipschitz continuity first: For any two $H_1,H_2\in\mathbb H_{M_0,M}$, there is one $C_*$, s.t
$$
|\Upsilon(H_1(\cdot),t)-\Upsilon(H_2(\cdot),t)|
\leq C_*\|H_1-H_2\|_\infty.
$$

First,
$$
\begin{aligned}
&|\Upsilon(H_1(\cdot),t)-\Upsilon(H_2(\cdot),t)|
\\
={}&
\left|
\int_0^t
\left[
\frac{\mathbb E[Z^c\exp\{X_1(s)\}\,dN^1(s)]}
{a(s,H_1(\cdot))}
-
\frac{\mathbb E[Z^c\exp\{X_2(s)\}\,dN^1(s)]}
{a(s,H_2(\cdot))}
\right]
\right|
\\
\leq{}&
\int_0^t
\frac{
\mathbb E[
|Z^c(\exp\{X_1(s)\}-\exp\{X_2(s)\})|
\,dN^1(s)]
}{
|a(s,H_1(\cdot))|
}
\\
&+
\int_0^t
\left|
\frac1{a(s,H_1(\cdot))}
-\frac1{a(s,H_2(\cdot))}
\right|
\mathbb E[|Z^c\exp\{X_2(s)\}|\,dN^1(s)],
\end{aligned}
$$
where $X_k(s)=\int_0^{s^-}D(u)\,dH_k(u)$.

Recall that
$$
\begin{aligned}
\|X_k\|_\infty
\leq\mathcal V(X_k)
&\leq\int_0^\tau|D(u)|\,|dH_k|(u)
\leq\mathcal V(H_k)\leq M
\\
|\exp\{X_1(s)\}-\exp\{X_2(s)\}|
&\leq e^M|X_1(s)-X_2(s)|
\end{aligned}
$$
And
$$
\begin{aligned}
|X_1(s)-X_2(s)|
={}&
\left|\int_0^{s^-}D(u)\,d(H_1-H_2)(u)\right|
\\
\leq{}&
|D(s^-)(H_1-H_2)(s^-)|
+
\left|\int_0^{s^-}(H_1-H_2)(u^-)\,dD(u)\right|
\\
\leq{}&
\|H_1-H_2\|_\infty
+
\|H_1-H_2\|_\infty K_D.
\end{aligned}
$$
So the first term has an upper bound
$$
\begin{aligned}
&\int_0^t
\frac{
\mathbb E[
|Z^c(\exp\{X_1(s)\}-\exp\{X_2(s)\})|
\,dN^1(s)]
}{
|a(s,H_1(\cdot))|
}
\\
&\qquad\leq
\frac{
2M_{\max}e^M(1+K_D)\|H_1-H_2\|_\infty
}{\nu}.
\end{aligned}
$$
For the second term,
$$
\int_0^t
\mathbb E[|Z^c\exp\{X_2(s)\}|\,dN^1(s)]
\leq 2M_{\max}\exp\{M\}
$$
And
$$
\begin{aligned}
&\left|
\frac1{a(s,H_1(\cdot))}
-\frac1{a(s,H_2(\cdot))}
\right|
\\
\leq{}&
\frac1{\nu^2}
|a(s,H_2(\cdot))-a(s,H_1(\cdot))|
\\
\leq{}&
\frac1{\nu^2}
\left|
\mathbb E[
Z^c\check Y^1(s)
(\exp\{X_2(s)\}-\exp\{X_1(s)\})D(s)]
\right|
\\
\leq{}&
\frac{2M_{\max}e^M(1+K_D)}{\nu^2}
\|H_1-H_2\|_\infty.
\end{aligned}
$$
Therefore, we established the Lipschitz continuity.  Since $\xi(\tilde B_{n_j}),\xi(\tilde B)\in \mathbb H_{M_0,M}$ and truncation $\xi$ is 1-Lipschitz, we have
$$
\begin{aligned}
|\Upsilon(\xi(\tilde B_{n_j})(\cdot),t)
-\Upsilon(\xi(\tilde B)(\cdot),t)|
&\leq
C_*\|\xi(\tilde B_{n_j})-\xi(\tilde B)\|_\infty
\\
&\leq
C_*\|\tilde B_{n_j}-\tilde B\|_\infty
\rightarrow 0
\end{aligned}
$$
Let $n_j\rightarrow\infty$. All four terms vanish, and we get $\tilde B(t)=\Upsilon(\xi(\tilde B)(\cdot),t)$.

To eliminate $\xi$, define
$$
\tau^*=\inf\{t\in[0,\tau]:|\tilde B(t)|\geq M_0\}.
$$
If $\tau^*\leq\tau$, then in $t\in[0,\tau^*]$, there is
$$
\tilde B(t)=\Upsilon(\tilde B(\cdot),t).
$$

In the discussion above, we enlarge $|X_1(s)-X_2(s)|$ to $(1+K_D)||H_1-H_2||_{\infty}$. However, we can get another bound $(1+K_D)\sup_{u\in[0,s]}|H_1(u)-H_2(u)|$. Using this bound and repeating the above process, we know that there exists a constant $C<\infty$ such that
$$
|\Upsilon(H_1,t)-\Upsilon(H_2,t)|
\leq C\int_0^t
\sup_{u\in[0,s]}|H_1(u)-H_2(u)|\,dF(s),
$$
where $dF(s)=E[dN^1(s)]$. Since $\xi(\tilde B)=\tilde B$ on $[0,\tau^*]$, let $H_1 =\xi(\tilde B) =  \tilde B, H_2=B_D$, we know
$$
| \tilde B(t) - B_D(t)|\leq C\int_0^t \sup_{u\in [0,s]}|\tilde B(u)-B_D(u)|dF(s)
$$
Define $h(t)=\sup_{u\in [0,t]}|\tilde B(u)-B_D(u)|.$  For $t\in[0,\tau^*]$, we have
$$
h(t)\leq C\int_0^t h(s)\,dF(s)
\leq CC_0\int_0^t h(s)\,ds,
$$
where the second inequality follows from
Assumption~\ref{assumption:IV_uniform_assumption4}.
By Gronwall's inequality, $h(t)=0$ on $[0,\tau^*]$.
Therefore,
$$
\tilde B(t)=B_D(t),\qquad t\in[0,\tau^*].
$$
 But this means $|B_D(\tau^*)|=|\tilde B(\tau^*)|=M_0$. Recall that $\|B_D\|_\infty= M^o<M_0$. So this leads to contradiction. This means $\|\tilde B\|_\infty\leq M_0$. So we can eliminate $\xi$.
$$
\tilde B(t)=\Upsilon(\tilde B(\cdot),t),\qquad t\in[0,\tau]
$$
By Assumption~\ref{assumption:IV_uniform_assumption3}, we know $\tilde B(t)=B_D(t)$. This means $\|\tilde B_{n_j}-B_D\|_\infty\rightarrow 0$ and contradicts the assumption that $\liminf_{n\rightarrow\infty}\|\tilde B_n-B_D\|_\infty>0$. Therefore, step 3 is established.

\paragraph{Step 4}

By step 3, $\|\tilde B_n-B_D\|_\infty\rightarrow 0$. Given that $||B_D ||_{\infty}=M^o<M_0$, for sufficiently large $n$, $\|\tilde B_n\|_\infty\leq M_0$. This means when $n$ is large,
$$
\xi(\tilde B_n)=\tilde B_n.
$$
In other words, $\hat B_D=\tilde B_n$. This establishes
$$
    \sup_{t\in[0,\tau]}|\hat B_D(t)-B_D(t)| \xrightarrow{\text{a.s.}} 0
    \quad \text{as } n\to\infty .
$$

Finally, under suitable smoothness conditions, since
$\hat\theta\xrightarrow{P}\theta$, a Taylor expansion gives
\[
\sup_{t\in[0,\tau]}
|\hat B_D(t;\hat\theta)-\hat B_D(t;\theta)|
=o_P(1).
\]
Combining this with the oracle consistency establishes
\[
\sup_{t\in[0,\tau]}
|\hat B_D(t;\hat\theta)-B_D(t)|
\xrightarrow{P}0.
\]

\subsubsection{Proof of Proposition~\ref{prop:IV_weakly_convergence}}
\label{app:IV_weakly_convergence}
We aim to provide a sum of i.i.d representation of $\sqrt{n}(\hat B_D(t)-B_D(t) )$ heuristically. Start from $\hat B_D(t;\theta)-B_D(t;\theta)$, where the nuisance parameters are fixed at the oracle values. The following $\hat B_D(t)$ denotes $\hat B_D(t;\theta)$.
$$
d \hat B_D(t)-dB_D(t) =\frac{ \sum_i Z_i^c
\exp\{\int_0^{t^-}D_i(s)d\hat B_D(s)\}
[dN_i^1(t)-\check Y_i^1(t)D_i(t)dB_D(t) ]}{nA_n(t,\hat B_D(\cdot))}.
$$
Recall the proof in step 3 of \ref{app:IV_uniformly_consistent}, we have
\[
\sup_{s}
\left|
a(s,H_1)-a(s,H_2)
\right|
\leq
2M_{\max}e^M
\left\{
1+\mathcal V(D)
\right\}
\|H_1-H_2\|_\infty.
\]

Then by Lemma~\ref{lemma:IV_uniform}, we have
$$
\sup_{s} |A_n(s,\hat B_D(\cdot))- a(s,B_D(\cdot))|\leq \sup_{s} |A_n(s,\hat B_D(\cdot))- a(s,\hat B_D(\cdot))| + \sup_{s} |a(s,\hat B_D(\cdot))- a(s,B_D(\cdot))| \overset{a.s}{\longrightarrow} 0 
$$

\iffalse
Under suitable regularity conditions ensuring that
the integrated denominator-replacement error is uniformly
$o_p(n^{-1/2})$, integrating over $[0,t]$ yields
$$
\hat B_D(t)-B_D(t)=\int_0^t \frac{\sum_i Z_i^c
\exp\{\int_0^{s^-}D_i(u)d\hat B_D(u)\}
[dN_i^1(s)-\check Y_i^1(s) D_i(s)dB_D(s) ] }{n a(s,B_D(\cdot))} + o_p(n^{-1/2}).
$$

Integrating over $[0,t]$ yields
$$
\hat B_D(t)-B_D(t)=\int_0^t \frac{\sum_i Z_i^c
\exp\{\int_0^{s^-}D_i(u)d\hat B_D(u)\}
[dN_i^1(s)-\check Y_i^1(s) D_i(s)dB_D(s) ] }{n a(s,B_D(\cdot))} + o_p(\frac{1}{\sqrt {n}}).
$$
\fi

We focus on the standardized error. 
By the strong uniform consistency of $\hat B_D$ and
an application of Slutsky's theorem, under suitable
regularity conditions, we have

$$
\begin{aligned}
V_n(t)&:= \sqrt n (\hat B_D(t)-B_D(t) )\\
& = \frac 1{\sqrt n} \sum_i\int_0^t\frac{ Z_i^c
\exp\{\int_0^{s^-}D_i(u)d\hat B_D(u)\}
[dN_i^1(s)-\check Y_i^1(s) D_i(s)dB_D(s) ] }{A_n(s,\hat B_D(\cdot))}\\
& = \frac 1{\sqrt n} \sum_i\int_0^t\frac{ Z_i^c
\exp\{\int_0^{s^-}D_i(u)d\hat B_D(u)\}
[dN_i^1(s)-\check Y_i^1(s) D_i(s)dB_D(s) ] }{a(s,B_D(\cdot))} + o_p(1)\\
& = \frac 1{\sqrt n}\sum_i \int_0^tH_i(s,\hat B_D)dR_i(s) + o_p(1),\\
\end{aligned}
$$
where
$$
dR_i(t)=dN_i^1(t)-\check Y_i^1(t)D_i(t)dB_D(t),
$$
and
$$
H_i(s,\hat B_D) = \frac{Z_i^c
\exp\{\int_0^{s^-}D_i(u)d\hat B_D(u)\} }{\mathbb E [Z_i^c
\check Y_i^1(s)
\exp\{\int_0^{s^-}D_i(u)dB_D(u)\}D_i(s)]}.
$$

Under suitable regularity conditions ensuring that
the integrated Taylor remainder is uniformly $o_p(1)$,
a Taylor expansion yields

$$
\begin{aligned}
V_n(t)& = \frac 1{\sqrt n}\sum_i\int_0^tH_i(s,\hat B_D)dR_i(s) + o_p(1)\\
& = \frac 1{\sqrt n}\sum_i\int_0^t[H_i(s,B_D) + \dot H_i(s,B_D)[\hat B_D-B_D] + r_{i,n}(s)]dR_i(s) + o_p(1)\\
& = \frac 1{\sqrt n}\sum_i \int_0^t[H_i(s,B_D) + \dot H_i(s,B_D)[\hat B_D-B_D] ]dR_i(s) + o_p(1)\\
\end{aligned}
$$

Define $W_n(t)=\frac 1{\sqrt n}\sum_i\int_0^t H_i(s,B_D)dR_i(s)$. Then do some tricks on the derivative term. Define $G_{1,i}(s^-; B_D)=D_i(s^-)B_D(s^-), G_{2,i}(s^-; B_D)=-\int_0^{s^-}B_D(u^-)dD_i(u)$, and then we have

\[
\begin{aligned}
\int_0^{s^-} D_i(u)dB_D(u)
&= D_i(s^-)B_D(s^-)
-\int_0^{s^-}B_D(u^-)dD_i(u)\\
&=G_{1,i}(s^-; B_D) + G_{2,i}(s^-; B_D).
\end{aligned}
\]

Then
$$
\begin{aligned}
G_{1,i}(s^-;\hat B_D) - G_{1,i}(s^-;B_D) &=D_i(s^-)(\hat B_D(s^-)-B_D(s^-) ) 
 =\frac 1{\sqrt n}D_i(s^-)V_n(s^-)\\
 G_{2,i}(s^-;\hat B_D) - G_{2,i}(s^-;B_D) &=-\int_0^{s^-}[\hat B_D(u^-) - B_D(u^-)] dD_i(u)
 =-\frac 1{\sqrt n}\int_0^{s^-}V_n(u^-)dD_i(u)\\
\end{aligned}
$$
By chain rule,
$$
H_i(s;\hat B_D) - H_i(s;B_D)=\frac{\partial H_i(s;B_D)}{\partial G_{1,i}(s^-)}(\frac{D_i(s^-)V_n(s^-)}{\sqrt n} ) - \frac{\partial H_i(s;B_D)}{\partial G_{2,i}(s^-)}(\frac{\int_0^{s^-}V_n(u^-)dD_i(u)}{\sqrt n}) +r_{i,n}(s)
$$
Plugging in this result and applying Fubini theorem yields
$$
\begin{aligned}
V_n(t) &= W_n(t) + \frac 1n \int_0^t V_n(s^-)\sum_i \frac{\partial H_i(s;B_D)}{\partial G_{1,i}(s^-) }D_i(s^-) dR_i(s) \\
&\qquad \qquad - \frac 1n\int_0^t \sum_i \frac{\partial H_i(s;B_D) }{\partial G_{2,i}(s^-) }(\int_0^{s^-}V_n(u^-)dD_i(u)) dR_i(s) +o_p(1)\\
&= W_n(t) + \frac 1n \int_0^t V_n(s^-)\sum_i \frac{\partial H_i(s;B_D) }{\partial G_{1,i}(s^-) }D_i(s^-) dR_i(s)\\
&\qquad \qquad - \frac 1n\int_0^t V_n(s^-)\sum_i dD_i(s) (\int_{s^+}^{t}\frac{\partial H_i(u;B_D) }{\partial G_{2,i}(u^-) }dR_i(u) )+ o_p(1)\\
&= W_n(t) + \frac 1n \int_0^t V_n(s^-)\sum_i [\frac{\partial H_i(s;B_D)}{\partial G_{1,i}(s^-) }D_i(s^-) dR_i(s)-dD_i(s) \int_{s^+}^{t}\frac{\partial H_i(u;B_D) }{\partial G_{2,i}(u^-) }dR_i(u) ]+ o_p(1).\\
& = W_n(t) + \int_0^t V_n(s^-)\frac 1n \sum_{i=1}^n dK_i(s;t) + o_p(1),
\end{aligned}
$$
where
$$
dK_i(s;t)=\frac{\partial H_i(s;B_D)}{\partial G_{1,i}(s^-) }D_i(s^-) dR_i(s)-dD_i(s) \int_{s^+}^{t}\frac{\partial H_i(u;B_D) }{\partial G_{2,i}(u^-) }dR_i(u).
$$

 This yields a Volterra equation. Under the
regularity conditions specified below, it can be written as
$$
V_n(t)= \int_0^t\mathcal F_n(s,t)dW_n(s) + o_p(1)
$$
where we define $d\bar K_n(s;t) = \frac 1n \sum_i dK_i(s;t)$, and
$$
\mathcal F_n(s,t) = 1 + \sum_{m=1}^\infty \int_{s<u_1<\cdots<u_m\leq t} d\bar K_n(u_1;u_2)\cdots d\bar K_n(u_m;t),
$$
where $u_{m+1}=t$ is understood in the $m$-th term.

A sufficient condition is that there is a finite, non-decreasing function $\Lambda_n$, s.t uniformly over $t$, there is

$$
|d\bar K_n(s;t) |\leq d\Lambda_n(s), \quad \Lambda_n(\tau)<\infty.
$$

Then the $m$-th term can be bounded as follows:

$$
|\int_{s<u_1<\cdots<u_m\leq t} d\bar K_n(u_1;u_2)\cdots d\bar K_n(u_m;t) |\leq \frac{|\Lambda_n(t)-\Lambda_n(s)|^m }{m!}
$$

So $\mathcal F_n(s,t)$ absolutely and uniformly converges, namely,

$$
|\mathcal F_{n}(s,t)|\leq \exp\{\Lambda_n(t)-\Lambda_n(s)\}.
$$

We notice that 

$$
H_i(s,B_D) = \frac{Z_i^c
\exp\{\int_0^{s^-}D_i(u)d B_D(u)\} }{\mathbb E [Z_i^c
\check Y_i^1(s)
\exp\{\int_0^{s^-}D_i(u)dB_D(u)\}D_i(s)]}= \frac{Z_i^c
\exp\{G_{1,i}(s^-;B_D)+G_{2,i}(s^-;B_D) \} }{\mathbb E [Z_i^c
\check Y_i^1(s)
\exp\{\int_0^{s^-}D_i(u)dB_D(u)\}D_i(s)]}.
$$

So we obtain 

$$
\frac{\partial H_i(s;B_D)}{\partial G_{1,i}(s^-) } = \frac{\partial H_i(s;B_D) }{\partial G_{2,i}(s^-) }=H_i(s;B_D),
$$
and
$$
dK_i(s;t)=H_i(s;B_D)D_i(s^-) dR_i(s)-dD_i(s) \int_{s^+}^{t}H_i(u;B_D)dR_i(u).
$$

We take the total variation to obtain upper bound.
$$
\begin{aligned}
    |dK_i(s;t) | &\leq |H_i(s;B_D)||D_i(s^-)| |dR_i|(s)+|dD_i|(s) \int_{s^+}^{t}|H_i(u;B_D)||dR_i|(u)\\
    &\leq |H_i(s;B_D)||D_i(s^-)| |dR_i|(s)+|dD_i|(s) \int_{s^+}^{\tau}|H_i(u;B_D)||dR_i|(u)\\
\end{aligned}
$$

So a suitable $\Lambda_n(t)$ is constructed as $\int_0^t d\Lambda_n(s)$, where
$$
d\Lambda_n(s) = \frac 1n \sum_{i=1}^n\Bigl[|H_i(s;B_D)||D_i(s^-)| |dR_i|(s)+|dD_i|(s) \int_{s^+}^{\tau}|H_i(u;B_D)||dR_i|(u)\Bigr]
$$

Since $D_i$ is binary with a bounded number of jumps,
$N_i^1$ has at most one jump, and $\check Y_i^1$
is uniformly bounded, under the preceding
bounded-variation and positivity conditions,
we have
\[
\begin{aligned}
\Lambda_n(\tau)
&\leq \frac1n\sum_{i=1}^n
\|H_i(\cdot;B_D)\|_\infty
\mathcal V(R_i)
\left\{
\|D_i\|_\infty+\mathcal V(D_i)
\right\}\\
&=O(1).
\end{aligned}
\]
Consequently, $\Lambda_n(\tau)=O_p(1)$,
and the Volterra resolvent is stochastically bounded. Define 
$$
K_i(r;t)=\int_{(0,r]}dK_i(s;t),
\qquad
K_0(r;t)=\mathbb E[K_i(r;t)],
$$
and
\[
\bar K_n(r;t)=\frac1n\sum_{i=1}^n K_i(r;t).
\]
Suppose that
\[
\sup_{0\le r\le t\le \tau}
\left|
\bar K_n(r;t)-K_0(r;t)
\right|
\xrightarrow{p}0.
\]

Suppose that the limiting kernel $K_0(\cdot;t)$
has uniformly bounded variation over $t\in[0,\tau]$.
Under suitable regularity conditions ensuring
the continuity of the Volterra resolvent with
respect to the underlying kernel, the corresponding
resolvent kernels satisfy

\[
\sup_{0\le s\le t\le\tau}
\left|
\mathcal F_n(s,t)-\mathcal F_0(s,t)
\right|
\xrightarrow{p}0,
\]
where \(\mathcal F_n\) and \(\mathcal F_0\) denote the resolvent kernels associated with \(d\bar K_n\) and \(dK_0\), respectively.

Finally, under suitable regularity conditions ensuring
that
\[
\sup_{t\in[0,\tau]}
\left|
\int_0^t
\{\mathcal F_n(s,t)-\mathcal F_0(s,t)\}
dW_n(s)
\right|
=o_p(1),
\]
we obtain the following asymptotic linear
representation under oracle nuisance:
$$
\begin{aligned}
V_n(t)&=\int_0^t \mathcal F_0(s,t)dW_n(s) + o_p(1)\\
&=\frac {1}{\sqrt n}\sum_i\int_0^t \mathcal F_0(s,t)H_i(s,B_D)dR_i(s) + o_p(1)\\
&=\frac {1}{\sqrt n}\sum_i\epsilon_i^B(t) + o_p(1),\\
\end{aligned}
$$
where
$$
\epsilon_i^B(t) = \int_0^t \mathcal F_0(s,t)H_i(s;B_D)(dN_i^1(s)- \check Y_i^1(s)D_i(s)dB_D(s) ).
$$

Now calculate full influence function of $\hat B_D(t;\hat \theta)$:
$$
\begin{aligned}
&\sqrt n(\hat B_D(t;\hat \theta) -B_D(t;\theta) )\\
=&\sqrt n(\hat B_D(t;\hat \theta) - \hat B_D(t;\theta)) + \sqrt n (\hat B_D(t;\theta)-B_D(t;\theta) )\\
=&\sqrt n(\hat B_D(t;\hat \theta) - \hat B_D(t;\theta)) + \frac{1}{\sqrt n}\sum_i \epsilon_i^B(t)+o_p(1) \\
=& (\frac{\partial B_D(t;\theta)}{\partial \theta }+o_p(1)) \sqrt n(\hat \theta - \theta) + \frac{1}{\sqrt n}\sum_i \epsilon_i^B(t)+o_p(1) \\
=& \frac{\partial B_D(t;\theta)}{\partial \theta }\sqrt n(\hat \theta - \theta) + \frac{1}{\sqrt n}\sum_i \epsilon_i^B(t)+o_p(1) \\
=&  \frac{1}{\sqrt n}\sum_i \epsilon_i(t)+o_p(1), \\
\end{aligned}
$$
where
$$
\epsilon_i(t):=\frac{\partial B_D(t;\theta)}{\partial \theta }\epsilon_i^{\theta}(t) +  \epsilon_i^B(t).
$$

With this i.i.d. representation, similar arguments
as in \citet{Lin2000} and \citet{Martinussen2017}
can be adopted to establish the convergence of
$V_n(t)$ in distribution to a mean-zero Gaussian process.

In practice, we recursively update $\frac{\partial B_D(t;\theta)}{\partial \theta }\epsilon_i^{\theta}(t)$ and $\epsilon_i^B(t)$ to calculate the influence function, and construct asymptotic variance by the mean of squared $\hat \epsilon_i(t)$.

\section{Additional details on the simulation studies}

\subsection{Recursive rules for CIFs}
\label{app:Recursive rules for CIFs}

We imposed the stronger joint ignorability condition
\[
\bar D_K
\perp
\left\{
\bigl(\tilde T(\bar d),V(\bar d)\bigr):
\bar d\in\{0,1\}^K
\right\}
\mid X.
\]
Thus, the conditioning on the observed treatment history in the SNCSTM can be
removed. For cause $j$, the structural model implies, for $t\geq t_k$,
\[
\frac{
1-
P\left\{
\tilde T(\bar d_k,\bar 0)\leq t,
V(\bar d_k,\bar 0)=j
\mid
\tilde T(\bar d_k,\bar 0)\geq t_k,X
\right\}
}{
1-
P\left\{
\tilde T(\bar d_{k-1},\bar 0)\leq t,
V(\bar d_{k-1},\bar 0)=j
\mid
\tilde T(\bar d_{k-1},\bar 0)\geq t_k,X
\right\}
}
=
\exp\left\{
-\beta_jd_k\bigl(\min(t,t_{k+1})-t_k\bigr)
\right\}.
\]

By no anticipation, the two treatment paths agree up to $t_k$, so their
cause-specific CIFs at $t_k$ are identical. For the two-cause setting,
\[
\begin{aligned}
&1-
P\left\{
\tilde T(\bar d_k,\bar 0)\leq t,
V(\bar d_k,\bar 0)=j
\mid
\tilde T(\bar d_k,\bar 0)\geq t_k,X
\right\}\\
&\qquad=
\frac{
1-F_{3-j}(t_k;\bar d_{k-1},\bar 0\mid X)
-F_j(t;\bar d_k,\bar 0\mid X)
}{
1-F_1(t_k;\bar d_{k-1},\bar 0\mid X)
-F_2(t_k;\bar d_{k-1},\bar 0\mid X)
},
\end{aligned}
\]
whereas
\[
\begin{aligned}
&1-
P\left\{
\tilde T(\bar d_{k-1},\bar 0)\leq t,
V(\bar d_{k-1},\bar 0)=j
\mid
\tilde T(\bar d_{k-1},\bar 0)\geq t_k,X
\right\}\\
&\qquad=
\frac{
1-F_{3-j}(t_k;\bar d_{k-1},\bar 0\mid X)
-F_j(t;\bar d_{k-1},\bar 0\mid X)
}{
1-F_1(t_k;\bar d_{k-1},\bar 0\mid X)
-F_2(t_k;\bar d_{k-1},\bar 0\mid X)
}.
\end{aligned}
\]

The common denominator therefore cancels, yielding
\[
\frac{
1-F_{3-j}(t_k;\bar d_{k-1},\bar 0\mid X)
-F_j(t;\bar d_k,\bar 0\mid X)
}{
1-F_{3-j}(t_k;\bar d_{k-1},\bar 0\mid X)
-F_j(t;\bar d_{k-1},\bar 0\mid X)
}
=
\exp\left\{
-\beta_jd_k\bigl(\min(t,t_{k+1})-t_k\bigr)
\right\}.
\]

Solving for $F_j(t;\bar d_k,\bar 0\mid X)$ gives
\[
\begin{aligned}
F_j(t;\bar d_k,\bar 0\mid X)
={}&F_j(t;\bar d_{k-1},\bar 0\mid X)\\
&+
\left\{
1-F_{3-j}(t_k;\bar d_{k-1},\bar 0\mid X)
-F_j(t;\bar d_{k-1},\bar 0\mid X)
\right\}\\
&\qquad\times
\left[
1-
\exp\left\{
-\beta_jd_k\bigl(\min(t,t_{k+1})-t_k\bigr)
\right\}
\right].
\end{aligned}
\]
\subsection{Simulation Results}
\label{app:More_simulation_results}

Tables~\ref{tab:sim_07}, \ref{tab:sim_15}, and \ref{tab:sim_60}
summarize the simulation results for event rates of $7\%$, $15\%$,
and $60\%$, respectively. Across these scenarios, both the SI and
IV settings exhibited generally modest absolute relative biases.
Mean estimated standard errors closely matched the Monte Carlo
standard deviations, and empirical coverage remained near the
nominal 95\% level, with mild undercoverage in some scenarios.
Monte Carlo standard deviations decreased as the sample size
increased, indicating improved precision.

\begin{table}[htbp]
\centering
\small
\setlength{\tabcolsep}{4pt}
\caption{Simulation results under the SI and IV settings when the event rate $r = 7\%$. Results are based on 10,000 Monte Carlo replications. \textbf{Perc. Bias}: absolute relative bias. \textbf{MCSD}: Monte Carlo standard deviation of point estimates. \textbf{Ave. SE}: mean estimated standard error. \textbf{95\% CP}: empirical coverage of pointwise 95\% confidence intervals.}
\label{tab:sim_07}
\begin{tabular}{@{}lll cccc @{\hspace{1.5em}} cccc@{}}
\toprule
 & & & \multicolumn{4}{c}{SI setting} & \multicolumn{4}{c}{IV setting} \\
\cmidrule(lr){4-7} \cmidrule(l){8-11}
$n$ & Cause & Metric
 & $\hat B_D(0.5)$ & $\hat B_D(1)$ & $\hat B_D(1.5)$ & $\hat\beta$
 & $\hat B_D(0.5)$ & $\hat B_D(1)$ & $\hat B_D(1.5)$ & $\hat\beta$ \\
\midrule
\multirow{8}{*}{$1{,}000$} & \multirow{4}{*}{1}
   & Perc. Bias & 0.33\% & 0.11\% & 0.11\% & 0.06\% & 0.98\% & 0.69\% & 0.77\% & 0.81\% \\
 & & MCSD       & 0.007 & 0.012 & 0.015 & 0.010 & 0.026 & 0.034 & 0.040 & 0.027 \\
 & & Ave. SE    & 0.007 & 0.011 & 0.015 & 0.010 & 0.026 & 0.034 & 0.040 & 0.027 \\
 & & 95\% CP    & 94.8\% & 95.5\% & 95.4\% & 95.2\% & 95.2\% & 95.3\% & 95.0\% & 94.9\% \\
\addlinespace[2pt]
 & \multirow{4}{*}{2}
   & Perc. Bias & 0.01\% & 0.54\% & 0.42\% & 0.47\% & 2.53\% & 0.85\% & 0.95\% & 0.83\% \\
 & & MCSD       & 0.007 & 0.010 & 0.013 & 0.009 & 0.022 & 0.029 & 0.034 & 0.023 \\
 & & Ave. SE    & 0.006 & 0.010 & 0.013 & 0.008 & 0.022 & 0.029 & 0.034 & 0.023 \\
 & & 95\% CP    & 94.2\% & 95.2\% & 95.2\% & 95.1\% & 95.7\% & 95.5\% & 95.2\% & 95.3\% \\
\midrule
\multirow{8}{*}{$5{,}000$} & \multirow{4}{*}{1}
   & Perc. Bias & 0.09\% & 0.12\% & 0.05\% & 0.04\% & 0.46\% & 0.79\% & 1.16\% & 1.11\% \\
 & & MCSD       & 0.003 & 0.005 & 0.007 & 0.004 & 0.012 & 0.015 & 0.018 & 0.012 \\
 & & Ave. SE    & 0.003 & 0.005 & 0.007 & 0.004 & 0.012 & 0.015 & 0.018 & 0.012 \\
 & & 95\% CP    & 94.8\% & 94.8\% & 94.7\% & 94.7\% & 95.2\% & 94.7\% & 94.5\% & 94.4\% \\
\addlinespace[2pt]
 & \multirow{4}{*}{2}
   & Perc. Bias & 0.07\% & 0.26\% & 0.46\% & 0.44\% & 1.69\% & 0.56\% & 0.21\% & 0.25\% \\
 & & MCSD       & 0.003 & 0.004 & 0.006 & 0.004 & 0.010 & 0.013 & 0.015 & 0.010 \\
 & & Ave. SE    & 0.003 & 0.005 & 0.006 & 0.004 & 0.010 & 0.013 & 0.015 & 0.010 \\
 & & 95\% CP    & 94.9\% & 95.0\% & 94.8\% & 94.9\% & 95.1\% & 94.9\% & 95.1\% & 95.1\% \\
\midrule
\multirow{8}{*}{$10{,}000$} & \multirow{4}{*}{1}
   & Perc. Bias & 0.27\% & 0.27\% & 0.01\% & 0.02\% & 0.71\% & 0.27\% & 0.26\% & 0.27\% \\
 & & MCSD       & 0.002 & 0.004 & 0.005 & 0.003 & 0.008 & 0.011 & 0.012 & 0.008 \\
 & & Ave. SE    & 0.002 & 0.004 & 0.005 & 0.003 & 0.008 & 0.011 & 0.012 & 0.008 \\
 & & 95\% CP    & 95.2\% & 94.6\% & 94.7\% & 94.7\% & 94.8\% & 95.5\% & 95.1\% & 95.1\% \\
\addlinespace[2pt]
 & \multirow{4}{*}{2}
   & Perc. Bias & 0.36\% & 0.56\% & 0.47\% & 0.47\% & 1.77\% & 1.51\% & 1.31\% & 1.31\% \\
 & & MCSD       & 0.002 & 0.003 & 0.004 & 0.003 & 0.007 & 0.009 & 0.010 & 0.007 \\
 & & Ave. SE    & 0.002 & 0.003 & 0.004 & 0.003 & 0.007 & 0.009 & 0.011 & 0.007 \\
 & & 95\% CP    & 94.9\% & 95.3\% & 95.2\% & 95.1\% & 95.1\% & 95.0\% & 95.2\% & 95.2\% \\
\bottomrule
\end{tabular}
\end{table}

\begin{table}[htbp]
\centering
\small
\setlength{\tabcolsep}{4pt}
\caption{Simulation results under the SI and IV settings when the event rate $r = 15\%$. Results are based on 10,000 Monte Carlo replications. \textbf{Perc. Bias}: absolute relative bias. \textbf{MCSD}: Monte Carlo standard deviation of point estimates. \textbf{Ave. SE}: mean estimated standard error. \textbf{95\% CP}: empirical coverage of pointwise 95\% confidence intervals.}
\label{tab:sim_15}
\begin{tabular}{@{}lll cccc @{\hspace{1.5em}} cccc@{}}
\toprule
 & & & \multicolumn{4}{c}{SI setting} & \multicolumn{4}{c}{IV setting} \\
\cmidrule(lr){4-7} \cmidrule(l){8-11}
$n$ & Cause & Metric
 & $\hat B_D(0.5)$ & $\hat B_D(1)$ & $\hat B_D(1.5)$ & $\hat\beta$
 & $\hat B_D(0.5)$ & $\hat B_D(1)$ & $\hat B_D(1.5)$ & $\hat\beta$ \\
\midrule
\multirow{8}{*}{$1{,}000$} & \multirow{4}{*}{1}
   & Perc. Bias & 0.06\% & 0.25\% & 0.25\% & 0.30\% & 0.32\% & 0.11\% & 1.04\% & 0.95\% \\
 & & MCSD       & 0.011 & 0.018 & 0.023 & 0.015 & 0.039 & 0.051 & 0.061 & 0.041 \\
 & & Ave. SE    & 0.011 & 0.017 & 0.023 & 0.015 & 0.038 & 0.051 & 0.061 & 0.041 \\
 & & 95\% CP    & 94.8\% & 95.0\% & 95.0\% & 94.9\% & 95.2\% & 95.5\% & 95.4\% & 95.5\% \\
\addlinespace[2pt]
 & \multirow{4}{*}{2}
   & Perc. Bias & 1.51\% & 0.68\% & 0.40\% & 0.41\% & 2.84\% & 1.51\% & 0.39\% & 0.63\% \\
 & & MCSD       & 0.009 & 0.015 & 0.019 & 0.013 & 0.032 & 0.043 & 0.051 & 0.034 \\
 & & Ave. SE    & 0.009 & 0.015 & 0.020 & 0.013 & 0.032 & 0.042 & 0.051 & 0.034 \\
 & & 95\% CP    & 95.0\% & 95.4\% & 95.3\% & 95.3\% & 95.5\% & 95.5\% & 95.3\% & 95.4\% \\
\midrule
\multirow{8}{*}{$5{,}000$} & \multirow{4}{*}{1}
   & Perc. Bias & 0.24\% & 0.28\% & 0.03\% & 0.01\% & 1.02\% & 0.83\% & 0.74\% & 0.74\% \\
 & & MCSD       & 0.005 & 0.008 & 0.010 & 0.007 & 0.017 & 0.022 & 0.027 & 0.018 \\
 & & Ave. SE    & 0.005 & 0.008 & 0.010 & 0.007 & 0.017 & 0.022 & 0.027 & 0.018 \\
 & & 95\% CP    & 95.3\% & 94.9\% & 95.0\% & 94.9\% & 95.1\% & 95.2\% & 94.9\% & 94.9\% \\
\addlinespace[2pt]
 & \multirow{4}{*}{2}
   & Perc. Bias & 0.28\% & 0.03\% & 0.02\% & 0.02\% & 1.84\% & 1.48\% & 1.27\% & 1.30\% \\
 & & MCSD       & 0.004 & 0.007 & 0.009 & 0.006 & 0.014 & 0.019 & 0.023 & 0.015 \\
 & & Ave. SE    & 0.004 & 0.007 & 0.009 & 0.006 & 0.014 & 0.019 & 0.022 & 0.015 \\
 & & 95\% CP    & 94.8\% & 95.0\% & 94.9\% & 94.8\% & 94.8\% & 94.8\% & 94.8\% & 94.8\% \\
\midrule
\multirow{8}{*}{$10{,}000$} & \multirow{4}{*}{1}
   & Perc. Bias & 0.19\% & 0.25\% & 0.14\% & 0.15\% & 1.08\% & 0.65\% & 0.48\% & 0.50\% \\
 & & MCSD       & 0.003 & 0.006 & 0.007 & 0.005 & 0.012 & 0.016 & 0.019 & 0.013 \\
 & & Ave. SE    & 0.003 & 0.006 & 0.007 & 0.005 & 0.012 & 0.016 & 0.019 & 0.013 \\
 & & 95\% CP    & 95.0\% & 94.9\% & 95.2\% & 95.1\% & 95.2\% & 95.1\% & 95.3\% & 95.2\% \\
\addlinespace[2pt]
 & \multirow{4}{*}{2}
   & Perc. Bias & 0.05\% & 0.14\% & 0.06\% & 0.06\% & 0.14\% & 0.17\% & 0.04\% & 0.02\% \\
 & & MCSD       & 0.003 & 0.005 & 0.006 & 0.004 & 0.010 & 0.013 & 0.016 & 0.011 \\
 & & Ave. SE    & 0.003 & 0.005 & 0.006 & 0.004 & 0.010 & 0.013 & 0.016 & 0.011 \\
 & & 95\% CP    & 94.6\% & 95.0\% & 95.2\% & 95.1\% & 94.8\% & 94.9\% & 94.7\% & 94.5\% \\
\bottomrule
\end{tabular}
\end{table}

\begin{table}[htbp]
\centering
\small
\setlength{\tabcolsep}{4pt}
\caption{Simulation results under the SI and IV settings when the event rate $r = 60\%$. Results are based on 10,000 Monte Carlo replications. \textbf{Perc. Bias}: absolute relative bias. \textbf{MCSD}: Monte Carlo standard deviation of point estimates. \textbf{Ave. SE}: mean estimated standard error. \textbf{95\% CP}: empirical coverage of pointwise 95\% confidence intervals.}
\label{tab:sim_60}
\begin{tabular}{@{}lll cccc @{\hspace{1.5em}} cccc@{}}
\toprule
 & & & \multicolumn{4}{c}{SI setting} & \multicolumn{4}{c}{IV setting} \\
\cmidrule(lr){4-7} \cmidrule(l){8-11}
$n$ & Cause & Metric
 & $\hat B_D(0.5)$ & $\hat B_D(1)$ & $\hat B_D(1.5)$ & $\hat\beta$
 & $\hat B_D(0.5)$ & $\hat B_D(1)$ & $\hat B_D(1.5)$ & $\hat\beta$ \\
\midrule
\multirow{8}{*}{$1{,}000$} & \multirow{4}{*}{1}
   & Perc. Bias & 0.28\% & 0.16\% & 0.22\% & 0.14\% & 0.08\% & 0.31\% & 0.40\% & 0.42\% \\
 & & MCSD       & 0.022 & 0.042 & 0.068 & 0.039 & 0.079 & 0.115 & 0.165 & 0.103 \\
 & & Ave. SE    & 0.022 & 0.042 & 0.066 & 0.039 & 0.078 & 0.114 & 0.165 & 0.103 \\
 & & 95\% CP    & 95.4\% & 95.2\% & 94.6\% & 94.5\% & 95.4\% & 95.7\% & 95.5\% & 95.5\% \\
\addlinespace[2pt]
 & \multirow{4}{*}{2}
   & Perc. Bias & 0.21\% & 0.24\% & 0.54\% & 0.32\% & 1.38\% & 1.14\% & 1.44\% & 1.35\% \\
 & & MCSD       & 0.019 & 0.035 & 0.054 & 0.032 & 0.065 & 0.091 & 0.125 & 0.081 \\
 & & Ave. SE    & 0.019 & 0.035 & 0.053 & 0.032 & 0.064 & 0.092 & 0.126 & 0.081 \\
 & & 95\% CP    & 94.7\% & 94.9\% & 94.6\% & 94.8\% & 95.5\% & 95.8\% & 95.9\% & 96.0\% \\
\midrule
\multirow{8}{*}{$5{,}000$} & \multirow{4}{*}{1}
   & Perc. Bias & 0.24\% & 0.07\% & 0.03\% & 0.03\% & 0.43\% & 0.36\% & 0.20\% & 0.26\% \\
 & & MCSD       & 0.010 & 0.019 & 0.030 & 0.017 & 0.034 & 0.050 & 0.071 & 0.045 \\
 & & Ave. SE    & 0.010 & 0.019 & 0.030 & 0.017 & 0.034 & 0.050 & 0.072 & 0.045 \\
 & & 95\% CP    & 94.8\% & 95.3\% & 95.0\% & 95.2\% & 95.3\% & 95.1\% & 95.3\% & 95.2\% \\
\addlinespace[2pt]
 & \multirow{4}{*}{2}
   & Perc. Bias & 0.32\% & 0.21\% & 0.17\% & 0.18\% & 0.99\% & 0.52\% & 0.22\% & 0.37\% \\
 & & MCSD       & 0.009 & 0.016 & 0.024 & 0.014 & 0.028 & 0.041 & 0.055 & 0.036 \\
 & & Ave. SE    & 0.008 & 0.016 & 0.024 & 0.014 & 0.028 & 0.041 & 0.055 & 0.036 \\
 & & 95\% CP    & 94.8\% & 94.8\% & 95.2\% & 95.1\% & 95.0\% & 94.8\% & 95.1\% & 94.9\% \\
\midrule
\multirow{8}{*}{$10{,}000$} & \multirow{4}{*}{1}
   & Perc. Bias & 0.07\% & 0.12\% & 0.01\% & 0.03\% & 0.16\% & 0.51\% & 0.32\% & 0.35\% \\
 & & MCSD       & 0.007 & 0.013 & 0.021 & 0.012 & 0.024 & 0.036 & 0.051 & 0.032 \\
 & & Ave. SE    & 0.007 & 0.013 & 0.021 & 0.012 & 0.024 & 0.036 & 0.051 & 0.032 \\
 & & 95\% CP    & 95.5\% & 95.4\% & 95.3\% & 95.2\% & 95.4\% & 95.1\% & 94.7\% & 94.9\% \\
\addlinespace[2pt]
 & \multirow{4}{*}{2}
   & Perc. Bias & 0.05\% & 0.13\% & 0.04\% & 0.01\% & 1.26\% & 0.66\% & 1.00\% & 0.99\% \\
 & & MCSD       & 0.006 & 0.011 & 0.017 & 0.010 & 0.020 & 0.029 & 0.039 & 0.025 \\
 & & Ave. SE    & 0.006 & 0.011 & 0.017 & 0.010 & 0.020 & 0.029 & 0.039 & 0.025 \\
 & & 95\% CP    & 95.0\% & 95.0\% & 95.1\% & 95.0\% & 95.0\% & 95.1\% & 95.1\% & 95.1\% \\
\bottomrule
\end{tabular}
\end{table}

\clearpage
\section{Sensitivity analysis for the case study}
\label{app:sensitivity analysis for the case study}

\begin{figure}[h]
    \centering
    \includegraphics[width=0.95\linewidth]{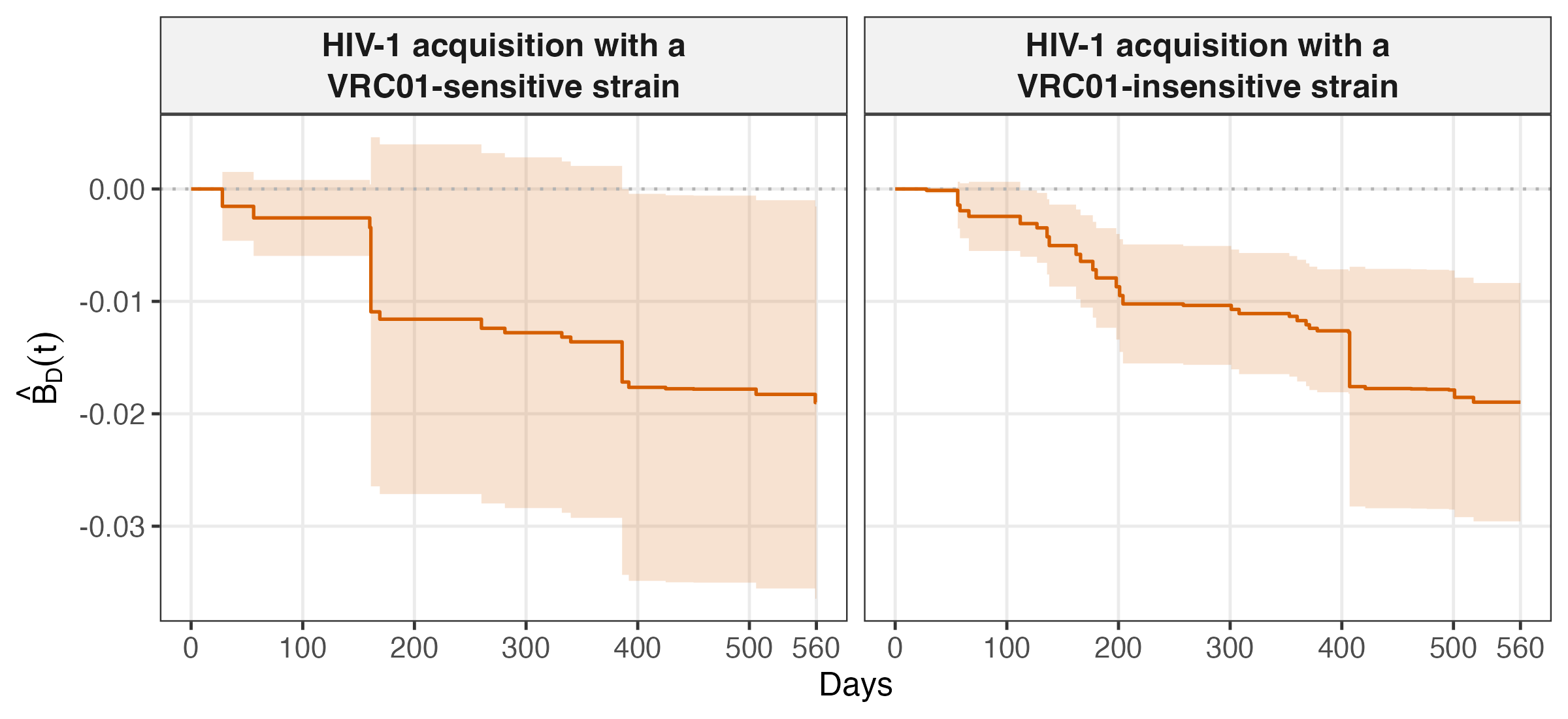}
    \caption{Estimated cumulative effects of oral PrEP on HIV-1 acquisition with a VRC01-sensitive strain (left panel) and VRC01-insensitive strain (right panel). Shaded region indicates pointwise 95\% confidence intervals.}
    \label{fig:B_D_si_ftc}
\end{figure}

\begin{figure}[h]
    \centering
    \includegraphics[width=0.95\linewidth]{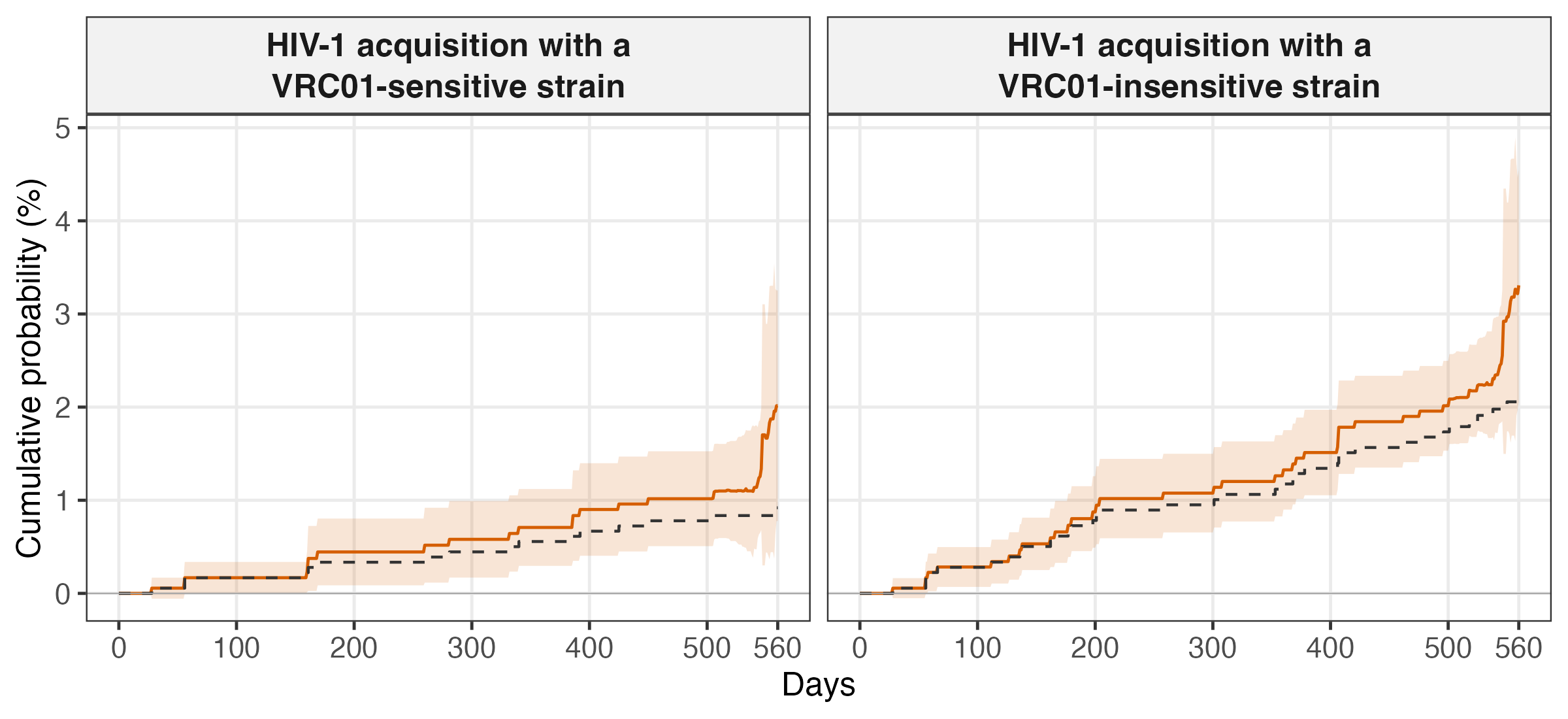}
    \caption{Orange solid lines: estimated counterfactual cause-specific cumulative incidence functions under no oral PrEP uptake. Shaded region indicates pointwise 95\% confidence intervals. Black dashed lines: observed cause-specific cumulative incidence functions estimated using the Fine--Gray model without removing the oral PrEP uptake.}
    \label{fig:cumulative_si_ftc}
\end{figure}

In a sensitivity analysis, we use the last FTC-TP measurement within each interval $[t_k,t_{k+1})$ and define $D_k=1$ if it exceeds the LLOQ of $0.1\,\mathrm{pmol/sample}$, and $D_k=0$ otherwise. 
In a second sensitivity analysis, we define $D_1$ using the FTC-TP measurement on day~0 and, for $k\geq 2$, define $D_k$ using the last FTC-TP measurement in $(t_{k-1},t_k]$. All other analysis settings remain the same. Figure~\ref{fig:B_D_si_ftc} shows $\hat{B}_D(t)$, the estimated cumulative effect of oral PrEP on the acquisition of VRC01-sensitive (left panel) and VRC01-insensitive (right panel) HIV-1 strains, with shaded bands indicating 95\% pointwise confidence intervals. If we further assume a constant additive effect on the subdistribution hazard, that is, $B_D(t)=\beta t$, the estimated effect of oral PrEP is $-3.41 \times 10^{-5}$ per day (95\% CI: $-6.57 \times 10^{-5}$ to $-0.24 \times 10^{-5}$) on the acquisition of VRC01-sensitive strains and $-3.44 \times 10^{-5}$ per day (95\% CI: $-5.36 \times 10^{-5}$ to $-1.52 \times 10^{-5}$) on the acquisition of VRC01-insensitive strains. Both estimates are negative and their confidence intervals do not contain zero. Figure~\ref{fig:cumulative_si_ftc} shows the estimated ``blipped-down'' cause-specific CIFs under no oral PrEP uptake, pooled across the VRC01 arms, alongside the observed cause-specific CIFs estimated from the Fine--Gray model without removing the effect of oral PrEP uptake. For both causes, the ``oral PrEP-free'' cumulative incidences are generally higher than the observed ones, again consistent with a protective effect of oral PrEP.

\end{document}